\documentclass[acmtog, nonacm, authorversion]{acmart}

\usepackage{microtype}
\usepackage{xspace}
\usepackage{bbm}
\usepackage{multirow}
\usepackage{soul}
\usepackage{tabularx}

\copyrightyear{2019}
\acmYear{2019}
\setcopyright{acmlicensed}
\acmConference[Siggraph Asia 2019]{Siggraph Asia 2019}{}{}
\acmBooktitle{}
\acmPrice{}
\acmDOI{}
\acmISBN{}

\newcommand{\eric}[1]{{\color{pink}[Eric: #1]}}

\newcommand{\name}{\textsc{StructureNet}\xspace}
\newcommand{\nary}{\textit{n}-ary\xspace}

\newcommand{\titlecap}[2]{\textbf{#1} #2}

\DeclareMathOperator*{\argmin}{arg\,min}

\newcommand{\eg}{\textit{e.g. }}
\newcommand{\ie}{\textit{i.e. }}

\acmSubmissionID{0240}

\begin{document}

% \title{StructureNet: Learning to Generate 3D Shape Structures with Hierarchical Graph Networks}
\title{\name: Hierarchical Graph Networks for 3D Shape Generation}

\author{Kaichun Mo}
% \email{kaichun@cs.stanford.edu}
\orcid{0000-0003-4365-5050}
\authornote{joint first authors}
\affiliation{%
  \institution{Stanford University}
}

\author{Paul Guerrero}
% \email{paul.guerrero@ucl.ac.uk}
\orcid{}
\authornotemark[1]
\affiliation{%
  \institution{University College London}
}

\author{Li Yi}
% \email{ericyi@stanford.edu}
\orcid{}
\affiliation{%
  \institution{Stanford University}
}

\author{Hao Su}
% \email{haosu@eng.ucsd.edu}
\orcid{}
\affiliation{%
  \institution{University of California, San Diego}
}

\author{Peter Wonka}
% \email{peter.wonka@kaust.edu.sa}
\orcid{}
\affiliation{%
  \institution{KAUST}
}

\author{Niloy J. Mitra}
% \email{n.mitra@cs.ucl.ac.uk}
\orcid{}
\affiliation{%
  \institution{University College London}
}
\affiliation{%
  \institution{Adobe Research}
}

\author{Leonidas J. Guibas}
% \email{guibas@cs.stanford.edu}
\orcid{}
\affiliation{%
  \institution{Stanford University}
}
\affiliation{%
  \institution{Facebook AI Research}
}

\authorsaddresses{Webpage: \href{https://cs.stanford.edu/~kaichun/structurenet/}{https://cs.stanford.edu/$\sim$kaichun/structurenet/}. \\
Emails: \href{mailto:kaichun@cs.stanford.edu}{kaichun@cs.stanford.edu}; \href{mailto:paul.guerrero@ucl.ac.uk}{paul.guerrero@ucl.ac.uk}; \href{mailto:ericyi@stanford.edu}{ericyi@stanford.edu}; \href{mailto:haosu@eng.ucsd.edu}{haosu@eng.ucsd.edu}; \href{mailto:peter.wonka@kaust.edu.sa}{peter.wonka@kaust.edu.sa}; \href{mailto:n.mitra@ucl.ac.uk}{n.mitra@ucl.ac.uk}; 
\href{mailto:guibas@cs.stanford.edu}{guibas@cs.stanford.edu}.
}

\begin{abstract}
The ability to generate novel, diverse, and realistic 3D shapes along with associated part semantics and structure is central to many applications requiring high-quality 3D assets or large volumes of realistic training data. A key challenge towards this goal is how to accommodate diverse shape variations, including both continuous deformations of parts as well as structural or discrete alterations which add to, remove from, or modify the shape constituents and compositional structure. Such object structure can typically be organized into a hierarchy of constituent object parts and relationships, represented as a hierarchy of \nary graphs. We introduce \name, a hierarchical graph network which (i)~can directly encode shapes represented as such \nary graphs; (ii)~can be robustly trained on large and complex shape families; and (iii)~can be used to generate a great diversity of realistic structured shape geometries. Technically, we accomplish this by drawing inspiration from recent advances in graph neural networks to propose an order-invariant encoding of \nary graphs, considering jointly both part geometry and inter-part relations during network training. We extensively evaluate the quality of the learned latent spaces for various shape families and show significant advantages over baseline and competing methods. The learned latent spaces enable several structure-aware geometry processing applications, including shape generation and interpolation, shape editing, or shape structure discovery directly from un-annotated images, point clouds, or partial scans.
\end{abstract}


%
% Keywords. The author(s) should pick words that accurately describe the work being
% presented. Separate the keywords with commas.

\begin{teaserfigure}
   \includegraphics[width=\textwidth]{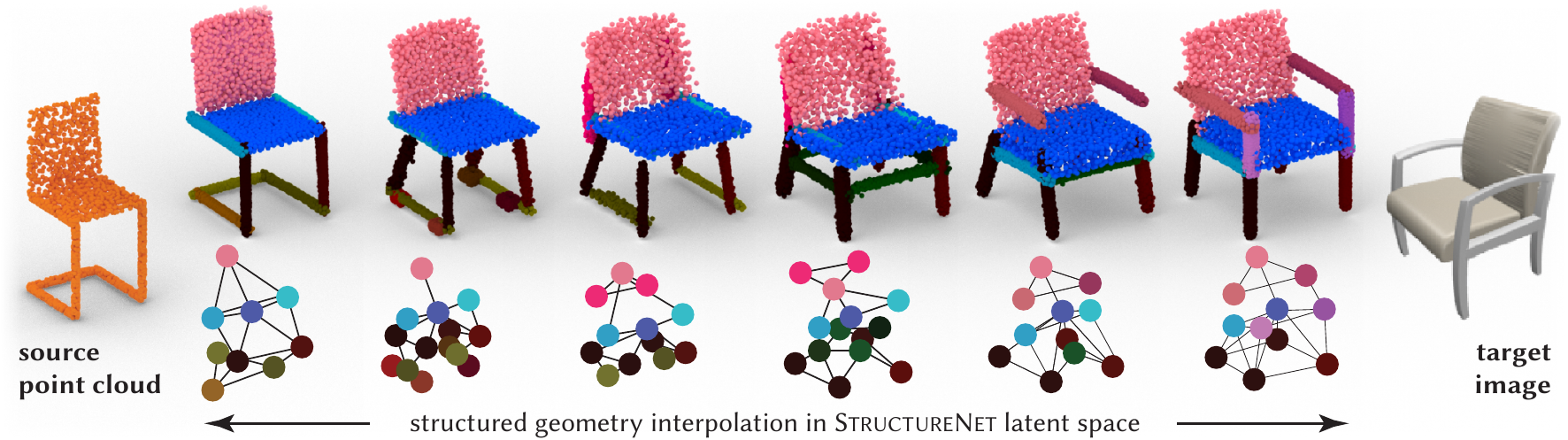}
  \caption{\name is a hierarchical graph network that produces a unified latent space to encode structured models with both continuous geometric and discrete structural variations. In this example, we projected an un-annotated point cloud (left) and un-annotated image (right) into the learned latent space yielding semantically segmented point clouds structured as a hierarchy of graphs.
  %hierarchical graph.
  The shape interpolation in the latent space also produces structured point clouds (top) including their corresponding graphs (bottom). Edges correspond to specific part relationships that are modeled by our approach. For simplicity, here we only show the graphs without the hierarchy. Note how the base of the chair morphs via functionally plausible intermediate
  %intermediate commonly occurring
  configurations, or the chair back transitions from a plain back to a back with arm-rests. 
  }
  \label{fig:teaser}
\end{teaserfigure}

\maketitle

\section{Introduction}

%Given the growing volumes of geometric data (e.g., point clouds, images), discovering latent object structure (i.e., what are the constituent parts and how are they arranged) from raw unannotated input and utilizing them towards structure-aware synthesis and manipulation have received significant attention in the recent years. 

A long-standing problem in shape analysis and synthesis is how to build generative models that support the creation of new, diverse, and realistic shapes. 
A key challenge is to accommodate diverse shape variations, including both continuous deformations of parts as well as structural or discrete alterations which add, remove, or modify the shape substructures present. We seek a continuous latent space that can incorporate all this diversity~\cite{Hinton:1990:MPH:102418.102422} and is able to encode, for example, chairs with or without armrests, chairs having four legs or swivel bases, as well as high or low backs, thin or thick legs, etc. Such a latent space, in turn, enables many non-trivial applications including  generating shapes with both novel structure and geometry, discovering object structures from raw unannotated point clouds or images by `projecting' them to the learned latent space, manipulating shapes in a structure-aware fashion, etc. 

%\emph{interpolate} between structurally different shapes, \emph{discover the structure} of shapes given as un-annotated point clouds, \emph{denoise} shapes, and demonstrate several other applications.

\begin{figure}[t!]
    \centering
    \includegraphics[width=\linewidth]{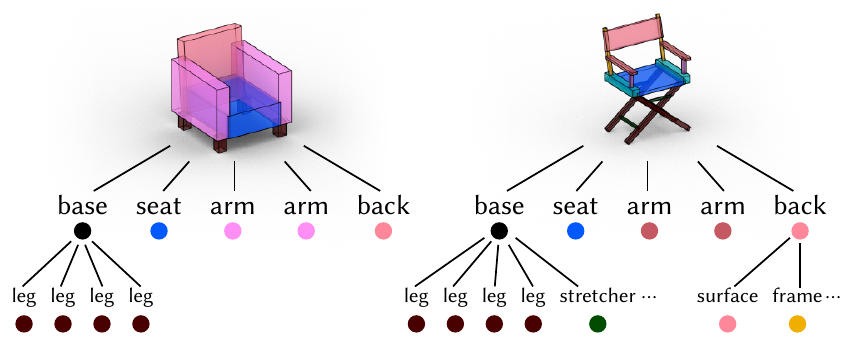}
    \caption{\titlecap{\textit{N}-ary part hierarchies.}{Shape parts can naturally be organized into \nary hierarchies. Here we show the part hierarchies of two shapes as defined by PartNet~\cite{mo2018partnet}. The top row shows oriented bounding boxes of leaf parts and the hierarchy is illustrated below. Hierarchy nodes have the same color as the corresponding part. Note how geometrically dissimilar shapes may have consistent hierarchies. Our shape representation captures this consistency.}}
    \label{fig:nary_illustration}
    \vspace{-3mm}
\end{figure}

One path to this goal is to represent shapes as \textit{structured objects}~\cite{Mitra:2014:SSP} comprising of a collection of parts that are organized according to part-level connectivity and inter-part relationships. Further, these parts can naturally be organized into \textit{hierarchies}~\cite{wang2011symmetry} encoded as \nary trees,
%possibly recursive,
with objects coming from the same shape family sharing similar hierarchies (see Figure~\ref{fig:nary_illustration}). It is important to note, however, that many semantically significant relationships in the geometry of 3D shapes, such as symmetries, can connect distant  nodes in the hierarchy, which may be spatially separated. These horizontal relations pose additional constraints on shape encoders. In this paper, we refer to such hierarchical \nary trees with horizontal connections as hierarchy of \nary graphs, or simply hierarchy of graphs. 
%families sharing same or similar hierarchies.  
Access to large volumes of 3D data (e.g., Turbosquid, 3DWareHouse) has now opened the possibility of learning latent shape spaces from data, aided by significant part annotation efforts within these databases. In the case of ShapeNet~\cite{chang2015shapenet}, both coarse-grained and fine-grained part annotations are available~\cite{yi2016scalable,mo2018partnet}.

A notable work that creates such generative models is the GRASS framework~\cite{grass:li:2017}. Inspired by recursive neural networks introduced in the context of natural language processing for encoding binary trees, GRASS further refines part-level object hierarchies into binary trees, and then recursively utilizes an encoder-decoder network~\cite{socher2011parsing} to build a latent space from which both a hierarchical structure and geometry at the leaves can be decoded. However, because of the binary constraint, GRASS has to additionally search over possible binarizations of the $n$-ary hierarchies found in objects, so that the binarized versions are consistent across objects in the same shape family. While this works nicely on small- to medium-sized datasets, the setup is difficult to train on large to very large shape families (e.g., PartNet~\cite{mo2018partnet}), as the task of finding a canonical binary tree representation becomes increasingly challenging (see Section~\ref{sec:results}).

We introduce \name, a hierarchical graph network %based on graph convolutions~\cite{xx} at each hierarchy level 
that directly encodes more general graphs with parents having a variable number of children and horizontal relationships between siblings.
%at different hierarchy levels.
\name relies on three main innovations: 
Firstly, by directly working with \nary graphs for object structures, we have fundamentally avoided unnecessary data variation that is introduced with binarization, thus can significantly simplify the learning task. 
Secondly, we achieve invariance with respect to part-level sibling ordering at both encoding and decoding time, by using symmetric functions (e.g., max-pooling) during encoding, and solving for a linear assignment problem to establish correspondences during decoding. 
%using hungarian matching
% to establish correspondence 
Finally, we make use of horizontal inter-part \textit{relationship edges} following a novel graph-based message passing protocol. These features enable us to robustly train \name on large to very large shape families so as to effectively  capture both structural and geometric variations. 

Building such a latent space for structured shapes has several advantages that can be exploited by various applications. 
%shapes' constituent parts and their structure has several advantages over generating a single atomic shape. 
First, the structure (i.e., part hierarchy and inter-part relationship) itself is useful for down-stream applications. Editors, for example, can edit, swap, or model parts individually, and make use of structural constraints such as symmetries. %Vision systems, such as robotic manipulators, also benefit from a more structured semantic description of shapes as they have specific actions associated with specific shape parts (e.g., grasping handle of a mug). 
Second, structure is often more consistent inside a shape category than geometry. Chairs, for example, usually have a seat, a backrest and a base at a coarse hierarchy level, even if there is large variability in the geometry of these parts (see Figure~\ref{fig:nary_illustration}). 
%Using this consistent structure makes it easier to learn a meaningful latent space, as we show experimentally. 
Finally, the ability to project raw unstructured shapes (e.g., images, point clouds, partial scans) onto such a latent space automatically induces structures on the raw input (i.e., provides a hierarchical part segmentation, capturing part-level contacts and/or symmetry relationships) and subsequently enables a diverse set of structure-aware manipulations. 

%The constructed latent spaces enables a variety of structure-aware applications, that can be grouped as follows: 
A broad range of applications is enabled by our hierarchical graph networks, which can be grouped as follows:
%These manipulations can be grouped as follows:
(i)~\textit{abstracting} raw inputs including point clouds, images, partial scans to obtain their structure; 
(ii)~\textit{creating} novel shapes in parameterized form or point cloud form based on a set of training shapes; 
(iii)~allowing \textit{structure-aware interpolation} between source and target shapes while displaying both topological and geometric variations; and finally, 
(iv)~\textit{structure-aware object manipulation} to smartly modify a part or replace a part. 
%editing or even supporting `shape algebra.'
%
For example, in Figure~\ref{fig:teaser},
the input source point clouds and target image are first independently abstracted, and then directly interpolated in a latent space that was learned on 5K 
% \kaichun{4871}
chairs from the PartNet dataset.
%
%Such a unified shape space parametrization naturally enables a variety of applications including \emph{generating} shapes with both novel structure and geometry, \emph{interpolating} between structurally different shapes, \emph{discovering the structure} of shapes given as unannotated point clouds or images, or  \emph{denoising} or \textit{completing} raw scans. 
%
% shape synthesis is a long standing problem; challenge is to get both valid topological and geometric structure; although generative models has been very successful in images (faces, indoor scenes, ..), large-scale 3D generative models have remained open
%
% Recently, latent spaces (PCA, AE, CAE) have been used/developed for geometry synthesis but keeping fixed connectivity/topology, but nothing exists to generate both topology/geometry.
%
% in the context of shape synthesis, modeling/synthesis has been proposed in restrictive setting (inverse procedural models, etc.) or using hand-crafted synthesis rules (xxxx). A notable exception is GRASS, but their network architecture makes unambigously capturing realistic structure difficult.
%
% importance of hierarchy and so far we have been missing large scale hierarchically organized data; talk about partnet as new data. Hence we study the problem of constructing a latent space that allows generating new shapes, both with topo and geo variations.
%
% We propose a first learning framework that produces samples provide points/semantically grouped/hierarchically organized. THe key to our approach ... + learning relationship edges.

% Discuss teaser + applications. We evaluate.
In summary, our contributions include: 
\begin{itemize}
    \item introducing an encoding for shape hierarchies of \nary part graphs that is general enough to allow for a consistent hierarchical representations of shapes within a category; 
    \item learning to encode and decode rich geometric relationships (e.g., adjacency, symmetry) between sibling parts, represented as edges in the hierarchical graphs, with graph neural networks to constrain realistic shape generation; 
    \item developing a generative model \name that allows shape synthesis for both box-structures and point clouds with diverse and valid geometrical and structural variations; 
    and \item illustrating the use of \name in both analysis and synthesis tasks, including shape reconstruction, novel shape generation, structure-aware shape interpolation, abstraction of un-annotated point clouds or images, and various shape editing applications.
\end{itemize}

\section{Related Work}
Our work is primarily related to structure-aware shape representations (see \cite{Mitra:2014:SSP}) that go beyond local geometry and depict shapes through the arrangement and relations between shape parts, as well as to 3D shape generative models that aim to model the variations of 3D shapes and to synthesize novel 3D shapes. Finally, our work is inspired by recent developments in neural networks for graph structured data. %We briefly review these works here.

\paragraph{Structure-Aware Shape Representations}
Understanding high-level shape structure such as parts and their relations is a central research topic in shape analysis. Recent work suggests that parts are a natural way to describe shapes~\cite{achlioptas2019shapeglot}. Most existing approaches focus on identifying shape parts \cite{golovinskiy2009consistent, sidi2011unsupervised, hu2012co, huang2011joint, kalogerakis2010learning, xie20143d, makadia2014learning, yi2016scalable, kalogerakis20173d}, or part parameters and relations \cite{muller2006procedural, chaudhuri2011probabilistic, kim2013learning, kalogerakis2012probabilistic, yumer2015semantic, fish2014meta, fish2016structure, sung2017complementme, ganapathi2018parsing}. These approaches are usually restricted in the complexity and variety of part layout they can handle. To be able to process complex structures frequently appearing in the real world, several methods parse object parts in hierarchies \cite{wang2011symmetry, van2013co, yi2017learning, mo2018partnet, yu2019partnet}. We also adapt a hierarchical structure representation for 3D shapes. However, we additionally augment the part hierarchy with `horizontal' relations so as to encode shape structure into a more general \nary graph. Our goal is to model and capture the distribution of such graphs in a shape collection. Structured 3D representations are also widely adopted for scene synthesis applications \cite{liu2014creating, li2019grains, zhao2016relationship}. These approaches usually target generating a scene hierarchy and rely on object retrieval to complete the scene, while we focus on object-level structure synthesis with corresponding part-level geometric details.

\begin{figure}[b!]
    \centering
    \includegraphics[width=\linewidth]{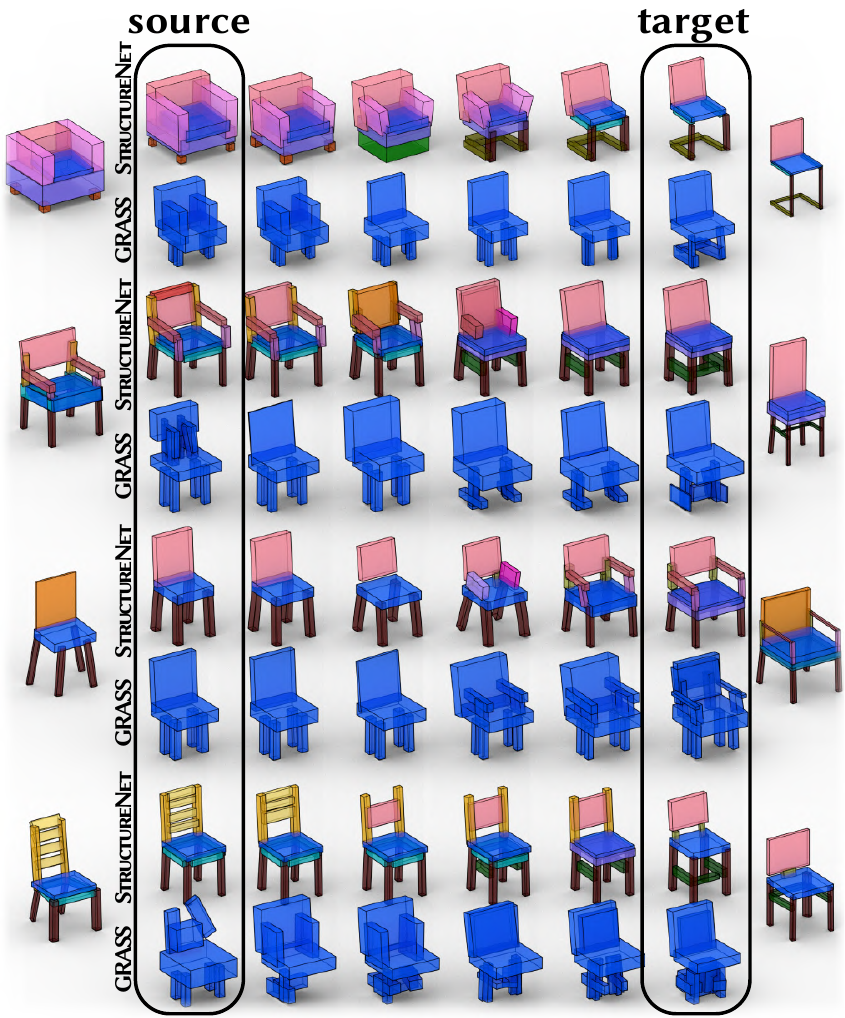}
    \caption{\titlecap{Interpolation compared to GRASS.}{We compare interpolations between several pairs of chairs using \name (colored boxes), and using GRASS (blue boxes). We interpolate between shapes from our test set shown on the left-most and right-most sides, after being reconstructed by both methods (marked as `source' and `target').
    %projected to the latent space of the respective method.
    %The left-most and right-most shapes are the original test set shapes we interpolate between, the corresponding reconstructions are marked as `source' and `target'.
    Our interpolations use a larger number of smaller structural changes to reach the target shape.}}
    \label{fig:interpolation_comparison}
    \vspace{-3mm}
\end{figure}

\paragraph{3D Deep Generative Models}
Recently, deep neural networks have been successfully leveraged to create  generative models for 3D shapes. Wu et al.~\shortcite{wu2016learning} learn to generate 3D shapes in a volumetric representation through a deep belief network. \cite{goodfellow2014generative} also use volumetric representations for 3D shapes but capture the distribution of objects through a generative adversarial network (GAN). To improve the generation quality, researchers have not only explored novel architecture designs for volumetric representations \cite{yan2016perspective, choy20163d, gwak2017weakly}, but also studied various 3D representations,
%for the generation purpose
such as point clouds \cite{fan2017point, achlioptas2017learning, li2018point}, multi-view depth maps \cite{arsalan2017synthesizing}, oct-tree representations \cite{tatarchenko2017octree, wang2018adaptive}, surface meshes \cite{sinha2017surfnet, groueix2018papier}, 
string-based shape synthesis~\cite{Kalojanov2019}, 
etc. These approaches, however,  focus on low-level geometry without considering the overall object structure in the generation process.

An alternate approach is to model structure along with geometry, which not only factorizes the complex distribution of 3D objects to facilitate learning but also makes the generation results more useful for downstream applications. Nash and Williams~\shortcite{nash2017shape} developed a variational auto-encoder (VAE) \cite{kingma2013auto} to learn a latent representations for 3D objects, where they could synthesize new shapes in a part-by-part manner. However, they represent shapes as ordered vectors and require one-to-one dense correspondences among training shapes, which is not easy to obtain for shapes with large topological differences. Wang et al.~\shortcite{wang2018global} first learn to synthesize voxel-based shape structures with parts and labels using a GAN on a set of segmented shapes, and then refine the geometry of each part through an auto-encoder. Wu et al.~\shortcite{wu2018structure} jointly learn and embed the geometry of parts and the pairwise relationship among parts using a VAE, where the geometry and structure features are intertwined in the encoder while disentangled in the decoder, to make the generation process structure-aware. The above approaches do not consider the hierarchical nature of object structures and simply focus on a flat arrangement of parts, making them less applicable to complex structures as defined by Mo et al.~\shortcite{mo2018partnet}. Tian et al.~\shortcite{tian2018learning} generate structured 3D object through executing a series of 3D shape programs, but do not provide a way to sample and generate such shape programs freely.

Most relevant to ours is GRASS~\cite{grass:li:2017}, which learns a distribution of binary symmetry hierarchies of shapes. Novel shape structure can be sampled and generated from the distribution.
%with an adversarial discriminator.
However, the required binary symmetry hierarchy can introduce arbitrary ordering in nodes and make the hierarchy inconsistent across shapes. 
Instead, \name allows direct handling of \nary graphs and further explicitly models relationship between parts in the graph. Hence it can be robustly trained on much larger datasets (see Figure~\ref{fig:interpolation_comparison} for a comparison).

%Our approach instead leverages the consistent \nary shape hierarchies from \cite{mo2018partnet}. We represent the shape structure as a hierarchy of graphs and develop a novel hierarchical graph network under a VAE framework for the generation purpose. %, which enables us to generate complex shape structures with high quality.

\paragraph{Neural Networks for Graph Structured Data}
The shape structure we aim to generate, which is essentially the layout of parts and the relationship among them, can be represented as a hierarchy of \nary graphs. A wide variety of works have explored how to design deep neural networks that can analyze graph structured data. To encode a tree structure, \cite{socher2011parsing, socher2012convolutional, grass:li:2017} use recursive neural network (RvNN) to sequentially collapse edges of the graph. In graph convolutional networks \cite{bruna2013spectral, defferrard2016convolutional, duvenaud2015convolutional, hamilton2017inductive, kipf2017semi, velivckovic2017graph, Xu:2019:GIN}, concepts from mature image CNNs are transferred to generic graphs. This approach has been widely applied in various shape processing methods where shapes are treated as mesh graphs \cite{boscaini2015learning, masci2015geodesic, yi2017syncspeccnn, wang2018dynamic}. With a similar motivation, we design a novel graph encoding framework that combines RvNNs with graph convolutions to process a shape represented as a hierarchy of graphs.

Our framework is also related to various graph neural networks for synthesis purposes. There is a significant amount of work from the natural language processing and program synthesis communities on tree-like graph generation \cite{socher2011parsing, vinyals2015grammar, dyer2016recurrent, maddison2014structured}, but most of these works are restricted to trees and not capable of handling more generic graphs. On the other hand, recent works like \cite{you2018graphrnn, li2018learning} aim at general graph generation and do not specialize their method to any domain. In comparison, our graph generation network is not restricted to trees, but is specialized to our shape representation that encodes both shape geometry and structure, and can thus fully leverage domain-specific knowledge.
\section{Overview}

We represent the hierarchy of shape parts as an \nary tree, where each node is a part or part assembly. Geometric relationships of parts, such as symmetries and adjacencies, are captured by additional edges between siblings of the tree, forming a graph among siblings. See Figure~\ref{fig:graph_hierarchy} for an example. Each node contains information about the geometry of the part, capturing in total the geometry of the shape.
In a given category, such as chairs, shapes tend to have consistent part hierarchies, as shown in the PartNet dataset~\cite{mo2018partnet}. Most chairs, for example, naturally decompose into backrest, seat, and base at the top hierarchy level. The \nary tree in our shape representation can directly capture this hierarchy, giving us a shape representation with a high degree of consistency between shapes of a given category. Section~\ref{sec:hog} describes our shape representation in detail.

This shape representation, including structure and geometry, is mapped into a latent space with a Variational Autoencoder. We introduce \emph{hierarchical graph networks} for the encoder and decoder, which are recursive networks~\cite{socher2011parsing} that perform graph convolutions~\cite{Xu:2019:GIN, kipf2017semi} at each recursion level. They are described in Section~\ref{sec:arch}.

This gives us a rich latent representation of both the structure and the geometry of shapes in a category, that can be used in several applications. In Section~\ref{sec:results} and the Supplementary, we demonstrate shape generation, interpolation, retrieval, editing, as well as the discovery of structure from unannotated images, point clouds, or raw scans.

\section{A Hierarchy of Graphs for Shape Structure}
\label{sec:hog}

\begin{figure}[h!]
    \centering
    \includegraphics[width=\linewidth]{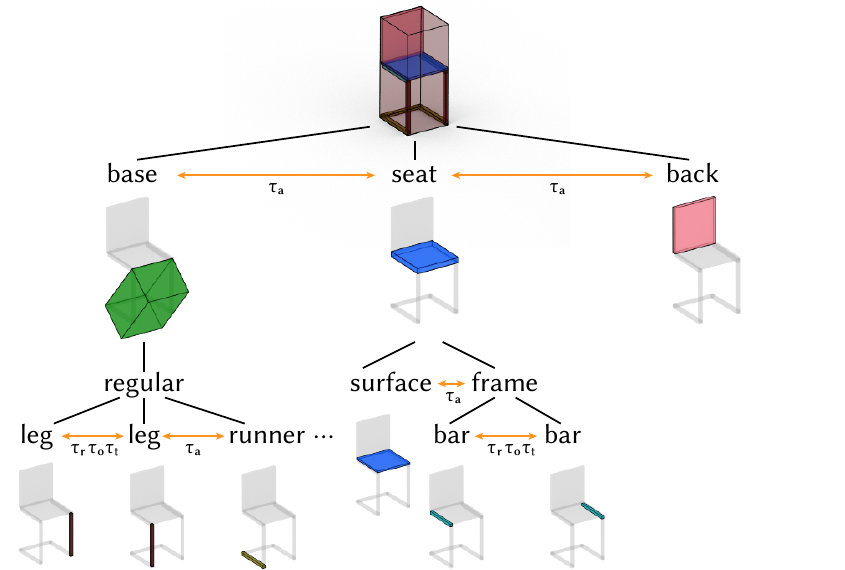}
    \caption{\titlecap{Shape Representation.}{Shapes are represented by their hierarchical decomposition into parts (black edges), with geometric relationships between siblings (orange arrows): adjacency ($\tau_{\mathrm{a}}$), translational symmetry ($\tau_{\mathrm{t}}$), reflective symmetry ($\tau_{\mathrm{r}}$), and rotational symmetry ($\tau_{\mathrm{o}}$). A pair of parts may have multiple relationships of different types. The part geometry can be represented as point clouds or oriented bounding boxes. Here we show the latter, colored by semantic (see Supplementary for a full list of semantics).}}
    \label{fig:graph_hierarchy}
    \vspace{-3mm}
\end{figure}

% rough symbol guidelines:
% capital letter: an object such as a shape or part (or a matrix, but that does not occur so often)
% bold capital letter (mathbf): a set of objects
% lower-case letter: a value or a function
We introduce a shape representation that captures both the geometry and structure of a shape, and is suitable for processing by our hierarchical graph network.
% that we will describe in the next section.
%
We assume that the shapes we work with can be decomposed into a meaningful set of parts. A \textit{shape} $S = (\mathbf{P}, \mathbf{H}, \mathbf{R})$ is then represented by a set of parts $\mathbf{P} = \{P_1, \dots, P_N\}$, describing the \textit{geometry} of the shape, and a \textit{structure} $(\mathbf{H}, \mathbf{R})$ that describes how these parts are organized and related to each other. The structure consists of two superimposed graphs: a \textit{hierarchical decomposition} $\mathbf{H}$ of the shape into parts, and a set of \textit{geometric relationships} $\mathbf{R}$ among the parts. Figure~\ref{fig:graph_hierarchy} illustrates an example.

\paragraph{Part representation} In our experiments, we support the use of two alternative representations for the geometry of a part $P_i$. Either, we represent a part's geometry with its minimum oriented bounding box $B_i = (c_i, q_i, r_i)$, where $c \in \mathbb{R}^3$ are the world coordinates of the box center, $q \in \mathbb{H}$ is the orientation of the box encoded as quaternion, and $r \in \mathbb{R}^3$ is the size of the box; or, we represent a part by a corresponding point cloud $A_i = \{x_1, \dots, x_k\}$, where $x \in \mathbb{R}^3$ are the world coordinates of a point.
In addition to geometry, each part also has a semantic label $l_i$, such as \texttt{back}, \texttt{seat}, or \texttt{base}, that is consistent across shapes of the same category. We refer readers to the supplementary material for the details of the consistent semantic hierarchies we use.

\paragraph{Hierarchical decomposition} The hierarchical decomposition starts with the entire shape as root, which is split into a set of constituent parts, such as \texttt{seat}, \texttt{back}, and \texttt{base} for chairs. These parts are then recursively decomposed into their constituent parts, until reaching the most fine-grained parts at the leafs. We represent this decomposition with a tree $(\mathbf{P}, \mathbf{H})$, where $\mathbf{P}$ are the shape parts and $\mathbf{H} \subset \mathbf{P}^2$ are directed edges from a parent part to all the children it is composed of. Note that a node may have \textit{any} number of children, and the tree need not be balanced, that is, paths to leaves from a node can differ significantly in length.

\paragraph{Geometric relationships} In addition to vertical composition in the hierarchy, parts may also be related `horizontally' by relationships such as \textit{adjacency} and/or \textit{symmetry}. These relationships can be crucial characteristics of shapes. Chairs, for example, often exhibit a reflective symmetry, as well as adjacency between several parts, and failing to take these relationships into account often results in the generation of unrealistic chairs. On the other hand, taking into account potential relationships between all pairs of parts in the hierarchy would require a number of relationships in the order of $\Theta(N^2)$, where $N$ is the total number of parts, and auto-encoding such a large set of relationships accurately poses significant difficulties. In our experiments, we found, however, that encoding all relationships is not necessary, as the most important relationships occur between siblings of the hierarchy (e.g., legs of a chair, drawers in a cabinet, armrests of a couch) and relationships between other parts of the hierarchy are usually less significant or indirectly implied via a chain of relations following the hierarchy tree. Thus, we choose to only capture geometric relationships between siblings in the hierarchy, significantly sparsifying the relationship graph.

We represent these relationships with additional undirected edges $\mathbf{R_i}$ between siblings $\{P_j, P_k\}$ among the children $\mathbf{C_i}$ of a parent part $P_i$.
We denote these edges as $(\{P_j, P_k\}, \tau)$. 
%
% \begin{equation}
% \mathbf{R_i} \subset \{\ (\{P_j, P_k\}, \tau)\ |\ (P_j, P_k, \tau) \in \mathbf{C_i}^2 \times \mathcal{T} \}.
% \end{equation}
%
Each edge has an associated relationship type $\tau$ from a list of possible relationship types $\mathcal{T}$. In our experiments, we use four relationship types: adjacency, reflective symmetry, rotational symmetry, and translational symmetry.
%e do not encode the relationship parameters in the edges, as they can be deduced from the geometry of the two connected parts.
These edges form a graph $(\mathbf{C_i}, \mathbf{R_i})$ among siblings, and our hierarchy effectively becomes a \textit{hierarchy of graphs}, where each shape part is expanded into a graph at the next lower level. We call each of these graphs an \nary graph, to emphasize the \nary nature of the hierarchy over these graphs.

% \begin{itemize}
%     \item shapes are represented by parts that are related by a structure
%     \item parts are represented as bounding boxes and optionally a latent feature vector of the part geometry
%     \item the structure is represented by two concepts: a hierarchical coarse-to-fine decomposition of the shape into its constituent parts, and geometric relationships between these parts. These two concepts are captured by a hierarchy of graphs.
%     %represented by graphs between siblings in the hierarchy.
%     Each part in the hierarchy represents a shape part, which is decomposed at the next lower level into a graph representing the part's constituent parts and their relationships.
% \end{itemize}
% % focus on the most important relationships, those between siblings

%%%%%%%%%%%%%%%%%%%%%%%%%%%%%%%%
%%%%%%%%%%%%%%%%%%%%%%%%%%%%%%%%
\section{Hierarchical Graph Networks}
\label{sec:arch}

% %The shape representation based on a hierarchy of graphs captures both geometry and structure of an object. 
% For any given shape category such as chairs, the constituent shapes form a subspace in our shape representation. However, the exact extent of this subspace is unknown and it is likely too complex to 
% %are believed to form a manifold.
% %\hao{believed to be a manifold? i feel that we just say what we did.}
% %However, the exact shape of this manifold is unknown and it is too complex to
% %there is no direct access to this manifold making it too complex to
% allow for direct manipulations like generation or interpolation. Our goal is to support sampling and navigation of this implicitly-defined 
% subspace, while losing as little geometric and structural fidelity as possible, by

\begin{figure*}[t!]
    \centering
    \includegraphics[width=\linewidth]{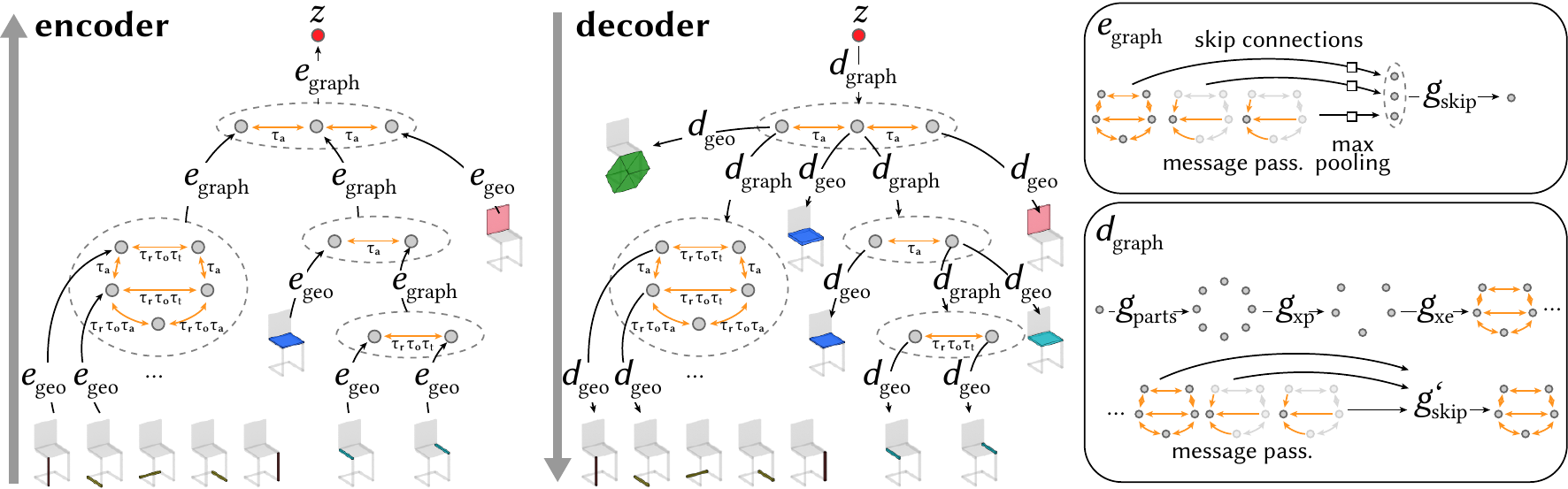}
    \caption{\titlecap{Hierarchical Graph Networks.}{Our variational autoencoder consists of two encoders and two decoders that both operate on our shape representation. The geometry encoder $e_{\text{geo}}$ encodes the geometry of a part into a fixed-length feature vector $f$, illustrated with a gray circle. The graph encoder $e_{\text{graph}}$ encodes the feature vectors of each part in a graph, and the relationships among parts, into a feature vector of the same size using graph convolutions. The graph encoder is applied recursively to obtain a feature vector $z$ that encodes the entire shape. The reverse process is performed by the graph and geometry decoders $d_{\text{graph}}$ and $d_{\text{geo}}$ to reconstruct the shape. The decoder also recovers the geometry of non-leaf nodes.}}
    \label{fig:arch}
    \vspace{-3mm}
\end{figure*}

We propose a novel Variational Autoencoder (VAE)~\cite{kingma2013auto} for our shape representation that can be used for generation, interpolation, and several other applications we will demonstrate in Section~\ref{sec:results}. 
%
% Our goal is to learn a reversible mapping to a lower-dimensional latent feature space with a Variational Autoencoder (VAE), that can be used for generation, interpolation, and several other applications we will demonstrate in Section~\ref{sec:results}. %~\cite{kingma2013auto}.
%
Our VAE consists of an encoder $e$, that maps a shape $S$ to a latent feature vector $z = e(S)$, and a decoder $d$, that maps the feature vector back to a shape, so that approximately $d(e(S)) \approx S$. As described later, we measure the quality of this approximation based on both geometric and structural similarity. We introduce \name or  \emph{hierarchical graph networks} as a new network architecture for the encoder and decoder that can efficiently encode and decode both the geometry and the structure of our shape representation. Figure~\ref{fig:arch} shows the network architecture.

\subsection{Encoder}
\label{sec:encoder}
% The encoder maps a shape represented as a hierarchy of \nary graphs to a latent feature vector $z$.
% To encode the geometry of a shape, we use a \emph{geometry encoder} $e_{\text{geo}}$ for each leaf part separately, giving a feature vector of fixed length that encodes the leaf part's geometry. To encode structure and semantics of a shape, we use a \emph{graph encoder} $e_{\text{graph}}$ that obtains a feature vector of fixed length for a parent part from its child graph. We recursively apply the graph encoder until we get a single feature vector $z$ for the root, as illustrated in Figure~\ref{fig:arch}. The feature size is set to $256$ in our experiments. The recursive nature of this approach is similar to previous work on structure encoding~\cite{grass:li:2017, li2019grains}, which in turn was inspired by natural language processing~\cite{socher2011parsing}. However, since we need to encode \nary graphs in each recursion, that may additionally have a variable number of parts, the architecture of both our encoder and especially our decoder are significantly different.
%Peter Version:
The encoder maps a shape represented as a hierarchy of \nary graphs to a latent feature vector $z$.
We set the dimensionality of the feature space to $256$, i.e. $z \in \mathbb{R}^{256}$. Each (leaf or intermediate) node $i$ in the tree $(\mathbf{P}, \mathbf{H})$ is also mapped to a feature vector $f_i \in \mathbb{R}^{256}$. The code z for a complete shape is simply the feature vector describing the root node $z = f_1$.
The encoder works recursively in a bottom-up manner using two types of encoders (See Fig.~\ref{fig:arch}). First, we compute feature vectors for the leaf nodes using a \emph{geometry encoder}. Then, we encode intermediate nodes using a \emph{graph encoder}.
The recursive nature of this approach is similar to previous work on structure encoding~\cite{grass:li:2017, li2019grains}, which in turn was inspired by natural language processing~\cite{socher2011parsing}. However, since we need to encode \nary graphs, that may additionally have a variable number of parts, the architecture of both our encoder and especially our decoder are significantly different.

% During our experiments, we observed that an increased recursion depth degrades the accuracy of our part geometry regression, analogous to the degradation described by the authors of ResNet~\cite{res}. Unlike a standard feedforward network, a recursive network is not amenable to skip connections, instead we have two strategies to deal with this inaccuracy. First, our more general hierarchy with a variable number of child parts per parent helps keeping the hierarchy shallower than in previous work. Second, we design each network to be as shallow as possible. Due to the recursion, the full function implemented by the network can still be highly complex and non-linear.

\paragraph{Geometry encoder} The geometry encoder $f_i = e_{\text{geo}}(P_i)$ encodes the geometric representation of a leaf node $P_i$ into a feature vector $f_i$. The geometric representation of a part can either be the bounding box $B_i$ of the part or its point cloud $A_i$. We employ specialized geometry encoders for each of these representations. The bounding box encoder consists of a single layer perceptron~(SLP). Since the part encoder is always followed by one or more applications of the graph encoder, a single layer is sufficient. Point clouds $A_i$ are first centered at their mean and uniformly scaled to have a unit bounding sphere. The center and  scale are then encoded using another SLP, while the normalized point cloud is encoded with a PointNet~\cite{qi2017pointnet}. Both of these outputs are merged and encoded by another SLP into the feature vector $f_i$. 

\paragraph{Graph encoder} The child graph of a part $P_i$ is given by $(\mathbf{C_i}, \mathbf{R_i})$, where $\mathbf{C_i}$ are the child parts and $\mathbf{R_i}$ their relationship edges. Each part $P_j \in \mathbf{C_i}$ in the child graph is represented by the concatenation of its feature vector and label $\hat{f}_j = (f_j, l_j)$, while relationship edges in $\mathbf{R_i}$ have as feature only their type $\tau$. Both the part label and edge type are encoded as one-hot vectors. Note that we do \textit{not} store the relationship parameters, such as the axis of a rotational symmetry. The graph encoder $f_i = e_{\text{graph}}(\{\hat{f}_j\ |\ P_j \in \mathbf{C_i}\}, \mathbf{R_i})$ encodes this child graph into a fixed-length feature vector $f_i$. 

%The graph encoder $f_i = e_{\text{graph}}(\{\hat{f}_j\ |\ P_j \in \mathbf{C_i}\}, \mathbf{R_i})$ encodes the child graph  of a part $P_i$ into a fixed-length feature vector $f_i$. Each part $P_j \in \mathbf{C_i}$ in the child graph is represented by the concatenation of its feature vector and label $\hat{f}_j = (f_j, l_j)$. $\mathbf{R_i}$ are the relationship edges of the child graph that have as feature only their type $\tau$. Both the part label and edge type are encoded as one-hot vectors.

The architecture of the graph encoder is inspired by the recent Graph Isomorphism Networks (GIN)~\cite{Xu:2019:GIN} and Dynamic Graph CNNs~\cite{wang2018dynamic}. To encode the child graph, we perform several iterations of message passing along the edges of the graph. In each iteration, a node aggregates features of its neighbors to compute an updated feature vector. Since we also have features for edges, we include them in each message that is passed over an edge. In each iteration $t$, a part's feature vector is updated with, 
\begin{equation}
    f^{(t)}_{j} = \frac{1}{M} \sum_{(\{P_{j}, P_{k}\}, \tau) \in \mathbf{R_i}} h^{(t)}\left(f^{(t-1)}_{j}, f^{(t-1)}_{k}, \tau\right),
\end{equation}
where $M$ is the number of neighbors for $P_j$ and $h^{(t)}$ are SLPs, one for each iteration;  $h^{(t)}$ encodes the message that is passed over an edge, consisting of the source part's feature vector $f^{(t-1)}_{k}$, the target part's feature vector $f^{(t-1)}_{j}$ and the edge type $\tau$. The messages from all neighboring parts are averaged to get the updated feature $f^{(t)}_{j}$.
Iterations start with $f^{(0)}_{j}=\hat{f}_{j}$ and we perform two iterations of message passing for each graph in our experiments. After message passing, the feature vector for the entire graph is computed by max-pooling over all child parts $P_j \in \mathbf{C_i}$:
\begin{equation}
    f^{(t)}_i = \max \{ f^{(t)}_{j} \}.
\end{equation}
Finally, we concatenate the graph feature vectors computed after each iteration, and pass them through another SLP $g$:
\begin{equation}
    f_i = g_{\mathrm{skip}}\left(f^{(0)}_i, f^{(1)}_i, f^{(2)}_i\right). 
\end{equation}
Note that this acts like skip connections for the iterations and allows the network to make use of features from all iterations.

\subsection{Decoder}
% constructed based on exists predictions, hierarchical hungarian matching only for loss (same effect)
The decoder transforms the root feature vector $z$ back into a shape represented as a hierarchy of graphs. It expands nodes in a top-down fashion. In each step, it first performs the reverse operation of the graph encoder, using the \emph{graph decoder} $d_{\text{graph}}$ to transform a latent code $f_i$ into its child graph. The decoder then transforms the resulting feature vector of each child back into the geometry representation of the child with the \emph{geometry decoder} $d_{\text{geo}}$. Unlike in the encoder, we decode the geometry of each part in the hierarchy, not only the leaf parts. This gives additional opportunity for supervision during training in the form of a reconstruction loss on the decoded intermediate geometry, as we will describe in Section~\ref{sec:losses}.
%. \leo{Perhaps say a little more about this, wrt to computing reconstruction losses.} 

\paragraph{Geometry decoder} We have two alternative decoders for the bounding box representation $B_i = d_{\text{geo}}(f_i)$ and the point cloud representation $A_i = d_{\text{geo}}(f_i)$ of a part. Both transform the feature vector of a part back to the part's geometry representation. The bounding box decoder is implemented as a multi-layer perceptron (MLP) with two layers, that transforms a feature vector $f_i$ to a bounding box $B_i = (c_i, q_i, r_i)$. The point cloud decoder obtains a normalized point cloud from the feature vector with a three-layer MLP, and the center and scale of the point cloud using an SLP. We pre-train the geometry encoder and decoder for point clouds, as a separate autoencoder for the point cloud geometry of shape parts. This gives us greatly increased training stability at the cost of a slightly decreased reconstruction accuracy. 

\paragraph{Graph decoder} The graph decoder transforms a parent feature vector $f_i$ back into the child graph $(\{\hat{f}_j\ |\ P_j \in \mathbf{C_i}\}, \mathbf{R_i}) = d_{\text{graph}}(f_i)$, where each child part $P_j \in \mathbf{C_i}$ is represented by a feature vector and its label $\hat{f}_j = (f_j, l_j)$. Since child graphs have a variable number of parts and edges, we always decode a fixed maximum number $n_p$ of child parts and all $n_p^2$ edges between them, together with a binary probability that a predicted part or edge exists in the child graph. Note that parts and their relations are simultaneously decoded. In our experiments, we use a maximum of $10$ parts. Parts and edges that are predicted not to exist in the graph are discarded.

We start by decoding initial feature vectors from the parent feature vector using an SLP $g_{\mathrm{parts}}$:
\begin{equation}
(\tilde{f}_1, \dots \tilde{f}_{n_p}) = g_{\mathrm{parts}}(f_i)
\end{equation}
for the maximum number of child parts $n_p$. 
To predict the existence of parts, we compute 
\begin{equation}
\label{eq:part_existence}
p_j = \sigma(g_{\text{xp}}(\tilde{f}_j)),
\end{equation}
where $p_j$ is the predicted probability that the child part $j$ exists, $\sigma$ is a sigmoid, and $g_{\text{xp}}$ is a single linear layer. Parts with $p_j < 0.5$ are discarded.

To predict the existence of edges, we can proceed similarly. Recall that the graph encoder accumulates information about the graph neighborhood in the feature of each part. Thus, we can recover the edges between a pair of parts based on their pair of feature vectors:
\begin{equation}
\label{eq:edge_existence}
(p_{(j_1, j_2, \tau_1)}, \dots, p_{(j_1, j_2, \tau_{|\mathcal{T}|})}) = \sigma(g_{\text{xe}}(\tilde{f}_{j_1}, \tilde{f}_{j_2})),
\end{equation}
where $p_{(j_1, j_2, \tau)}$ is the predicted probability that an edge between child parts $P_{j_1}$ and $P_{j_2}$ of type $\tau$ exists, and  $|\mathcal{T}|$ is the number of edge types. As $g_{\text{xe}}$, we use a two-layer MLP. Edges are discarded if any of the adjacent parts do not exist, or if $p_{(j_1, j_2, \tau)} < 0.5$.

We perform two iterations of message passing along the predicted edges, analogous to the message passing described in Section~\ref{sec:encoder}, starting with the initial feature vectors $\tilde{f}_j$ and resulting in the final child feature vectors $f_j$. Experimentally, we found the message passing to enable the parts to refine and coordinate their geometry based on the relationships described by the edges, like symmetry or adjacency. As a final step, we decode two additional values from the feature vectors $f_j$: the semantic labels $l_j$ and a probability $p^{\text{leaf}}_j$ that $P_j$ is a leaf part. If a part is predicted to be a leaf part, we do not attempt to predict the existence of children for the node, making it easier for the network to stop the recursion. We found that this helps convergence especially in the early stages of training.

\subsection{Training and Losses}
\label{sec:losses}
We train our VAE on a dataset $\mathbf{S}$ of shapes from a given category. We assume models in the dataset not to have parts with more than $n_p = 10$ children. Each shape is represented as a hierarchy of graphs with known structure. %We will describe our datasets in more detail in Section~\ref{sec:results}. 

Our goal is to train the encoder and decoder of our VAE to perform a reversible mapping of each shape $S$ to a feature vector $z$ in a latent space where manipulations of the shape such as generation and interpolation are easier. To learn this mapping, we use a loss that is composed of three parts: 
\begin{equation}
    \mathcal{L}_{\text{total}} = \mathbb{E}_{S \sim \mathcal{S}}\ [\mathcal{L}_r(S) + \mathcal{L}_{sc}(S) + \beta \mathcal{L}_v(S)],
\end{equation}
where $\mathcal{S}$ is the distribution of shapes in a category and $\mathbb{E}$ denotes the expected value. The \emph{reconstruction loss} $\mathcal{L}_r$ encourages reversibility of the mapping, the \emph{structure consistency loss} $\mathcal{L}_{sc}$ encourages consistency between the reconstructed parts and reconstructed relationship edges, and the traditional \emph{variational regularization} $\mathcal{L}_v$ of VAEs with regularization weight $\beta$ that encourages the manifold of shapes in latent space to be smooth and simple, see~\cite{kingma2013auto} for a description. We empirically set $\beta=0.05$ for our experiments. We now define the reconstruction loss and the consistency loss.

\paragraph{Reconstruction loss} The VAE is encouraged to learn a reversible mapping by training it with a reconstruction loss:
\begin{equation}
    \mathcal{L}_{r}(S) = q\big(S,\ d(e(S))\big),
\end{equation}
where $q$ is a distance metric between reconstructed shapes and ground truth shapes. The distance needs to be designed to provide good gradients to the encoder and decoder.

To compare two shapes $S$ and $S' = d(e(S))$, we first need to establish a correspondence between parts in the two shapes. Here we need to choose between two strategies: we could either encode and reconstruct the order of parts in $S$, or use an order-invariant encoder and establish a correspondence by matching the structure and geometry of the reconstructed shape to the input shape. 
We choose the second option, as we empirically found the order-invariant network produces superior performance (similar conclusion was reached for point cloud encoding~\cite{qi2017pointnet}). We compute a linear assignment of the parts in the two shapes separately for each child graph. Starting at the root, the assignment of parent parts determines which child graphs are matched at the next lower level. This gives an assignment $\mathbf{M} \subset \mathbf{P} \times \mathbf{P}'$ over all parts in the two shapes, where $\mathbf{P}'$ are the reconstructed parts. To train part and edge existence predictions, we include the reconstructed parts that are predicted not to exist in this assignment. Parts are matched based on their geometry representations. We define the geometry difference between parts with point cloud geometry as a squared version of the chamfer distance~\cite{Barrow:1977:Chamfer} between the point clouds:
\begin{equation}
q_{\text{geo}}(P_i, P_j) = q_{\mathrm{chs}}(A_i,\ A_j),
\end{equation}
with the squared version of the chamfer distance~\cite{fan2017point} defined as:
\begin{equation}
\begin{split}
q_{\text{chs}}(A_i, A_j)\ =\ & \frac{1}{|A_i|} \sum_{x_i \in A_i} \min_{x_j \in A_j} \|x_i - x_j\|^2_2\ + \\
&\frac{1}{|A_j|} \sum_{x_j \in A_j} \min_{x_i \in A_i} \|x_j - x_i\|^2_2.
\end{split}
\end{equation}
For the bounding box representation, we cannot directly take the difference of the box parameters, since the orientation and scale of the box representation is ambiguous (e.g., a bounding box can rotated by multiples of 90 degrees about any local axis and re-scaled to give the same bounding box). Instead, we take the chamfer distance between point samples on the boundaries of the two boxes:
\begin{equation}
\label{eq:box_dist}
q_{\text{geo}}(P_i, P_j) = q_{\mathrm{chs}}(T(B_i)\mathbf{U},\ T(B_j)\mathbf{U}),
\end{equation}
where $\mathbf{U}$ is a pre-computed set of samples on the unit cube, and $T(B_i)$ is a 4D transformation matrix that transforms the unit cube to the part's bounding box $B_i$. Since non-uniformly scaling the unit cube with $T(B_i)$ results in a non-uniform point density, $q_{\mathrm{chamfer}}$ weighs the transformed point samples with the area of the face they were sampled from~\cite{tulsiani2017learning}.
Based on the assignment $\mathbf{M}$, the distance $q$ between two shapes is composed of five loss terms, as described next. 

\paragraph{(i)~Geometry loss}
The \emph{geometry loss} measures the distance between the geometry of two parts:
\begin{equation}
    \mathcal{L}_{\text{geo}}(S, S') = \sum_{(P_i, P'_j) \in \mathbf{M}} q_{\text{geo}}(P_i, P'_j).
\end{equation}
Additionally, the geometry of unmatched parts is trained to be all zeros to make the linear assignment more robust.

\paragraph{(ii)~Normal loss}
The geometry loss works well in general, but it is less sensitive to the orientation of small bounding boxes, especially if they have the same size along some of their dimensions, such as a rotation of thin rods about their longest axis. To make the reconstruction of part geometry represented as bounding boxes more sensitive to the orientation of the boxes, we add the \emph{normal loss} that approximates the distance of the reconstructed box normals to the input box normals:
\begin{equation}
    \mathcal{L}_{\text{normal}}(S, S') = \sum_{(P_i, P'_j) \in \mathbf{M}} q_{\mathrm{chs}}(T(q_i)\,\mathbf{N},\ T(q_j)\,\mathbf{N}),
\end{equation}
where $\mathbf{N}$ are the six unique normals of the unit cube. The transformation $T(q_i)$ rotates these normals to the orientation $q_i$ of part $P_i$. As in Eq.~\ref{eq:box_dist}, we use the squared chamfer distance between the predicted and ground truth normals.
%\leo{This should be written more clearly.}

\paragraph{(iii)~Part existence loss}
The \emph{part existence loss} measures how accurately the existence of parts is reconstructed:
\begin{equation}
    \mathcal{L}_{\text{xp}}(S, S') = \sum_{P_j \in \mathbf{P}'} H(p_j,\  \mathbbm{1}_{\mathbf{P}'_{\mathbf{M}}}(P_j)),
\end{equation}
% \begin{equation}
%     \mathcal{L}_{\text{xp}} = \sum_{P_j \in \mathbf{P}'} H(p_j, k),\ \text{with}\ k = 
%     \begin{cases}
%         1 & \text{if}\ \exists P_i\ :\ (P_i, P_j) \in \mathbf{M} \\
%         0 & \text{otherwise},
%     \end{cases}
% \end{equation}
where $p_j$ is the predicted part existence probability defined in Eq.~\ref{eq:part_existence}, $\mathbbm{1}$ is the indicator function, $\mathbf{P}'$ are all parts in $S'$, and $\mathbf{P}'_{\mathbf{M}}$ is the subset that has a match in $S$. The cross entropy $H$ encourages existence for parts that have a match, and non-existence for all other parts.

\paragraph{(iv)~Edge existence loss}
The \emph{edge existence loss} measures how accurately the existence of edges is reconstructed:
\begin{equation}
    \mathcal{L}_{\text{xe}}(S, S') = \sum_{(\{P'_{j1}, P'_{j2}\}, \tau) \in \mathbf{R}'} H(p_{(j1, j2, \tau)},\ \mathbbm{1}_{\mathbf{R}'_{\mathbf{M}}}(\{P'_{j1}, P'_{j2}\}, \tau)),
\end{equation}
%
% \begin{gather}
%     \mathcal{L}_{\text{xp}} = \sum_{(\{P_{j1}, P_{j2}\}, \tau) \in \mathbf{R}'} H(p_{(j1, j2, \tau)}, k),\\
%     \text{with}\ k = 
%     \begin{cases}
%         1 & \text{if}\ \exists (P_{i1}, P_{i2})\ :\ 
%         \begin{aligned}[t]
%             & (P_{i1}, P_{j1}) \in \mathbf{M}\ \land\\
%             & (P_{i2}, P_{j2}) \in \mathbf{M}\ \land\\
%             & (\{P_{i1}, P_{i2}\}, \tau) \in \mathbf{R}
%         \end{aligned}
%          \\
%         0 & \text{otherwise},
%     \end{cases}
% \end{gather}
%
where $p_{(j1, j2, \tau)}$ is the predicted edge existence probability defined in Eq.~\ref{eq:edge_existence}, $\mathbf{R}'$ are all edges in $S'$, and $\mathbf{R}'_{\mathbf{M}}$ the subset that has a match of the same type $\tau$ in $S$. This loss encourages existence for edges that have a match, and non-existence otherwise.

\paragraph{(v)~Semantic loss}
The \emph{semantic loss} is the cross entropy between the reconstructed label probabilities and the input labels, given as one-hot vectors:
\begin{equation}
    \mathcal{L}_{\text{sem}}(S, S') = \sum_{(P_i, P'_j) \in \mathbf{M}} H(l_i, l_j),
\end{equation}
where $l_i$ and $l_j$ are the input and reconstructed labels of the matched parts.

\paragraph{(v)~Leaf loss}
Finally, the \emph{leaf loss} measures the accuracy of the leaf prediction $p^{\text{leaf}}_i$:
\begin{equation}
\mathcal{L}_{\text{leaf}}(S, S') = \sum_{(P_i, P'_j) \in \mathbf{M}} H(p^{\text{leaf}}_j,\ \mathbbm{1}_{\mathbf{P}_{\text{leaf}}}(P_i)),
\end{equation}
%\{P_i | \mathbf{C_i} = \varnothing\}
where $\mathbf{P}_{\text{leaf}}$ is the subset of parts in $S$ that are leafs. 

Finally, the distance $q$ between two shapes is the sum of these five losses:
\begin{equation}
    q(S, S') = \alpha \mathcal{L}_{\text{geo}} + \gamma \mathcal{L}_{\text{normal}} + \mathcal{L}_{\text{xp}} + \mathcal{L}_{\text{xe}} + \lambda \mathcal{L}_{\text{sem}} + \mathcal{L}_{\text{leaf}}.
\end{equation}
Empirically, we set $(\alpha, \gamma, \lambda) = (20, 10, 0.1)$ in all our experiments.

\paragraph{Structure consistency loss} 
Some types of errors in the reconstruction of parts are more severe than others. If a part is not in a geometric relationship with other parts, small reconstruction errors in the position, orientation or scale of the part are often less noticeable. However, if these errors break existing relationships, such as symmetry or adjacency relationships, even small errors can be much more apparent. Hence, we add a loss that encourages the reconstructed parts to be structurally consistent with the reconstructed geometric relationships of a shape -- a self-consistency constraint between the part relations and the part geometries which is an important aspect of our loss design. This can be understood as a constraint violation loss, where the relationships act as constraints. 

\begin{figure}[h!]
    \centering
    \includegraphics[width=\linewidth]{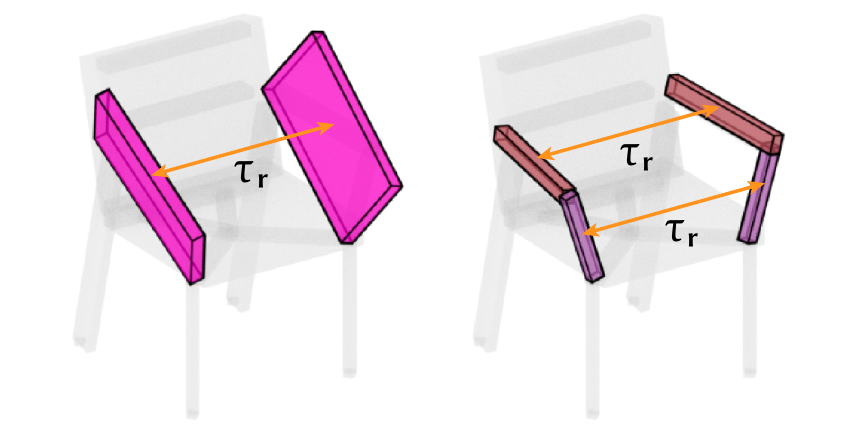}
    \caption{\titlecap{Relationships between subtrees.}{A relationship between two non-leaf parts also holds for their subtrees. The reflective symmetry $\tau_{\text{r}}$ of the parent parts on the left also holds for their children on the right.}}
    \label{fig:edges}
    \vspace{-3mm}
\end{figure}

Given a relationship edge $(P'_i, P'_j, \tau)$ of the reconstructed shape, we quantify how much the geometry of the parts $P'_i$ and $P'_j$ violates the relationship described by the edge. Additionally, a relationship between two parts should also hold for their subtrees. For example, a mirror symmetry between two parents should also constrain their two subtrees to be mirrored in the same way, see Figure~\ref{fig:edges} for an illustration. Hence, we also encourage the entire subtrees $\mathbf{D}_i$ and $\mathbf{D}_j$ of $P'_i$ and $P'_j$ to follow the same relationship.
We first define the point cloud representation of a subtree, including the root, as:
\begin{equation}
D_i = \bigcup_{P_k \in \mathbf{D}_i \cup \{P_i\}} T(B_k)\,\mathbf{U}.
\end{equation}
As in Eq.~\ref{eq:box_dist}, $\mathbf{U}$ is a pre-computed set of samples on the unit cube, and $T(B_k)$ is a 4D transformation matrix that transforms the unit cube to a part's bounding box $B_k$. When representing part geometries with point clouds, we directly use the union of the point clouds.

Next, we  introduce a loss for \emph{symmetries}, and a loss for \emph{adjacencies}.
For symmetries $\mathbf{R}_{\text{sym}}'$, we first compute the closest configuration of $P'_j$ relative to $P'_i$ that would not violate the relationship, and vice-versa for $P'_i$ relative to $P'_j$. We use the distance from that configuration as loss:
\begin{equation}
    \mathcal{L}_{\text{sym}}(S') = \sum_{(\{P'_{i}, P'_{j}\}, \tau) \in \mathbf{R}_{\text{sym}}'} q_{\text{chs}}\Big(D_j,\ \rho_{\tau}(B_i, B_j)D_i\Big),
\end{equation}
where $B_i$ and $B_j$ are the bounding boxes of the two parts and $\rho_{\tau}$ is a function that computes an affine transformation from $B_i$ to the closest configuration of $B_j$ that does not violate the relationship type $\tau$. When representing part geometries with point clouds, we first compute the oriented bounding box of the corresponding point clouds to obtain $B_i$ and $B_j$, respectively.

For adjacency realtions, $\mathbf{R}_{\text{adj}}'$, our loss is the minimum distance $q_{\text{min}}$ between the geometry representations of the leaf parts only:
\begin{equation}
    \mathcal{L}_{\text{adj}}(S') = \sum_{(\{P'_{i}, P'_{j}\}, \tau) \in \mathbf{R}_{\text{adj}}'} q_{\text{min}}(L_i,\ L_j),
\end{equation}
where $L_i$ and $L_j$ are subsets of $D_i$ and $D_j$, representing only the leaf parts of the subtrees.

Finally, the structure consistency loss is the sum of symmetry and adjacency losses: 
\begin{equation}
\label{eq:consistency}
\mathcal{L}_{sc}(S) = \mathcal{L}_{\text{sym}}(S') + \mathcal{L}_{\text{adj}}(S'),   
\end{equation}
where $S' = d(e(S))$ is the reconstructed shape.

% \begin{itemize}
%     \item For the symmetry nodes at the parent levels, we also enforce the computed symmetry transformation to hold for the all children nodes in the subtrees. For example, the symmetric relationships between two chair arms are defined at the coarsest level in the hierarchy. There could be children nodes (e.g. arm vertical bar, arm connector) for each arm. We do this by adding an additional loss for the subtree difference via the computed symmetry at the parent level. Mathematically, compute all points for the entire subtree boxes, then use the parent-level sym-params to compute their chamfer distance.
%     \item Fo
%     r the adjacency nodes at the parent levels, we enforce the leaf-node levels of boxes to be adjacent. For example, the chair back and chair seat has the adjacency relationship at the coarsest level. This means that at each subsequent children levels, the boxes should also hold this adjacency relationship. In our implementation, we enforce such property by adding an additional contact loss (min of chamfer distance) between all leaf-nodes of the two parent nodes.
% \end{itemize}
\section{Experiments}
\label{sec:results}

Accurately capturing structure in a smooth latent space gives us a simple and effective approach to
%Being able to infer structures from data allows us to solve
many shape understanding and synthesis problems. In this section, we present quantitative and qualitative evaluations to demonstrate the effectiveness of our hierarchical graph networks on 5 tasks: shape reconstruction, generation, interpolation, abstraction, and editing. Additional results are available in the Supplementary.

\paragraph{Data Preparation.}
We use PartNet~\cite{mo2018partnet} as the main dataset for all the experiments in the paper.
PartNet provides fine-grained and hierarchical part annotations with consistent semantic labels for 26,671 3D objects from 24 object categories. We use the three largest categories for our experiments: cabinets, chairs, and tables. In the Supplementary, we show results for three additional categories: vases, trashcans can and beds. Also in the Supplementary is a description of the semantic hierarchy in these categories. 
Since we have a maximum number of child parts per parent part $n_p = 10$, we remove shapes that have more than $10$ children in any of their parts. Note that this maximum could also be increased if needed, slightly increasing memory consumption\footnote{with our current settings, the memory consumption is $\sim 1$GB for a batch size of $32$.}.
Additionally, we remove shapes that have unlabeled parts. The remaining $4871$ chairs, $5099$ tables, and $862$ cabinets are divided into 
%We use a subset of models where all the parts are assigned semantic labels and have at most 10 children per parent\footnote{Our method can generalize to accommodate more parts per parent while we increase 10 to a bigger number.}. There are 4,871 chairs, 5,099 tables and 862 cabinets in our data.
training, validation and test sets using the data splits published in the PartNet dataset, which have a ratio of $7:1:2$.

In PartNet, shapes are represented as meshes that are divided into individual parts. Each shape in the dataset is scaled to be contained in the unit sphere. To obtain bounding boxes $B_i$ for each part, we fit an oriented minimum-volume bounding box to the mesh of each part. Point clouds $A_i$ are obtained by uniformly sampling the part's surface with $1000$ points. The part hierarchy $H$ is given explicitly in the dataset. To define geometric relationships $R$ between parts, we find symmetries using the method described by Wang et al.~\cite{Wang:2011:Sym}, and define two parts as adjacent if their smallest distance is below $0.05*\bar{r}$, where $\bar{r}$ is the average bounding sphere radius of the two parts.
On average, our shapes have $16.94$ parts, arranged in a hierarchy of average depth $3.59$, with an average number of $29.93$ relationship edges. Each part has a semantic label chosen from a list of labels specific to each category. The number of different labels ranges from $36$ for cabinets to $82$ for tables.

% tot_num, avg node, avg depth, avg edges
% chair: 4871 19.34099774173681 3.0385957708889344 65.6062410182714
% table: 5099 13.835457932928025 4.126691508138851 48.868405569719556
% cabinet: 862 21.680974477958237 3.5220417633410674 92.38283062645012

% chair: 3456 + 472 + 943 = 4871
% table: 3503 + 537 + 1059 = 5099
% cabinet: 621 + 79 + 162 =  862

%For the box experiments, we detect Oriented Bounding Boxes (OBBs) parametrized by 3 dimensions for box center, 4 dimensions for box quaternion rotation and 3 dimensions for the box scales along principle axes. For the point cloud experiments, we sample 1,000 points for each part as the representation.

%For the graph edges, we detect the pairwise inter-part adjacency and symmetry relationships among the sibling parts of the same parent node at all the hierarchy levels. Two parts are detected to be adjacent if the distance between the closest two points of the part point clouds is smaller than 0.05, relatively to the average part sizes of the two parts. We detect three types of symmetry relationships: translation symmetry, reflective symmetry and rotational symmetry. We compute the symmetric parameters analytically based on the two point clouds and draw a symmetry relationship between two parts if the fitting error is less than 0.05 relatively to the average scale of the two point clouds. We refer the readers to the supplementary materials for the implementation details of symmetry detection.

%\paragraph{Implementation.} We implement \name in PyTorch. 

% \subsection{Quantitative Experiments}

\subsection{Shape Reconstruction}
\label{sec:experiments_recon}
As a first experiment, we measure the reconstruction performance of our hierarchical graph networks to find out how accurately our latent space can represent the shapes in the test set. To get an accurate reconstruction performance, just for the experiments in this section, we train an non-variational autoencoder version of our network.
%, setting the variational regularization weight $\beta$ to zero and generating a single latent vector with the encoder instead of a distribution.
%our network without the variational regularization, setting the weight $\beta$ of the regularization to zero.
We use two groups of errors to measure reconstruction performance. Three reconstruction errors for the geometry, the hierarchy, and the relationship edges, and two structure consistency errors that measure the consistency of the reconstructed geometry with both the reconstructed and the input relationship edges.

The \emph{geometry reconstruction error} $E_{\mathbf{P}}$ is defined analogously to the geometry loss $\mathcal{L}_{\text{geo}}$, except that we use the non-squared chamfer distance. Since our shapes are contained in the unit sphere, this gives us more easily interpretable distance values in $[0, 2]$.

The \emph{hierarchy reconstruction error} $E_{\mathbf{H}}$ counts how many missing or unmatched parts are in the reconstructed hierarchy, using the assignment $\mathbf{M}$ between the input and reconstructed shapes described in Section~\ref{sec:losses}:
\begin{equation}
    E_{\mathbf{H}} = \frac{1}{|\mathbf{P}|} \Big(|\mathbf{P} \setminus \mathbf{P}_{\mathbf{M}}| + |\mathbf{P'} \setminus \mathbf{P'}_{\mathbf{M}}|\Big),
\end{equation}
where $\mathbf{P}$ and $\mathbf{P'}$ are the sets of parts in the input and reconstructed shapes, respectively. $\mathbf{P}_{\mathbf{M}}$ and $\mathbf{P}'_{\mathbf{M}}$ denote the corresponding subsets of matched parts.

The \emph{edge reconstruction error} $E_{\mathbf{R}}$ uses the assignment $\mathbf{M}$ to measure the precision and recall of the reconstructed edges, which we summarize in an error metric defined as one minus the $F_1$ score:
\begin{equation}
    E_{\mathbf{R}} = 1 - \left(2 \frac{e_p*e_r}{e_p+e_r}\right),
\end{equation}
where the precision and recall are defined as $e_p = |\mathbf{R}'_{\mathbf{M}}| / |\mathbf{R}'|$ and $e_r = |\mathbf{R}'_{\mathbf{M}}| / |\mathbf{R}|$, respectively.

The \emph{reconstructed consistency error} $E_{\text{rc}}$ measures the consistency of the reconstructed geometry with the reconstructed relationship edges. It is defined equivalent to $\mathcal{L}_{\text{sc}}$, except that we use the non-squared chamfer distance.

Similarly, the \emph{ground truth consistency error} $E_{\text{gc}}$ measures the consistency of the reconstructed geometry with the input relationship edges.

\begin{table}[t!]
\caption{\titlecap{Reconstruction performance on each shape category.}{We compare the reconstruction performance of \name on six shape categories. See the supplementary for a qualitative evaluation of the categories bed, trashcan, and vase. The first three columns show the box geometry, hierarchy, and edge reconstruction errors, respectively. The consistency of the reconstructed shapes with the reconstructed relationship edges (recon) and the ground truth relationship edges (gt) is shown in the last two columns. Bed comes in last due to severe undersampling of the shape category, with only $54$ training shapes. Our performance is best for chairs, which have a more balanced variety of training shapes than the other categories.}}
\begin{tabular}{@{}lrrrrrr@{}}
\toprule
\multirow{2}{*}{}              & \multicolumn{3}{c}{reconstruction error} & \phantom{abc} & \multicolumn{2}{c}{consistency error} \\
\cmidrule{2-4} \cmidrule{6-7}
                               & \multicolumn{1}{c}{$E_{\mathbf{P}}$}  & \multicolumn{1}{c}{$E_{\mathbf{H}}$} & \multicolumn{1}{c}{$E_{\mathbf{R}}$} && \multicolumn{1}{c}{$E_{\text{rc}}$}          & \multicolumn{1}{c}{$E_{\text{gc}}$}           \\
\midrule
Bed              & 0.069   & 0.609    & 0.518   && 0.019   & 0.032      \\
Cabinet          & 0.066   & 0.461    & 0.386   && 0.021   & 0.027      \\
Chair            & 0.062   & 0.200    & 0.246   && 0.018   & 0.023      \\
Table            & 0.073   & 0.309    & 0.357   && 0.021   & 0.026      \\
Trashcan         & 0.083   & 0.073    & 0.110   && 0.014   & 0.015      \\
Vase             & 0.147   & 0.214    & 0.391   && 0.014   & 0.060      \\
\bottomrule
\end{tabular}
\label{tab:recon_quant}
\vspace{-3mm}
\end{table}

The reconstruction performance of \name on the three shape categories chair, table and cabinet is given in Table~\ref{tab:recon_quant}. In this experiment, we represent part geometry as oriented bounding boxes. The largest cause of reconstruction error we encountered when training our network is the unbalanced variety of shapes in each category. The datasets typically contain some sub-types of a category, such as square tables, more often than other sub-types, such as triangular tables. As a consequence, the network may have too few examples to learn a good representation of the more exotic shape varieties. The chair dataset is the most balanced among the categories, giving us lower reconstruction errors than for the other categories.
Additionally, the cabinet and bed datasets contain shapes with more complex structure on average, giving us higher hierarchy and edge reconstruction errors.

We compare our reconstruction performance to two baselines: GRASS~\cite{grass:li:2017} as a state-of-the-art structure-aware shape generation method, and an autoencoder based on PointNet++~\cite{qi2017pointnet++}, as a state-of-the-art point cloud autoencoder that holistically encodes the point cloud of a shape without using any explicit structure.

\begin{figure}[t]
    \centering
    \includegraphics[width=\linewidth]{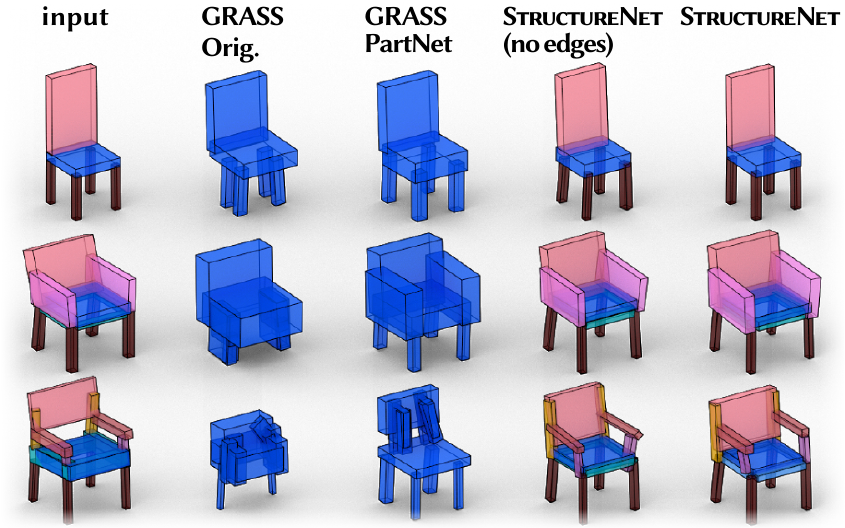}
    \vspace{3mm}
    \begin{tabular}{@{}lcr@{}}
    \toprule
                       &\phantom{}& geometry reconstruction error $E_{\mathbf{P}}$      \\
    \midrule
    GRASS Orig.        && 0.103           \\
    GRASS PartNet      && 0.082           \\
    \name (no edges)   && 0.065           \\
    \name              && \textbf{0.061}  \\ 
    \bottomrule
    \end{tabular}
    \vspace{-7mm}
    \caption{\titlecap{Reconstruction compared to GRASS.}{In the top three rows, we show reconstructions of the left-most shape using the two variants of GRASS described in Section~\ref{sec:experiments_recon}, and then reconstructions using a version of \name that does not use edges, and finally using our full method. Due to our more consistent shape representation, our method scales better to datasets with the size of PartNet. Using edges additionally improves part relationships such as the symmetries between the armrests, as seen in the third row.
    This is confirmed by the reconstruction error over the whole dataset, shown in the table below.}}
    \label{fig:box_recon}
\end{figure}

\paragraph{Comparison to GRASS}
The comparison to GRASS is shown in Figure~\ref{fig:box_recon}. A qualitative comparison is shown in the top three rows, and a quantitative comparison on the chair dataset in the table below. The oriented bounding box of each leaf part is illustrated as transparent box, colored according to its semantic (see the Supplementary for the full semantic tree of each category). GRASS results are colored uniformly, since semantics are not available. We measure only the geometry reconstruction error, since the structure of GRASS is not directly comparable to our shape representation. GRASS requires a binarization of shape hierarchies that are naturally $n$-ary. A binarization that is consistent between all shapes in a category is difficult to find, and this difficulty grows with the number and variety of shapes in a category. On our dataset, which has $\sim$10 times the size of the dataset originally used in GRASS, this task becomes too difficult. For this reason, the authors provided us with two modified version of their method, that each sacrifice some generality for a reduced number of possible binarizations. The first version, which we call \emph{GRASS Orginal}, reduces generality and possible binarizations by a small amount, resulting in a method very similar to the original GRASS. Results for this method are  shown in the second column of Figure~\ref{fig:box_recon} and the first row of the table. Due to the large number of possible binarizations, the performance is low. The second version, which we call \emph{GRASS PartNet} uses the semantic hierarchy of PartNet to significantly reduce the number of possible binarizations. This increases the performance of GRASS, as shown in the third column of the figure and the second row of the table. 
Our approach, on the other hand, can encode and decode $n$-ary hierarchies directly, leading to a more consistent representation of the structure that gives us a significant improvement in reconstruction performance, as shown in column four and row 3 of the table. Relationship edges provide and additional boost to the reconstruction performance, by ensuring that symmetries that are present in the input are maintained in the reconstruction. 
Note that we evaluate this comparison on a reduced subset of $4031$ chairs, since the GRASS authors reported that their pipeline failed to produce results for the remaining $840$ chairs in our dataset.

\begin{figure}[t]
    \centering
    \includegraphics[width=\linewidth]{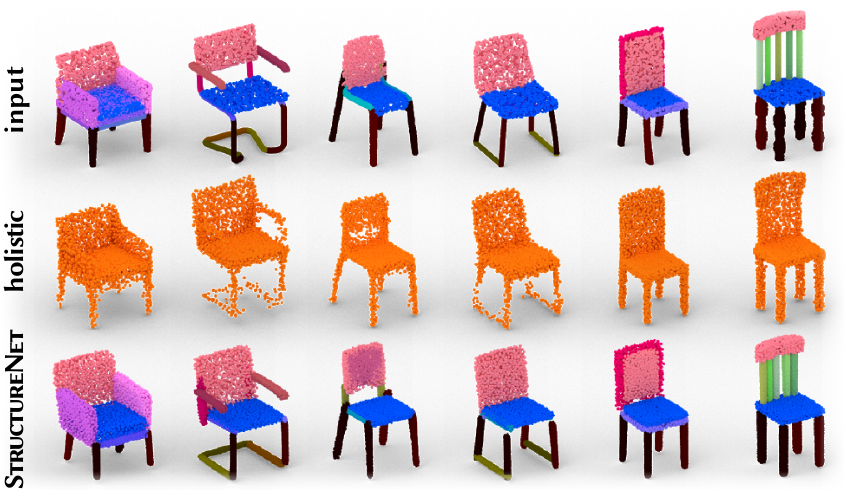}
    \caption{\titlecap{Reconstruction compared to a holistic approach.}{We reconstruct the point cloud representation of the shapes in the top row with a holistic approach (orange) that does not use structure, and compare to \name. The holistic approach suffers from noise that results in a loss of detail, whereas reconstruction errors in our approach take the form of slightly modified chair structures, such as the hole added to the backrest in the third column. However, the functional realism of the shape is usually preserved and the explicit structure allows us to preserve significantly more detail.}}
    \label{fig:pc_recon}
    \vspace{-3mm}
\end{figure}

\paragraph{Comparison to a holistic autoencoder}

We train an autoencoder based on PointNet++~\cite{qi2017pointnet++} and PointSetGen~\cite{fan2017point} to compare the effects of encoding geometry only to our structure-aware latent space.
%encoding structure in our latent space to encoding geometry only.
PointNet++ is used as encoder, followed by a point cloud decoder network proposed in PointSetGen. We train both this autoencoder and \name on the chair dataset, with part geometry represented as point clouds. For the PointNet++ autoencoder, we merge the geometry of the leaf parts into a single point cloud for the shape. Results are shown in Figure~\ref{fig:pc_recon} where points are visualized as small spheres, colored by their semantic in the same way as the bounding boxes in Figure~\ref{fig:box_recon}. Note that the holistic results have significantly more noise, making it harder to recover details. Since we encode structure, errors in our approach instead take the form of slight modifications to the structure and layout of parts, such as added hole in the backrest of the chair in the third column, or the slightly modified arrangement of bars in the backrest of the chair in the last column. The structure, however, tends to remain realistic and since we represent the point cloud of each part separately, individual parts are sharper an details are better preserved.

%We train an Auto-encoder (AE) for the point-cloud version of \name. We sample 1,000 points per part to represent the detailed part geometry. We first train a Part AE that embeds all the part point clouds to a latent space. Then, we train the backbone network with the pretrained Part Encoder and Part Decoder parameters fixed. 

%Compared to the holistic point cloud auto-encoder methods, \name reconstructs shapes by first predicting the structural hierarchy and then all leaf node part geometry. In this way, we can generate the shape with discrete structure, sharp part boundaries and detailed part geometry. Figure~\ref{fig:pc_recon} shows our reconstruction results, with comparison to a baseline holistic point cloud auto-encoder network. For the baseline method, we use PointNet++~\cite{qi2017pointnet++} as the point cloud encoder and the vanilla point cloud decoder network proposed in~\cite{fan2017point}. We describe the architecture details in the supplementary material. Holistic method is optimized in a part-agnostic way so that the part details are not reconstructed and it tends to provide continuous and blurry connections between the discrete part structures. 

%Table~\ref{tab:ablation} shows some abalation study that we leave out some of the network components: \paul{todo here!}

\subsection{Shape Generation}
A straight-forward application of our hierarchical graph network is shape generation. In the remainder of Section~\ref{sec:results}, we use a VAE with the variational regularization weight $\beta = 0.05$. This gives us a dense and smooth distribution of shapes in latent space that we can draw from to generate new samples of shapes, including geometry and structure. We show both qualitative results and a quantitative comparison to GRASS for this application.

\begin{figure*}[p]
    \centering
    \includegraphics[width=\textwidth]{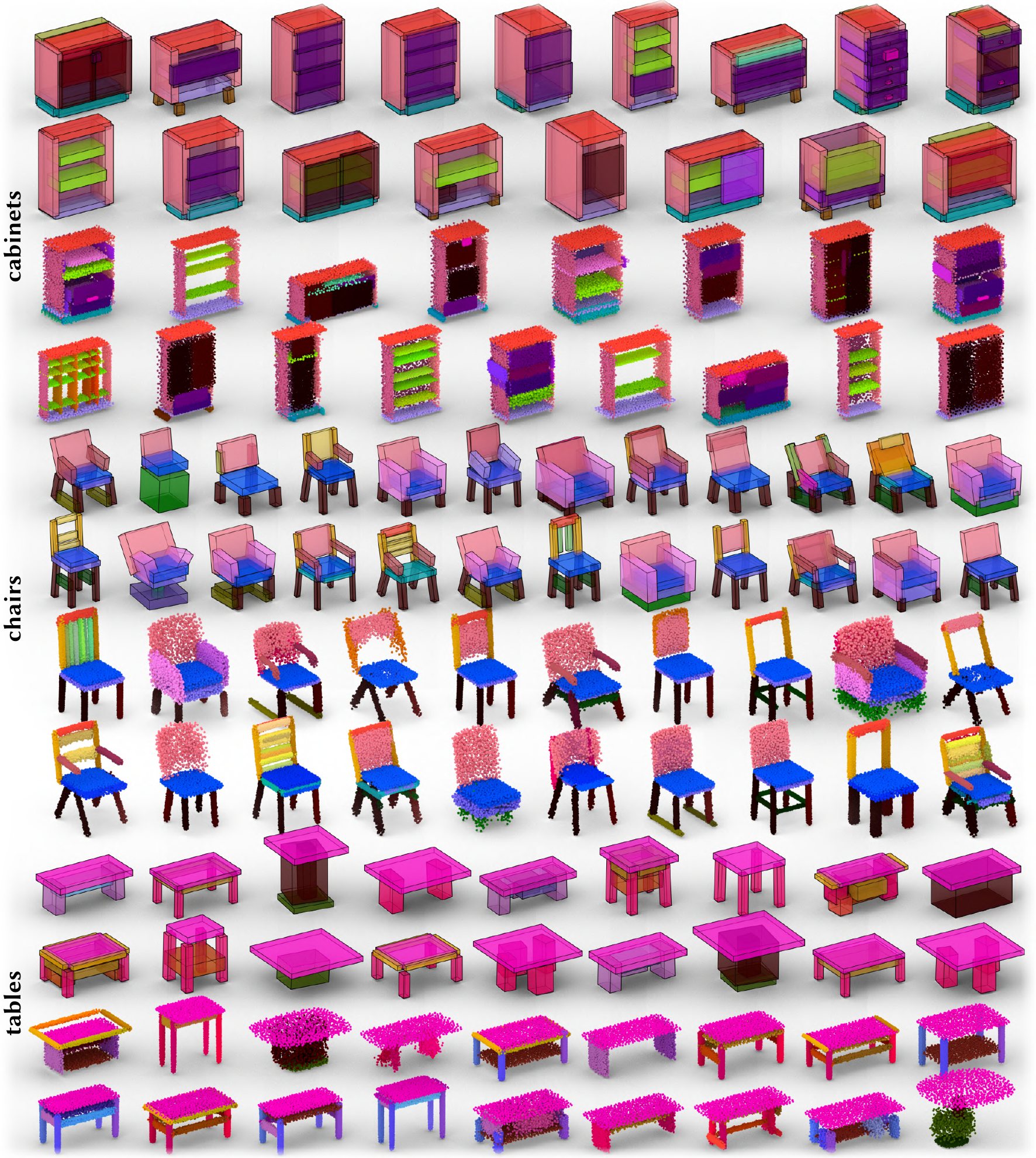}
    \caption{\titlecap{Generated shapes.}{We show shapes in all categories decoded from random latent vectors, including shapes with bounding box geometry, and shapes with point cloud geometry. Parts are colored according to semantics, see the Supplementary for the full semantic hierarchy  for each category. Since we explicitly encode shape structure in our latent representation, the generated shapes have a large variety of different structures.} }
    \label{fig:generation}
    \vspace{-3mm}
\end{figure*}

\paragraph{Qualitative evaluation}
Several examples of generated shapes are presented in Figure~\ref{fig:generation}, using both the bounding box representation and the point cloud representation for part geometry. Our results show a large variety in structure and part geometry, with a layout of individual parts that is functionally plausible. For each shape, we generate our full shape representation, including the geometry of individual parts, the hierarchical decomposition of these parts, symmetry and adjacency relationship edges between siblings, and part semantics. This rich high-level representation of the shapes is useful for several applications, some of which we will present in the following sections.

\begin{figure}[h!]
    \centering
    \includegraphics[width=\linewidth]{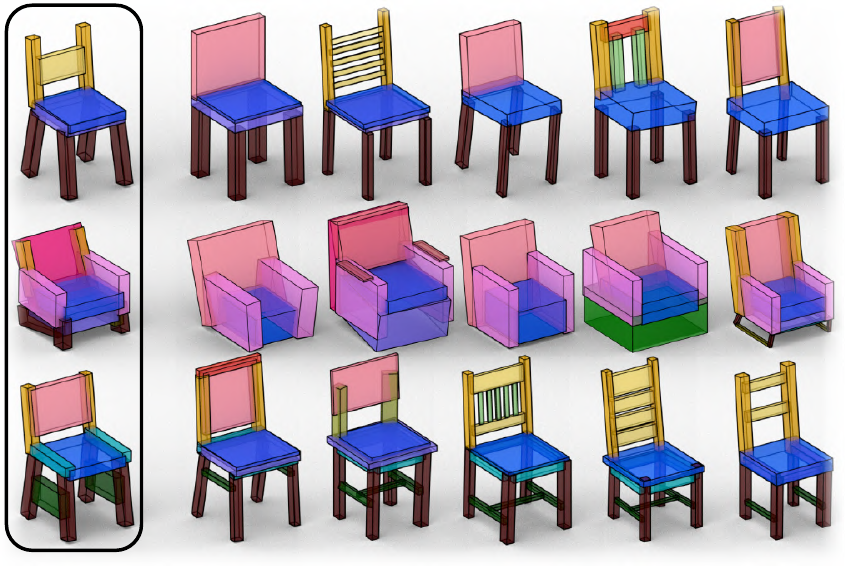}
    \caption{\titlecap{Novelty of generated shapes.}{In the first column we show three generated shapes, and on the right the five closest matches in the training set, measured with the chamfer distance. Our generated structure and geometry is different from the shapes in the training set.}}
    \label{fig:closest_train}
    \vspace{-3mm}
\end{figure}

\paragraph{Shape novelty and overfitting}
To evaluate the novelty of our generated shapes, we show the top-five closest training samples to several generated chairs in Figure~\ref{fig:closest_train}, using the chamfer distance as metric. We can see that the generated shapes are quite different in both geometry and structure from the closest matches, suggesting little overfitting to the training set.

\begin{table}[t!]
\caption{\titlecap{Shape generation compared to GRASS.} We compare the shape distribution learned by the two version of GRASS described in Section~\ref{sec:experiments_recon} to our method without edges and to our full method, using two metrics that measure how close the shapes are to the data distribution (quality) and how much of the data distribution is covered by the generated shapes (coverage). We report the scores relative to our method, higher numbers indicate better performance. Results show that our latent distribution better captures the data distribution.}
\begin{tabular}{@{}lrr@{}}
\toprule
                               &    rel. quality    &     rel. coverage     \\ \midrule
GRASS Orig.   &    0.714   &      0.818             \\
GRASS Partnet &    0.788    &      0.818            \\
\name (no edges)                 &    0.984     &      0.989            \\ 
\name              &    \textbf{1.0000}     &      \textbf{1.0000}           \\ \bottomrule
\end{tabular}
\label{tab:box_gen}
\vspace{-3mm}
\end{table}

\paragraph{Quantitative evaluation}
Quantitatively, the goal of this application is to cover as much of the data distribution as possible, while at the same time, avoiding unrealistic chairs that are distant from the main mass of the data distribution. We quantify this goal with two metrics.
The \emph{quality}, of a generated shape set is measured by the average closest distance to any data sample, while the \emph{coverage} is measured as the average closest distance from each data sample to a generated sample.
\begin{equation}
\begin{split}
    \text{quality} \coloneqq& \sum_{S' \in \mathcal{S}_G} \min_{S \in \mathcal{S}} d_S(S', S) \text{ and} \\
    \text{coverage} \coloneqq& \sum_{S \in \mathcal{S}} \min_{S' \in \mathcal{S}_G} d_S(S', S),
\end{split}
\end{equation}
where $\mathcal{S}$ is the training set, $\mathcal{S}_G$ is a set of generated shapes, and $d_S$ is the chamfer distance between the point representations of two shapes. To compare with GRASS using these metrics, we compute a set of $1000$ shapes using both \name and GRASS, and compute their quality and coverage. We show results relative to the performance of \name in Table~\ref{tab:box_gen} (\ie \name scores divided by the method scores). We see an improvement over GRASS in both quality and coverage of the generated shapes.

\subsection{Shape Interpolation}
\label{sec:experiments_interp}
\begin{figure*}[t]
    \centering
    \includegraphics[width=\textwidth]{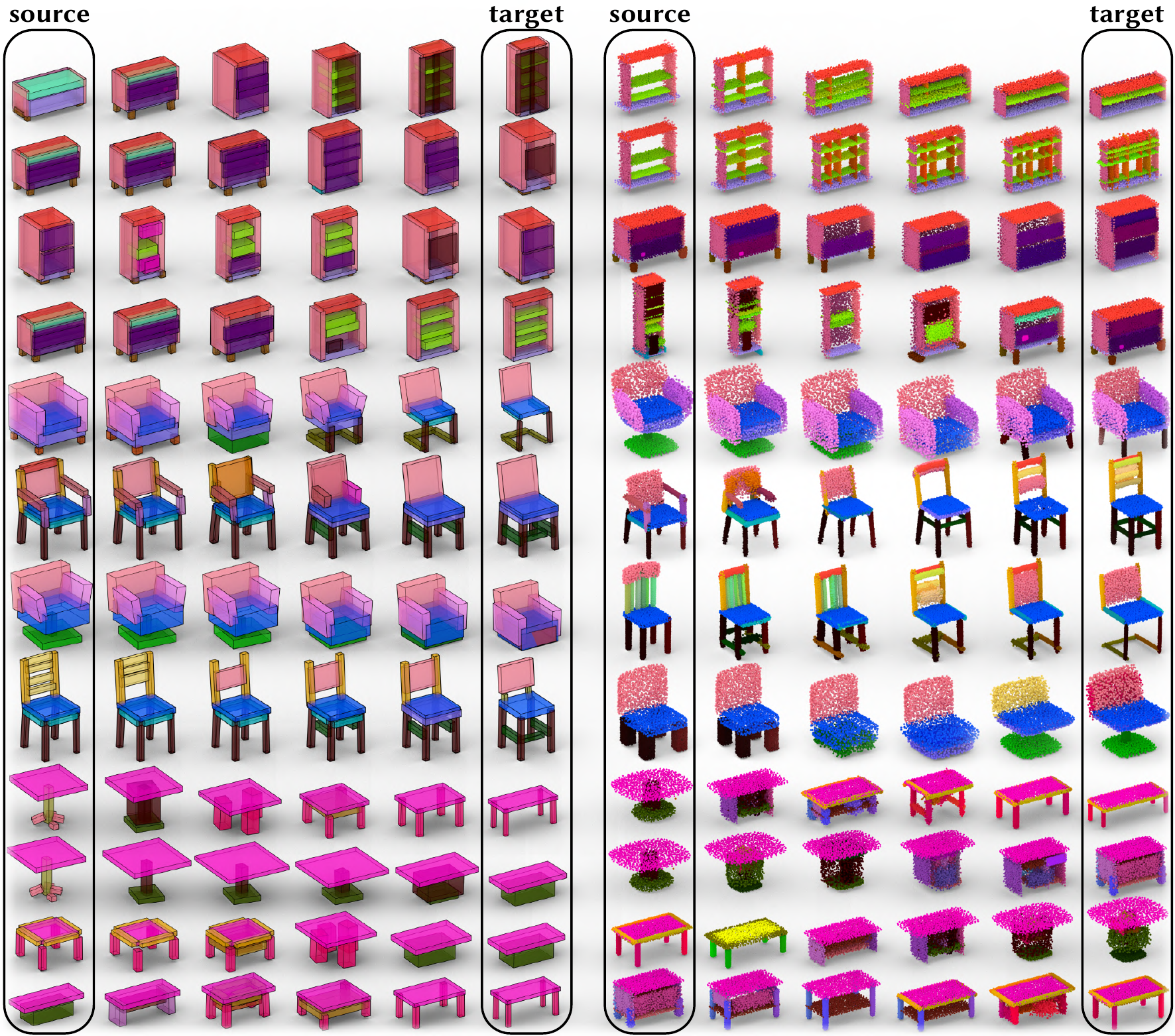}
    \caption{\titlecap{Interpolation of shapes.}{We show interpolations between a source shape and a target shape from all categories. Interpolations are symmetric, so source and target are interchangeable. Interpolations between shapes with box geometry is shown on the left side, and point cloud geometry on the right side. Note how each interpolation is a smooth transition between two different structures that preserves functional plausibility in each step. In the interpolations between shelves with different numbers of boards, for example (bottom right), the number of boards is gradually increased/decreased in each step, and each step is a functional shelf.}}
    \label{fig:interpolation}
    \vspace{-3mm}
\end{figure*}

We further examine the quality of our learned latent space with interpolations between shapes, visualizing samples along line segments in the space. We show several examples of interpolations in Figure~\ref{fig:interpolation}. The left half of the figure shows interpolations between shapes with bounding box geometry, the right half between shapes with point cloud geometry. Note how the structure changes in small intuitive steps between the source and target of an interpolation. For example, in row $6$ on the left side, the backrest of the chair is simplified part by part, while on the base, bars between the legs are added in multiple steps, and armrests are simplified to have fewer parts before disappearing. At the same time, each of the steps represents a valid and functional chair. We find it interesting to see in which way the network learned to arrange shape configurations in latent space, especially since these arrangements often seem to correspond to our own intuition. For example, the pedestal base of the chair on the top right is first made smaller, before completely disappearing, and reappearing again as 4 separate legs that increase in size. Similarly, the transition from shelves with few boards to shelves with many boards near the bottom right of the figure, transitions by increasing or decreasing the number of boards step by step.

\paragraph{Comparison to GRASS}
We provide a qualitative comparison of these interpolations to GRASS PartNet in Figure~\ref{fig:interpolation_comparison}. First, we see issues with the reconstruction accuracy of grass, but looking at the interpolations only, we see that some interpolation of the structure, such as the reduction of the number of legs of the chairs happens much less gradual than in our interpolations, with fewer, larger steps.

\begin{figure}[t!]
    \centering
    \includegraphics[width=\linewidth]{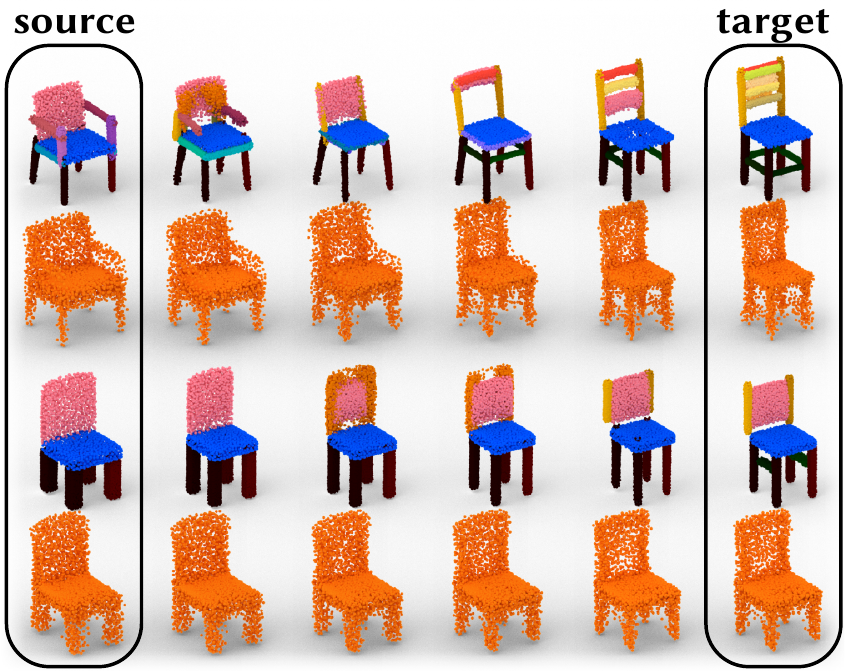}
    \caption{\titlecap{Interpolation compared to a holistic approach.}{We compare our interpolation (colored) to a holistic approach (orange) that encodes shapes as point clouds without any structure. Explicitly encoding structure gives us a sequence of small structural changes in the intermediate steps, whereas the holistic approach produces no significant structural changes.
    %operates by shrinking parts until they disappear (see the armrests in the first example),
    Additionally, our per-part geometry is cleaner than the per-part geometry in the holistic approach, where it is hard to identify detailed parts.}}
    \label{fig:pc_interp}
    \vspace{-3mm}
\end{figure}

\paragraph{Comparison to holistic interpolation}
Figure~\ref{fig:pc_interp} compares \name for point cloud geometry to interpolating without structure using the same holistic point cloud network described in Section~\ref{sec:experiments_recon}, but using a VAE instead of an autoencoder. The VAE is necessary to obtain a smooth latent space suitable for interpolation, but makes the shapes much noisier than for the autoencoder, whereas our point clouds do not suffer as much from this switch to a VAE. Interpolations are smooth for the holistic VAE, but lack interesting transitions between structural details. Parts such as the armrests in the second row gradually disappear, whereas our armrests get replaced by simpler, but still functionally valid variants before being removed.

%Traversing on the manifold space should give us realistic transitions. We show interpolation results in this section. We conduct experiments using the VAE versions of the networks.

% \paragraph{Box-paramterized Shapes} Figure~\ref{fig:interpolation} for example interpolations. We can see discrete structural and topological transitions accompanying the continuous geometry changes.

% We compare the interpolation results to GRASS. We use GRASS PartNet as it performs much better than GRASS Orig in this experiment. We observe that for easier interpolation, both ours and GRASS presents good results. However, for more complicated shapes, GRASS fails to reconstruct the correct structures for the two end-point shapes. Our method gives more structural variation while doing interpolation.

% \paragraph{Point Cloud Shapes} Figure~\ref{fig:pc_interp} shows example interpolations.

% \begin{figure*}[t]
%     \centering
%     \includegraphics[width=\textwidth]{images/interpolation_pc.pdf}
%     \caption{Interpolation of point cloud shapes from different categories.}
%     \label{fig:interpolation_pc}
%     \vspace{-3mm}
% \end{figure*}

% We compare our method to the holistic point cloud VAE network. We see that xxx

\begin{figure}[t]
    \centering
    \includegraphics[width=\linewidth]{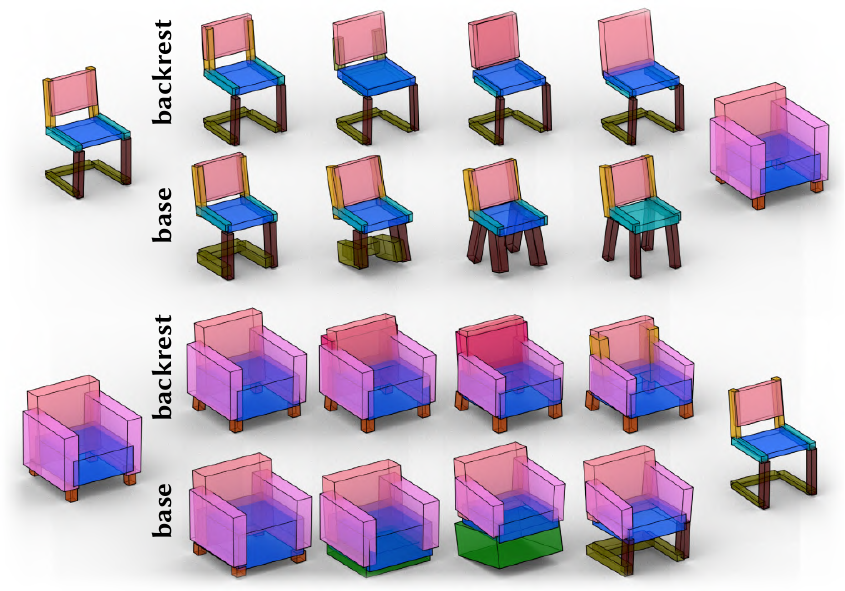}
    \caption{\titlecap{Part Interpolation.}{We interpolate either only the backrest (first row) or only the base (second row) between the chairs on the left and right side. Intermediate shapes preserve structural plausibility of the interpolated result mainly through geometric differences to the target part, but faithfully interpolate the structure. We observe that these interpolations are not necessarily symmetric: the base interpolations follow different paths to be compatible with the different back styles.}}
    \label{fig:part_interp}
    \vspace{-3mm}
\end{figure}

\paragraph{Partial interpolation}
Since shape structures in a category are consistent, we can do interpolations between corresponding parts of a shape. We perform the partial interpolation by taking the encoded part latent vector $f_i$ of a shape, interpolating it with the corresponding part latent vector of another shape, and then re-encoding the shape with the interpolated part latent vector. This ensures plausibility of the resulting shape by effectively projecting the shape to the learned manifold of shapes.
In Figure~\ref{fig:part_interp}, we interpolate either only the backrest or only the base for each of two chairs. The structure of the interpolated part changes to resemble the target part, while the other parts of the shape remain largely unchanged. We can also see that the final step of the interpolation does not fully reach the target part, because the network ensures the plausibility of the resulting chair. For example, in the second interpolation, the short legs of the sofa would result in an implausible shape when attached to the chair. Thus, the interpolation depends on the context of the part.
%For this reason, we can also observe that interpolations are no longer symmetric.
Even though the geometry is not the same, the \emph{structure} of the chair bases is interpolated correctly and resembles the target structure at the final interpolation step. 
%\kaichun{As Niloy suggested, we should also mention the base interpolation are different A to B and B to A. This is because it needs to consider the context of the other parts.}

%A structure that is consistent between different shapes allows us to do part-based interpolation.

%Our template is consistent. So it's easy to switch and interpolate parts among two quite different shapes.
%We can't just direct copy and paste, which will cause the gap and style inconsistency. We need to interpolate the part feature in the encoding stage, mix up with the other part features to a bottleneck, and then finally decode the entire shape as a whole.

%Figure~\ref{fig:part_interp} shows the interpolation between two different shapes. Part interpolation successfully interpolate the chair back and chair base respectively while keeping the other parts unchanged. The interpolation needs to consider the other parts as context to keep the middle shapes on the interpolation path to be realistic. We notice two different interpolation path for the base interpolation between the two shapes, since the contexts of chair back and arm styles are different.

\subsection{Shape Abstraction}
\label{sec:shape_abstraction}
%\todo{why abstraction is important?}

Discovering higher-level structure in un-annotated point clouds and images are long-standing vision and graphics problems. Our rich latent representation provides us with an approach to tackle these problems. We can encode images, point clouds, and shapes into a common latent space, using separately trained encoders for images and un-annotated point clouds. Given a trained \name autoencoder $d(e(S))$ for the box representation of our shapes, we train additional encoders to take images $I$ and un-annotated point clouds $O$ to the same latent space. From the latent space, we can use our pre-trained decoder $d$ to recover a shape $S'$ that is similar to the shape represented with the input image or point cloud, but has all the information of our shape representation.

\begin{figure}[t]
    \centering
    \includegraphics[width=\linewidth]{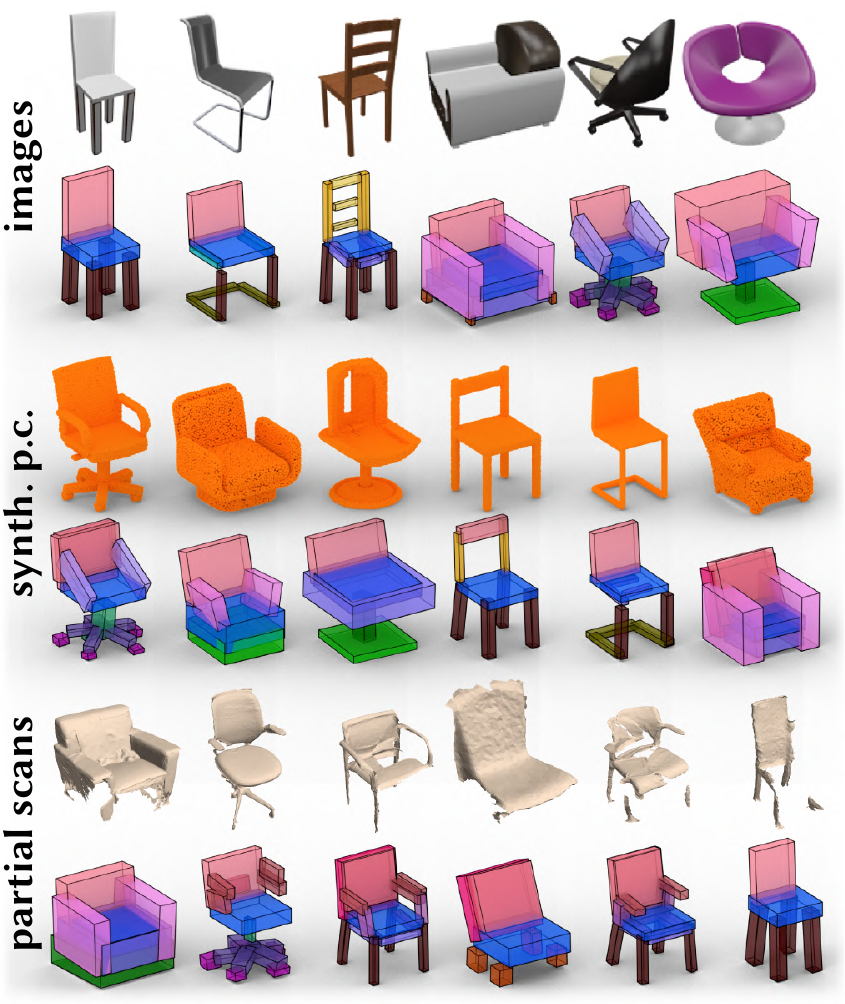}
    \caption{\titlecap{Image and point cloud abstraction.}{Images, synthetic point clouds, and real-world scans from ScanNet~\cite{dai2017scannet} are embedded into our learned latent space, allowing us to effectively recover a full shape description that matches the raw input.}}
    \label{fig:abstraction}
    \vspace{-3mm}
\end{figure}

\paragraph{Image abstraction}
The image encoder $e_I$ is a ResNet18~\cite{he2016deep} that was pre-trained on ImageNet~\cite{deng2009imagenet}. Inspired by the joint embedding approach of \cite{Li:2015:JES:2816795.2818071}, we refine this encoder on a dataset of images rendered from the shapes in our training set. We render the shapes with textures obtained from ShapeNet~\cite{chang2015shapenet}, from $24$ random angles around the their up vector, from a random elevation between $25$ and $30$ degrees, and a random distance between $1.2$ and $2.0$ times the bounding sphere radius. For each image, we additionally have the corresponding latent vector in the latent space of the trained autoencoder $d(e(S))$. We train the image encoder to map each image to the latent representation of the shape it was rendered from $\mathcal{L}_I = e(S) - e_I(\text{render}(S, \theta))$, where $\theta$ are the random camera parameters. We test on images generated from shapes in our validation and test sets, examples are shown in Figure~\ref{fig:abstraction}, top. While the proportions of objects are not completely accurate, the overall shape, and even many of the details are represented accurately in the recovered shape, suggesting that the joint embedding successfully aligns structurally similar images and shapes in the latent space.

% \begin{figure}[t!]
%     \centering
%     \includegraphics[width=\linewidth]{images/pc2struct.png}
%     \caption{Point Cloud Abstraction.}
%     \label{fig:pc2struct}
%     \vspace{-3mm}
% \end{figure}

% \begin{figure}[t!]
%     \centering
%     \includegraphics[width=\linewidth]{images/scannet.png}
%     \caption{ScanNet Shape Abstraction.}
%     \label{fig:scannet}
%     \vspace{-3mm}
% \end{figure}

\paragraph{Point cloud abstraction}
The point cloud encoder $e_O$ is implemented as PointNet++~\cite{qi2017pointnet++} and similar to the image encoder, we train the network to encode $10$k points obtained from each shape in our training set into the latent representation of the corresponding shape with $\mathcal{L}_O = e(S) - e_O(\text{sample}(S))$. Examples of abstractions computed for point clouds in our test set are shown in Figure~\ref{fig:abstraction}, rows 3 and 4. We also tested our approach on real-world scans obtained from ScanNet~\cite{dai2017scannet}. 
Even though the statistics of the point distribution on real-world scans is likely to be different from our synthetic point sets, and the scans are missing large regions, our shape abstraction can recover good matches for each of the point clouds (see Figure~\ref{fig:abstraction}, last two rows).

\subsection{Shape Editing}
Edits performed on a shape in a traditional 3D editor do not take into account the plausibility of the resulting shape. Our learned latent space gives us a definition of shape plausibility.
%, regions covered by the variational prior distribution represent plausible shapes.
Edits of a shape that preserve plausibility can thus be performed by finding the shape in our latent space that best satisfies the given edit.
Below we present two simple shape editing applications based on this approach.

%We edit individual boxes in a generated object and re-encode the edited object to consistently update the remaining boxes. (For example, if we change the size of one box that is in a symmetry relation with other boxes, the other boxes need to change accordingly.) This requires training with an additional edit preservation loss, that makes sure the edited box(es) are changed as little as possible during re-encoding.

%Alternatively, we can explicitly optimize the non-edited boxes to satisfy the symmetry relationships.

%Since all the shapes are explained by the same hierarchy, we can replace the corresponding parts or subtrees of parts.

%\paragraph{Edge Enforcement}
%We make use the decoded edges to make the legs more symmetric, or the back bars more symmetric. Also true for the adjacency relationships.

%\paragraph{Latent Space Arithmetics}
%Adding the difference between two shapes to a third shape (shape analogies), etc.

%\begin{figure}[t!]
%    \centering
%    \includegraphics[width=\linewidth]{images/shape_algebra.png}
%    \caption{Shape Analogies.}
%    \label{fig:shape_algebra}
%    \vspace{-3mm}
%\end{figure}

\paragraph{Structure-preserving part edits}
In our latent space, shapes with similar structure are located close to each other. We can preserve the structure of a shape during editing by working with shapes that are close to the original shape in latent space.
Starting from a shape with bounding box geometry, we edit one of its boxes, by translating, non-uniformly scaling, or rotating it. We then optimize for a shape in our latent space that is as close as possible to the original shape, while also satisfying the box edit:
\begin{equation}
    \argmin_{z}\ \left( \left\|z-z^*\right\|^2_2 + q_{\mathrm{chs}}(T(B_e(z))\mathbf{U},\ T(B^t_e)\mathbf{U}) + \mathcal{L}_{sc}(d(z)) \right).
\end{equation}
The first term is the squared distance between the latent vector of the edited shape $z$ and the latent vector of the original shape $z^*$. The second term is minimized by shapes that satisfy the edit, using the squared chamfer distance between the configuration of the edited box $B_e$ in $z$ and its target configuration $B^t_e$. Since the box edit is likely to break existing symmetries, we also specifically optimize for a shape $S = d(z)$ that is consistent with its relationships using the loss $\mathcal{L}_{sc}$, as defined in Eq.~\ref{eq:consistency}.

\begin{figure}[t]
    \centering
    \includegraphics[width=\linewidth]{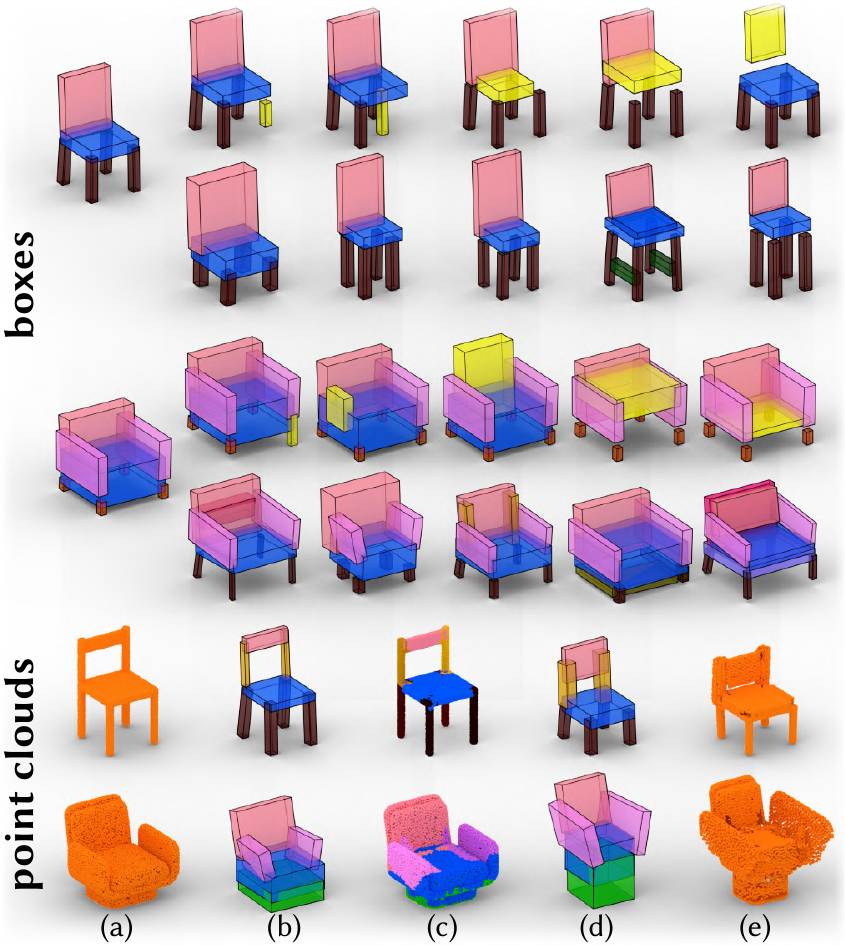}
    \caption{\titlecap{Structure-aware part editing.}{We show editing results on two shapes with box geometry (first four rows) and two shapes with point cloud geometry (two bottom rows). For the two shapes with box geometry, we perform five different edits each, one edit per column. The edited box is highlighted in yellow, and the result is shown below. We see that the other boxes in the shape are adjusted to maintain shape plausibility. For the two shapes with point cloud geometry, we show intermediate results for one edit each. From left to right, these are (a) the original point cloud; (b) the predicted box abstraction; (c) the induced segmentation; (d) edited boxes; and (e) the induced edit of the point-cloud.}}
    \label{fig:part_editing}
    \vspace{-3mm}
\end{figure}

% Starting from a shape abstraction, we can modify one part box, and optimize for a latent code that is as close to the original shape code and also satisfies the box editing. We remove the encoder network and use the decoder network of a pretrained \name to solve an optimization problem to traverse on the manifold. Fixing the decoder parameters, we solve the following objective function:
% \begin{equation}
%     \min_{x} \left\|x-x^*\right\|^2 + q_{\mathrm{chamfer}}(T(B(x))\mathbf{U},\ T(B_t)\mathbf{U}) + \mathcal{L}_c(S(x), S(x^*))
% \end{equation}
% where $x^*$ is the latent code of the starting shape, $B(x)$ is the decoded box parameter for the box being edited, $B_t$ is the target box parameter, $T(\dots)$ is a 4D transformation matrix that transforms the unit cube point cloud $U$, $S(\dots)$ is the decoder function that generates the shape boxes, and $\mathcal{L}_c$ is the loss defined in Eq.\ref{eq:consistency}.

Figure~\ref{fig:part_editing} shows $5$ edits on each of the two shapes shown on the top left-hand side. Edited boxes are marked in yellow in the top row of each shape, and the result of the edit is shown in the bottom row. Since results only use shapes that can be found in our latent space, the results maintain plausibility of the shapes, adjusting the other parts in the shape as needed. Since we optimize for proximity to the original shape in latent space, the resulting shapes have similar structure, with a few minor exceptions, such as the added bars between the chair legs in the fourth column. This experiment suggests that structural differences are more distant in the latent space than geometric differences. This can also be verified by examining the visualization of the latent space provided in the Supplementary.
%Figure~\ref{fig:embedding}.

% some editing results. The edited boxes being are highlighted in yellow. We solve an optimization problem to traverse on the learned manifold to find a shape that is both realistic (on the manifold) and satisfying the editing effect of the yellow box.

\paragraph{Point Cloud Editing}
We can extend this editing approach to unannotated point clouds using the abstraction approach described in Section~\ref{sec:shape_abstraction}. The abstraction of the point cloud induces an instance segmentation, where each point is assigned to the closest bounding box. After editing the boxes with the method described above, we can update the subset of points corresponding to each box with the same transformation applied to the box, giving us an edited point cloud. The bottom two rows of Figure~\ref{fig:part_editing} show these steps in two examples edits. The edited point clouds show some artifacts due to the hard boundaries between different segments, but closely resemble the edited boxes. In the future, we could augment this method with either soft assignments of points to boxes (where boxes act similar to bones in character animation), or a segment refinement step, where the segment boundaries are optimized to coincide with surface discontinuities of the shape.

% \begin{figure}[t!]
%     \centering
%     \includegraphics[width=\linewidth]{images/pc_editing.png}
%     \caption{\titlecap{Point cloud editing results.}{We show (a) original shape pc; (b) the predicted box abstraction; (c) the induced segmentation; (d) box deformation; and (e) the induced point-cloud editing.}}
%     \label{fig:pc_editing}
%     \vspace{-3mm}
% \end{figure}

%We can also perform editing on the point cloud shapes. We can first run a shape abstraction network to abstract the shape, then we can use the box as deformation controller to deform the input point clouds. 

%Figure~\ref{fig:pc_editing} shows the editing for two point cloud shapes. We first generate the box abstraction, the we induce a segmentation on the input point clouds based on the affinities to the predicted boxes. Then, we transform the points associated to one box to the box local coordinate system. We can deform the boxes by adjusting the box parameters, e.g. translating the box locations, enlarging the sizes, or rotating the boxes, and the point clouds will deform as the box parameters change. We achieve the final editing effects on point cloud inputs by editing the boxes.

%\subsection{Shape Denoising}
%We train a denoising autoencoder for objects with parts represented as bounding boxes and optionally point clouds if time permits.

%\subsection{Shape Completion}
%We train an autoencoder for shape completion (partial shapes in, completed shapes out) with parts represented as bounding boxes and optionally point clouds if time permits.

\subsection{Limitations and Failure Cases}
\begin{figure}[t!]
    \centering
    \includegraphics[width=\linewidth]{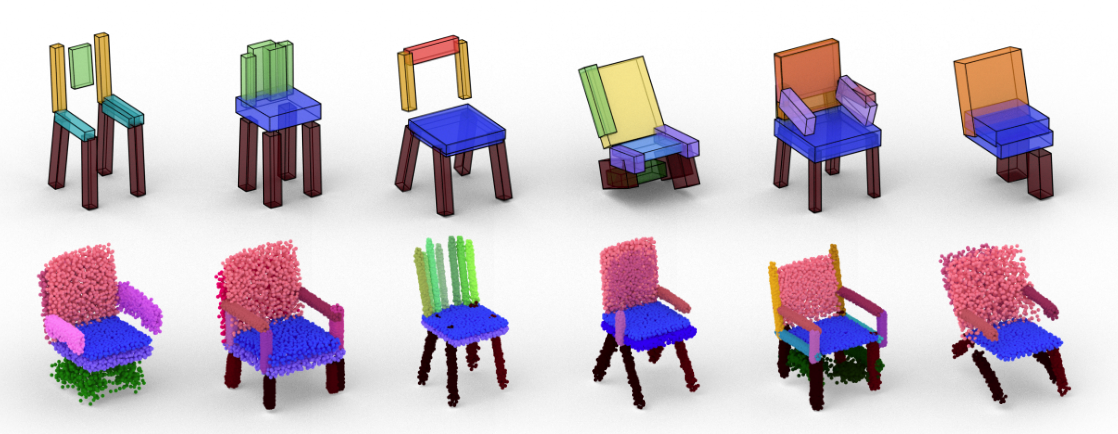}
    \caption{\titlecap{Failure cases analysis.}{We present several failure cases we observed for box-shape and point cloud generation. We see discrete errors such as missing parts (\eg first row, first column), duplicate parts (\eg second row, second column), detached parts (\eg first row, third column), asymmetric parts (\eg second row, first column) and fuzzy point cloud generation (\eg second row, fifth column).}}
    \label{fig:limitation}
    \vspace{-3mm}
\end{figure}

We discuss several limitations and failure cases: 
(i)~\name, being a data-driven method, naturally inherits any data sampling biases in the datasets (e.g., shape families with very few examples such as pingpong tables). 
(ii)~Even though our empirical experiments demonstrate good performance on adjacency recovery and symmetry enforcement, the inferred latent space may contain models with detached parts or asymmetric parts, especially for datasets that contain exotic, poorly represented shape variants.
(iii)~We restrict the maximum sibling count to $n_p = 10$, and hence cannot encode shapes with more than 10 childs in any given part (the full shape can have a much larger number of parts). The memory cost is quadratic in $n_p$, although, at our current setting, this is still far from being the most memory-consuming component (our current consumption is approx. $1-2$ GB).
%Simply increasing $n_p$ is not a desirable option as the number of possible edges, being quadratic, quickly becomes unmanageable. 
%
(iv)~Noise that would make a point cloud more blurry in holistic generation methods instead affects the structure in our method. Strong noise may result in missing parts, duplicate parts, detached parts, etc., although the structure, being discrete, is quite robust to this type of noise. See the bed category in the Supplementary for an example of structural noise.
%Reconstructed or generated shapes, though having many desirable properties such as sharp part geometry and boundary, may still be erroneous 
(v)~Structure-aware point cloud generation is still a new topic requiring further research. In our experiments, to stabilize  network training, we pretrain and freeze the part point cloud networks, which increases training robustness at the cost of failing to recover fine-grained geometry details (e.g.,  details on chair legs and chair backs).
Figure~\ref{fig:limitation} shows different failure cases of \name.

%-- Our edges are soft constraints and are not fully guaranteed. 
%-- our paper is just the first step. PartNet is challenging that it has a large variation of styles, data is inbalanced, some shapes are crazily difficult (containing >100 parts for a shape), for reconstruction and generation, there are rooms to improve.
%-- We have Discrete errors, like missing a part, predicting duplicate parts
%-- PC exps have a lot of room to improve. We fix the pretrained PC VAE and fails to recover the fine geometry details. 

\section{Conclusion}

We have presented \name as a VAE that directly encodes shape structure and geometry represented as a hierarchy of \nary graphs. We have achieved this by proposing a recursive/hierarchical encoder-decoder architecture that simultaneously considers both geometry of parts, either as oriented bounding boxes or point clouds, and inter-part structures capturing adjacency and symmetry relations.
Our key technical novelty is the handling of \nary graphs by (i) explicitly predicting the presence or absence of parts or relationship edges; (ii) designing the encoder and the decoder to be invariant of the ordering of siblings across instances of the \nary graphs; and (iii) introducing novel losses that enforce consistency between geometry and structure at all levels of the hierarchy.
The learned \nary structural graph latent space, by jointly capturing geometric and structure, greatly simplifies several applications. For example, we can `enter' the latent space by \textit{projecting} un-annotated data (e.g., partial scans, point clouds, or images) onto the latent space; perform structure-aware edits on individual shapes; do hierarchy-preserving interpolation between multiple shapes, or generate novel and diverse variations by directly sampling the encoded latent shape. 

In future work we plan to (a) obtain user feedback to improve the estimated structure of un-annotated data and use this feedback to improve the latent embedding; (b) look at more nuanced representations of fine-grained part geometry and its correlations with structural motifs; (c) estimate point-level correspondences between raw shape data from images or point clouds and the learned models so as to transfer textures and other appearance information to the model and more generally study style-content factorizations; and (d) extend these ideas from individual objects to entire scenes, where the objects now become the parts and we now focus on object relationships in scenes.

% \todo{Future work:}
% 1. {Improved supervision data} project unannotated data, get then structured, user interaction to improve the 'annotations', re-learn latent space;
% 2. {Texture transfer} image to pointcloud; unify images-scans-pc and provide new options for denoising, completion, .. generally correspondence transfer
% 3. {Island, bridge, ... } object level and hierarchy of scenes; is it possible to link across model families? explore style-content separation. 

\begin{acks}
This project was supported by a Vannevar Bush Faculty Fellowship, NSF grant RI-1764078, NSF grant CCF-1514305, a Google Research award, an ERC Starting Grant (SmartGeometry StG-2013-335373), ERC PoC Grant (SemanticCity), Google Faculty Awards, Google PhD Fellowships, Royal Society Advanced Newton Fellowship, KAUST OSR number CRG2017-3426 and gifts from Adobe, Autodesk and Qualcomm.
We especially thank Kun Liu, Peilang Zhu, Yan Zhang, and Kai Xu for the help preparing binary symmetry hierarchies~\cite{wang2011symmetry,grass:li:2017} on PartNet~\cite{mo2018partnet}. We also thank the anonymous reviewers for their fruitful suggestions.
\end{acks}

\bibliographystyle{ACM-Reference-Format}
\bibliography{sgen}


\begin{thebibliography}{80}

%%% the Web address.

\ifx \showCODEN    \undefined \def \showCODEN     #1{\unskip}     \fi
\ifx \showDOI      \undefined \def \showDOI       #1{#1}\fi
\ifx \showISBNx    \undefined \def \showISBNx     #1{\unskip}     \fi
\ifx \showISBNxiii \undefined \def \showISBNxiii  #1{\unskip}     \fi
\ifx \showISSN     \undefined \def \showISSN      #1{\unskip}     \fi
\ifx \showLCCN     \undefined \def \showLCCN      #1{\unskip}     \fi
\ifx \shownote     \undefined \def \shownote      #1{#1}          \fi
\ifx \showarticletitle \undefined \def \showarticletitle #1{#1}   \fi
\ifx \showURL      \undefined \def \showURL       {\relax}        \fi
% invisible to TeX
\providecommand\bibfield[2]{#2}
\providecommand\bibinfo[2]{#2}
\providecommand\natexlab[1]{#1}
\providecommand\showeprint[2][]{arXiv:#2}

\bibitem[\protect\citeauthoryear{Achlioptas, Diamanti, Mitliagkas, and
  Guibas}{Achlioptas et~al\mbox{.}}{2018}]%
        {achlioptas2017learning}
\bibfield{author}{\bibinfo{person}{Panos Achlioptas}, \bibinfo{person}{Olga
  Diamanti}, \bibinfo{person}{Ioannis Mitliagkas}, {and}
  \bibinfo{person}{Leonidas Guibas}.} \bibinfo{year}{2018}\natexlab{}.
\newblock \showarticletitle{Learning representations and generative models for
  3d point clouds}.
\newblock \bibinfo{journal}{\emph{International Conference on Machine Learning
  (ICML)}} (\bibinfo{year}{2018}).
\newblock


\bibitem[\protect\citeauthoryear{Achlioptas, Fan, Hawkins, Goodman, and
  Guibas}{Achlioptas et~al\mbox{.}}{2019}]%
        {achlioptas2019shapeglot}
\bibfield{author}{\bibinfo{person}{Panos Achlioptas}, \bibinfo{person}{Judy
  Fan}, \bibinfo{person}{Robert Hawkins}, \bibinfo{person}{Noah Goodman}, {and}
  \bibinfo{person}{Leonidas Guibas}.} \bibinfo{year}{2019}\natexlab{}.
\newblock \showarticletitle{ShapeGlot: Learning Language for Shape
  Differentiation}.
\newblock \bibinfo{journal}{\emph{Proceedings of the IEEE International
  Conference on Computer Vision (ICCV)}} (\bibinfo{year}{2019}).
\newblock


\bibitem[\protect\citeauthoryear{Arsalan~Soltani, Huang, Wu, Kulkarni, and
  Tenenbaum}{Arsalan~Soltani et~al\mbox{.}}{2017}]%
        {arsalan2017synthesizing}
\bibfield{author}{\bibinfo{person}{Amir Arsalan~Soltani},
  \bibinfo{person}{Haibin Huang}, \bibinfo{person}{Jiajun Wu},
  \bibinfo{person}{Tejas~D Kulkarni}, {and} \bibinfo{person}{Joshua~B
  Tenenbaum}.} \bibinfo{year}{2017}\natexlab{}.
\newblock \showarticletitle{Synthesizing 3d shapes via modeling multi-view
  depth maps and silhouettes with deep generative networks}. In
  \bibinfo{booktitle}{\emph{Proceedings of the IEEE conference on computer
  vision and pattern recognition (CVPR)}}. \bibinfo{pages}{1511--1519}.
\newblock


\bibitem[\protect\citeauthoryear{Barrow, Tenenbaum, Bolles, and Wolf}{Barrow
  et~al\mbox{.}}{1977}]%
        {Barrow:1977:Chamfer}
\bibfield{author}{\bibinfo{person}{H.~G. Barrow}, \bibinfo{person}{J.~M.
  Tenenbaum}, \bibinfo{person}{R.~C. Bolles}, {and} \bibinfo{person}{H.~C.
  Wolf}.} \bibinfo{year}{1977}\natexlab{}.
\newblock \showarticletitle{Parametric Correspondence and Chamfer Matching: Two
  New Techniques for Image Matching}. In \bibinfo{booktitle}{\emph{Proceedings
  of the 5th International Joint Conference on Artificial Intelligence - Volume
  2}} \emph{(\bibinfo{series}{IJCAI'77})}. \bibinfo{publisher}{Morgan Kaufmann
  Publishers Inc.}, \bibinfo{address}{San Francisco, CA, USA},
  \bibinfo{pages}{659--663}.
\newblock
\urldef\tempurl%
\url{http://dl.acm.org/citation.cfm?id=1622943.1622971}
\showURL{%
\tempurl}


\bibitem[\protect\citeauthoryear{Boscaini, Masci, Melzi, Bronstein, Castellani,
  and Vandergheynst}{Boscaini et~al\mbox{.}}{2015}]%
        {boscaini2015learning}
\bibfield{author}{\bibinfo{person}{Davide Boscaini}, \bibinfo{person}{Jonathan
  Masci}, \bibinfo{person}{Simone Melzi}, \bibinfo{person}{Michael~M
  Bronstein}, \bibinfo{person}{Umberto Castellani}, {and}
  \bibinfo{person}{Pierre Vandergheynst}.} \bibinfo{year}{2015}\natexlab{}.
\newblock \showarticletitle{Learning class-specific descriptors for deformable
  shapes using localized spectral convolutional networks}. In
  \bibinfo{booktitle}{\emph{Computer Graphics Forum}},
  Vol.~\bibinfo{volume}{34}. Wiley Online Library, \bibinfo{pages}{13--23}.
\newblock


\bibitem[\protect\citeauthoryear{Bruna, Zaremba, Szlam, and LeCun}{Bruna
  et~al\mbox{.}}{2014}]%
        {bruna2013spectral}
\bibfield{author}{\bibinfo{person}{Joan Bruna}, \bibinfo{person}{Wojciech
  Zaremba}, \bibinfo{person}{Arthur Szlam}, {and} \bibinfo{person}{Yann
  LeCun}.} \bibinfo{year}{2014}\natexlab{}.
\newblock \showarticletitle{Spectral networks and locally connected networks on
  graphs}.
\newblock \bibinfo{journal}{\emph{International Conference on Learning
  Representations (ICLR)}} (\bibinfo{year}{2014}).
\newblock


\bibitem[\protect\citeauthoryear{Chang, Funkhouser, Guibas, Hanrahan, Huang,
  Li, Savarese, Savva, Song, Su, et~al\mbox{.}}{Chang et~al\mbox{.}}{2015}]%
        {chang2015shapenet}
\bibfield{author}{\bibinfo{person}{Angel~X Chang}, \bibinfo{person}{Thomas
  Funkhouser}, \bibinfo{person}{Leonidas Guibas}, \bibinfo{person}{Pat
  Hanrahan}, \bibinfo{person}{Qixing Huang}, \bibinfo{person}{Zimo Li},
  \bibinfo{person}{Silvio Savarese}, \bibinfo{person}{Manolis Savva},
  \bibinfo{person}{Shuran Song}, \bibinfo{person}{Hao Su}, {et~al\mbox{.}}}
  \bibinfo{year}{2015}\natexlab{}.
\newblock \showarticletitle{Shapenet: An information-rich 3d model repository}.
\newblock \bibinfo{journal}{\emph{arXiv preprint arXiv:1512.03012}}
  (\bibinfo{year}{2015}).
\newblock


\bibitem[\protect\citeauthoryear{Chaudhuri, Kalogerakis, Guibas, and
  Koltun}{Chaudhuri et~al\mbox{.}}{2011}]%
        {chaudhuri2011probabilistic}
\bibfield{author}{\bibinfo{person}{Siddhartha Chaudhuri},
  \bibinfo{person}{Evangelos Kalogerakis}, \bibinfo{person}{Leonidas Guibas},
  {and} \bibinfo{person}{Vladlen Koltun}.} \bibinfo{year}{2011}\natexlab{}.
\newblock \showarticletitle{Probabilistic reasoning for assembly-based 3D
  modeling}. In \bibinfo{booktitle}{\emph{ACM Transactions on Graphics (TOG)}},
  Vol.~\bibinfo{volume}{30}. ACM, \bibinfo{pages}{35}.
\newblock


\bibitem[\protect\citeauthoryear{Choy, Xu, Gwak, Chen, and Savarese}{Choy
  et~al\mbox{.}}{2016}]%
        {choy20163d}
\bibfield{author}{\bibinfo{person}{Christopher~B Choy}, \bibinfo{person}{Danfei
  Xu}, \bibinfo{person}{JunYoung Gwak}, \bibinfo{person}{Kevin Chen}, {and}
  \bibinfo{person}{Silvio Savarese}.} \bibinfo{year}{2016}\natexlab{}.
\newblock \showarticletitle{3d-r2n2: A unified approach for single and
  multi-view 3d object reconstruction}. In \bibinfo{booktitle}{\emph{European
  conference on computer vision}}. Springer, \bibinfo{pages}{628--644}.
\newblock


\bibitem[\protect\citeauthoryear{Dai, Chang, Savva, Halber, Funkhouser, and
  Nie{\ss}ner}{Dai et~al\mbox{.}}{2017}]%
        {dai2017scannet}
\bibfield{author}{\bibinfo{person}{Angela Dai}, \bibinfo{person}{Angel~X.
  Chang}, \bibinfo{person}{Manolis Savva}, \bibinfo{person}{Maciej Halber},
  \bibinfo{person}{Thomas Funkhouser}, {and} \bibinfo{person}{Matthias
  Nie{\ss}ner}.} \bibinfo{year}{2017}\natexlab{}.
\newblock \showarticletitle{ScanNet: Richly-annotated 3D Reconstructions of
  Indoor Scenes}. In \bibinfo{booktitle}{\emph{Proc. Computer Vision and
  Pattern Recognition (CVPR), IEEE}}.
\newblock


\bibitem[\protect\citeauthoryear{Defferrard, Bresson, and
  Vandergheynst}{Defferrard et~al\mbox{.}}{2016}]%
        {defferrard2016convolutional}
\bibfield{author}{\bibinfo{person}{Micha{\"e}l Defferrard},
  \bibinfo{person}{Xavier Bresson}, {and} \bibinfo{person}{Pierre
  Vandergheynst}.} \bibinfo{year}{2016}\natexlab{}.
\newblock \showarticletitle{Convolutional neural networks on graphs with fast
  localized spectral filtering}. In \bibinfo{booktitle}{\emph{Advances in
  neural information processing systems}}. \bibinfo{pages}{3844--3852}.
\newblock


\bibitem[\protect\citeauthoryear{Deng, Dong, Socher, Li, Li, and Fei-Fei}{Deng
  et~al\mbox{.}}{2009}]%
        {deng2009imagenet}
\bibfield{author}{\bibinfo{person}{Jia Deng}, \bibinfo{person}{Wei Dong},
  \bibinfo{person}{Richard Socher}, \bibinfo{person}{Li-Jia Li},
  \bibinfo{person}{Kai Li}, {and} \bibinfo{person}{Li Fei-Fei}.}
  \bibinfo{year}{2009}\natexlab{}.
\newblock \showarticletitle{Imagenet: A large-scale hierarchical image
  database}. In \bibinfo{booktitle}{\emph{2009 IEEE conference on computer
  vision and pattern recognition}}. Ieee, \bibinfo{pages}{248--255}.
\newblock


\bibitem[\protect\citeauthoryear{Duvenaud, Maclaurin, Iparraguirre, Bombarell,
  Hirzel, Aspuru-Guzik, and Adams}{Duvenaud et~al\mbox{.}}{2015}]%
        {duvenaud2015convolutional}
\bibfield{author}{\bibinfo{person}{David~K Duvenaud}, \bibinfo{person}{Dougal
  Maclaurin}, \bibinfo{person}{Jorge Iparraguirre}, \bibinfo{person}{Rafael
  Bombarell}, \bibinfo{person}{Timothy Hirzel}, \bibinfo{person}{Al{\'a}n
  Aspuru-Guzik}, {and} \bibinfo{person}{Ryan~P Adams}.}
  \bibinfo{year}{2015}\natexlab{}.
\newblock \showarticletitle{Convolutional networks on graphs for learning
  molecular fingerprints}. In \bibinfo{booktitle}{\emph{Advances in neural
  information processing systems}}. \bibinfo{pages}{2224--2232}.
\newblock


\bibitem[\protect\citeauthoryear{Dyer, Kuncoro, Ballesteros, and Smith}{Dyer
  et~al\mbox{.}}{2016}]%
        {dyer2016recurrent}
\bibfield{author}{\bibinfo{person}{Chris Dyer}, \bibinfo{person}{Adhiguna
  Kuncoro}, \bibinfo{person}{Miguel Ballesteros}, {and} \bibinfo{person}{Noah~A
  Smith}.} \bibinfo{year}{2016}\natexlab{}.
\newblock \showarticletitle{Recurrent neural network grammars}.
\newblock \bibinfo{journal}{\emph{Proc. NAACL}} (\bibinfo{year}{2016}).
\newblock


\bibitem[\protect\citeauthoryear{Fan, Su, and Guibas}{Fan
  et~al\mbox{.}}{2017}]%
        {fan2017point}
\bibfield{author}{\bibinfo{person}{Haoqiang Fan}, \bibinfo{person}{Hao Su},
  {and} \bibinfo{person}{Leonidas~J Guibas}.} \bibinfo{year}{2017}\natexlab{}.
\newblock \showarticletitle{A point set generation network for 3d object
  reconstruction from a single image}. In \bibinfo{booktitle}{\emph{Proceedings
  of the IEEE conference on computer vision and pattern recognition}}.
  \bibinfo{pages}{605--613}.
\newblock


\bibitem[\protect\citeauthoryear{Fish, Averkiou, Van~Kaick, Sorkine-Hornung,
  Cohen-Or, and Mitra}{Fish et~al\mbox{.}}{2014}]%
        {fish2014meta}
\bibfield{author}{\bibinfo{person}{Noa Fish}, \bibinfo{person}{Melinos
  Averkiou}, \bibinfo{person}{Oliver Van~Kaick}, \bibinfo{person}{Olga
  Sorkine-Hornung}, \bibinfo{person}{Daniel Cohen-Or}, {and}
  \bibinfo{person}{Niloy~J Mitra}.} \bibinfo{year}{2014}\natexlab{}.
\newblock \showarticletitle{Meta-representation of shape families}.
\newblock \bibinfo{journal}{\emph{ACM Transactions on Graphics (TOG)}}
  \bibinfo{volume}{33}, \bibinfo{number}{4} (\bibinfo{year}{2014}),
  \bibinfo{pages}{34}.
\newblock


\bibitem[\protect\citeauthoryear{Fish, van Kaick, Bermano, and Cohen-Or}{Fish
  et~al\mbox{.}}{2016}]%
        {fish2016structure}
\bibfield{author}{\bibinfo{person}{Noa Fish}, \bibinfo{person}{Oliver van
  Kaick}, \bibinfo{person}{Amit Bermano}, {and} \bibinfo{person}{Daniel
  Cohen-Or}.} \bibinfo{year}{2016}\natexlab{}.
\newblock \showarticletitle{Structure-oriented networks of shape collections}.
\newblock \bibinfo{journal}{\emph{ACM Transactions on Graphics (TOG)}}
  \bibinfo{volume}{35}, \bibinfo{number}{6} (\bibinfo{year}{2016}),
  \bibinfo{pages}{171}.
\newblock


\bibitem[\protect\citeauthoryear{Ganapathi-Subramanian, Diamanti, Pirk, Tang,
  Niessner, and Guibas}{Ganapathi-Subramanian et~al\mbox{.}}{2018}]%
        {ganapathi2018parsing}
\bibfield{author}{\bibinfo{person}{Vignesh Ganapathi-Subramanian},
  \bibinfo{person}{Olga Diamanti}, \bibinfo{person}{Soeren Pirk},
  \bibinfo{person}{Chengcheng Tang}, \bibinfo{person}{Matthias Niessner}, {and}
  \bibinfo{person}{Leonidas Guibas}.} \bibinfo{year}{2018}\natexlab{}.
\newblock \showarticletitle{Parsing geometry using structure-aware shape
  templates}. In \bibinfo{booktitle}{\emph{2018 International Conference on 3D
  Vision (3DV)}}. IEEE, \bibinfo{pages}{672--681}.
\newblock


\bibitem[\protect\citeauthoryear{Golovinskiy and Funkhouser}{Golovinskiy and
  Funkhouser}{2009}]%
        {golovinskiy2009consistent}
\bibfield{author}{\bibinfo{person}{Aleksey Golovinskiy} {and}
  \bibinfo{person}{Thomas Funkhouser}.} \bibinfo{year}{2009}\natexlab{}.
\newblock \showarticletitle{Consistent segmentation of 3D models}.
\newblock \bibinfo{journal}{\emph{Computers \& Graphics}} \bibinfo{volume}{33},
  \bibinfo{number}{3} (\bibinfo{year}{2009}), \bibinfo{pages}{262--269}.
\newblock


\bibitem[\protect\citeauthoryear{Goodfellow, Pouget-Abadie, Mirza, Xu,
  Warde-Farley, Ozair, Courville, and Bengio}{Goodfellow et~al\mbox{.}}{2014}]%
        {goodfellow2014generative}
\bibfield{author}{\bibinfo{person}{Ian Goodfellow}, \bibinfo{person}{Jean
  Pouget-Abadie}, \bibinfo{person}{Mehdi Mirza}, \bibinfo{person}{Bing Xu},
  \bibinfo{person}{David Warde-Farley}, \bibinfo{person}{Sherjil Ozair},
  \bibinfo{person}{Aaron Courville}, {and} \bibinfo{person}{Yoshua Bengio}.}
  \bibinfo{year}{2014}\natexlab{}.
\newblock \showarticletitle{Generative adversarial nets}. In
  \bibinfo{booktitle}{\emph{Advances in neural information processing
  systems}}. \bibinfo{pages}{2672--2680}.
\newblock


\bibitem[\protect\citeauthoryear{Groueix, Fisher, Kim, Russell, and
  Aubry}{Groueix et~al\mbox{.}}{2018}]%
        {groueix2018papier}
\bibfield{author}{\bibinfo{person}{Thibault Groueix}, \bibinfo{person}{Matthew
  Fisher}, \bibinfo{person}{Vladimir~G Kim}, \bibinfo{person}{Bryan~C Russell},
  {and} \bibinfo{person}{Mathieu Aubry}.} \bibinfo{year}{2018}\natexlab{}.
\newblock \showarticletitle{A papier-m{\^a}ch{\'e} approach to learning 3d
  surface generation}. In \bibinfo{booktitle}{\emph{Proceedings of the IEEE
  conference on computer vision and pattern recognition}}.
  \bibinfo{pages}{216--224}.
\newblock


\bibitem[\protect\citeauthoryear{Gwak, Choy, Chandraker, Garg, and
  Savarese}{Gwak et~al\mbox{.}}{2017}]%
        {gwak2017weakly}
\bibfield{author}{\bibinfo{person}{JunYoung Gwak},
  \bibinfo{person}{Christopher~B Choy}, \bibinfo{person}{Manmohan Chandraker},
  \bibinfo{person}{Animesh Garg}, {and} \bibinfo{person}{Silvio Savarese}.}
  \bibinfo{year}{2017}\natexlab{}.
\newblock \showarticletitle{Weakly Supervised 3D Reconstruction with
  Adversarial Constraint}. In \bibinfo{booktitle}{\emph{3D Vision (3DV), 2017
  Fifth International Conference on 3D Vision}}.
\newblock


\bibitem[\protect\citeauthoryear{Hamilton, Ying, and Leskovec}{Hamilton
  et~al\mbox{.}}{2017}]%
        {hamilton2017inductive}
\bibfield{author}{\bibinfo{person}{Will Hamilton}, \bibinfo{person}{Zhitao
  Ying}, {and} \bibinfo{person}{Jure Leskovec}.}
  \bibinfo{year}{2017}\natexlab{}.
\newblock \showarticletitle{Inductive representation learning on large graphs}.
  In \bibinfo{booktitle}{\emph{Advances in Neural Information Processing
  Systems}}. \bibinfo{pages}{1024--1034}.
\newblock


\bibitem[\protect\citeauthoryear{He, Zhang, Ren, and Sun}{He
  et~al\mbox{.}}{2016}]%
        {he2016deep}
\bibfield{author}{\bibinfo{person}{Kaiming He}, \bibinfo{person}{Xiangyu
  Zhang}, \bibinfo{person}{Shaoqing Ren}, {and} \bibinfo{person}{Jian Sun}.}
  \bibinfo{year}{2016}\natexlab{}.
\newblock \showarticletitle{Deep residual learning for image recognition}. In
  \bibinfo{booktitle}{\emph{Proceedings of the IEEE conference on computer
  vision and pattern recognition}}. \bibinfo{pages}{770--778}.
\newblock


\bibitem[\protect\citeauthoryear{Hinton}{Hinton}{1990}]%
        {Hinton:1990:MPH:102418.102422}
\bibfield{author}{\bibinfo{person}{Geoffrey~E. Hinton}.}
  \bibinfo{year}{1990}\natexlab{}.
\newblock \showarticletitle{Mapping Part-whole Hierarchies into Connectionist
  Networks}.
\newblock \bibinfo{journal}{\emph{Artif. Intell.}} \bibinfo{volume}{46},
  \bibinfo{number}{1-2} (\bibinfo{date}{Nov.} \bibinfo{year}{1990}),
  \bibinfo{pages}{47--75}.
\newblock
\showISSN{0004-3702}
\urldef\tempurl%
\url{https://doi.org/10.1016/0004-3702(90)90004-J}
\showDOI{\tempurl}


\bibitem[\protect\citeauthoryear{Hu, Fan, and Liu}{Hu et~al\mbox{.}}{2012}]%
        {hu2012co}
\bibfield{author}{\bibinfo{person}{Ruizhen Hu}, \bibinfo{person}{Lubin Fan},
  {and} \bibinfo{person}{Ligang Liu}.} \bibinfo{year}{2012}\natexlab{}.
\newblock \showarticletitle{Co-segmentation of 3d shapes via subspace
  clustering}. In \bibinfo{booktitle}{\emph{Computer graphics forum}},
  Vol.~\bibinfo{volume}{31}. Wiley Online Library, \bibinfo{pages}{1703--1713}.
\newblock


\bibitem[\protect\citeauthoryear{Huang, Koltun, and Guibas}{Huang
  et~al\mbox{.}}{2011}]%
        {huang2011joint}
\bibfield{author}{\bibinfo{person}{Qixing Huang}, \bibinfo{person}{Vladlen
  Koltun}, {and} \bibinfo{person}{Leonidas Guibas}.}
  \bibinfo{year}{2011}\natexlab{}.
\newblock \showarticletitle{Joint shape segmentation with linear programming}.
  In \bibinfo{booktitle}{\emph{ACM transactions on graphics (TOG)}},
  Vol.~\bibinfo{volume}{30}. ACM, \bibinfo{pages}{125}.
\newblock


\bibitem[\protect\citeauthoryear{Ioffe and Szegedy}{Ioffe and Szegedy}{2015}]%
        {ioffe2015batch}
\bibfield{author}{\bibinfo{person}{Sergey Ioffe} {and}
  \bibinfo{person}{Christian Szegedy}.} \bibinfo{year}{2015}\natexlab{}.
\newblock \showarticletitle{Batch normalization: Accelerating deep network
  training by reducing internal covariate shift}.
\newblock \bibinfo{journal}{\emph{Proceedings of the 32nd International
  Conference on Machine Learning, PMLR 37:448-456, 2015.}}
  (\bibinfo{year}{2015}).
\newblock


\bibitem[\protect\citeauthoryear{Kalogerakis, Averkiou, Maji, and
  Chaudhuri}{Kalogerakis et~al\mbox{.}}{2017}]%
        {kalogerakis20173d}
\bibfield{author}{\bibinfo{person}{Evangelos Kalogerakis},
  \bibinfo{person}{Melinos Averkiou}, \bibinfo{person}{Subhransu Maji}, {and}
  \bibinfo{person}{Siddhartha Chaudhuri}.} \bibinfo{year}{2017}\natexlab{}.
\newblock \showarticletitle{3D shape segmentation with projective convolutional
  networks}. In \bibinfo{booktitle}{\emph{Proceedings of the IEEE Conference on
  Computer Vision and Pattern Recognition}}. \bibinfo{pages}{3779--3788}.
\newblock


\bibitem[\protect\citeauthoryear{Kalogerakis, Chaudhuri, Koller, and
  Koltun}{Kalogerakis et~al\mbox{.}}{2012}]%
        {kalogerakis2012probabilistic}
\bibfield{author}{\bibinfo{person}{Evangelos Kalogerakis},
  \bibinfo{person}{Siddhartha Chaudhuri}, \bibinfo{person}{Daphne Koller},
  {and} \bibinfo{person}{Vladlen Koltun}.} \bibinfo{year}{2012}\natexlab{}.
\newblock \showarticletitle{A probabilistic model for component-based shape
  synthesis}.
\newblock \bibinfo{journal}{\emph{ACM Transactions on Graphics (TOG)}}
  \bibinfo{volume}{31}, \bibinfo{number}{4} (\bibinfo{year}{2012}),
  \bibinfo{pages}{55}.
\newblock


\bibitem[\protect\citeauthoryear{Kalogerakis, Hertzmann, and Singh}{Kalogerakis
  et~al\mbox{.}}{2010}]%
        {kalogerakis2010learning}
\bibfield{author}{\bibinfo{person}{Evangelos Kalogerakis},
  \bibinfo{person}{Aaron Hertzmann}, {and} \bibinfo{person}{Karan Singh}.}
  \bibinfo{year}{2010}\natexlab{}.
\newblock \showarticletitle{Learning 3D mesh segmentation and labeling}.
\newblock \bibinfo{journal}{\emph{ACM Transactions on Graphics (TOG)}}
  \bibinfo{volume}{29}, \bibinfo{number}{4} (\bibinfo{year}{2010}),
  \bibinfo{pages}{102}.
\newblock


\bibitem[\protect\citeauthoryear{Kalojanov, Lim, Mitra, and Kobbelt}{Kalojanov
  et~al\mbox{.}}{2019}]%
        {Kalojanov2019}
\bibfield{author}{\bibinfo{person}{Javor Kalojanov}, \bibinfo{person}{Isaak
  Lim}, \bibinfo{person}{Niloy Mitra}, {and} \bibinfo{person}{Leif Kobbelt}.}
  \bibinfo{year}{2019}\natexlab{}.
\newblock \showarticletitle{{String-Based Synthesis of Structured Shapes}}.
\newblock \bibinfo{journal}{\emph{Computer Graphics Forum}}
  \bibinfo{volume}{38}, \bibinfo{number}{2} (\bibinfo{year}{2019}),
  \bibinfo{pages}{027--036}.
\newblock


\bibitem[\protect\citeauthoryear{Kim, Li, Mitra, Chaudhuri, DiVerdi, and
  Funkhouser}{Kim et~al\mbox{.}}{2013}]%
        {kim2013learning}
\bibfield{author}{\bibinfo{person}{Vladimir~G Kim}, \bibinfo{person}{Wilmot
  Li}, \bibinfo{person}{Niloy~J Mitra}, \bibinfo{person}{Siddhartha Chaudhuri},
  \bibinfo{person}{Stephen DiVerdi}, {and} \bibinfo{person}{Thomas
  Funkhouser}.} \bibinfo{year}{2013}\natexlab{}.
\newblock \showarticletitle{Learning part-based templates from large
  collections of 3D shapes}.
\newblock \bibinfo{journal}{\emph{ACM Transactions on Graphics (TOG)}}
  \bibinfo{volume}{32}, \bibinfo{number}{4} (\bibinfo{year}{2013}),
  \bibinfo{pages}{70}.
\newblock


\bibitem[\protect\citeauthoryear{Kingma and Welling}{Kingma and
  Welling}{2014}]%
        {kingma2013auto}
\bibfield{author}{\bibinfo{person}{Diederik~P Kingma} {and}
  \bibinfo{person}{Max Welling}.} \bibinfo{year}{2014}\natexlab{}.
\newblock \showarticletitle{Auto-encoding variational bayes}.
\newblock \bibinfo{journal}{\emph{International Conference on Learning
  Representations (ICLR)}} (\bibinfo{year}{2014}).
\newblock


\bibitem[\protect\citeauthoryear{Kipf and Welling}{Kipf and Welling}{2017}]%
        {kipf2017semi}
\bibfield{author}{\bibinfo{person}{Thomas~N. Kipf} {and} \bibinfo{person}{Max
  Welling}.} \bibinfo{year}{2017}\natexlab{}.
\newblock \showarticletitle{Semi-Supervised Classification with Graph
  Convolutional Networks}. In \bibinfo{booktitle}{\emph{International
  Conference on Learning Representations (ICLR)}}.
\newblock


\bibitem[\protect\citeauthoryear{Li, Zaheer, Zhang, Poczos, and
  Salakhutdinov}{Li et~al\mbox{.}}{2019b}]%
        {li2018point}
\bibfield{author}{\bibinfo{person}{Chun-Liang Li}, \bibinfo{person}{Manzil
  Zaheer}, \bibinfo{person}{Yang Zhang}, \bibinfo{person}{Barnabas Poczos},
  {and} \bibinfo{person}{Ruslan Salakhutdinov}.}
  \bibinfo{year}{2019}\natexlab{b}.
\newblock \showarticletitle{Point cloud gan}.
\newblock \bibinfo{journal}{\emph{In ICLR Workshop on Deep Generative Models
  for Highly Structured Data}} (\bibinfo{year}{2019}).
\newblock


\bibitem[\protect\citeauthoryear{Li, Xu, Chaudhuri, Yumer, Zhang, and
  Guibas}{Li et~al\mbox{.}}{2017}]%
        {grass:li:2017}
\bibfield{author}{\bibinfo{person}{Jun Li}, \bibinfo{person}{Kai Xu},
  \bibinfo{person}{Siddhartha Chaudhuri}, \bibinfo{person}{Ersin Yumer},
  \bibinfo{person}{Hao Zhang}, {and} \bibinfo{person}{Leonidas Guibas}.}
  \bibinfo{year}{2017}\natexlab{}.
\newblock \showarticletitle{GRASS: Generative Recursive Autoencoders for Shape
  Structures}.
\newblock \bibinfo{journal}{\emph{ACM Transactions on Graphics}}
  \bibinfo{volume}{36}, \bibinfo{number}{4} (\bibinfo{year}{2017}).
\newblock


\bibitem[\protect\citeauthoryear{Li, Patil, Xu, Chaudhuri, Khan, Shamir, Tu,
  Chen, Cohen-Or, and Zhang}{Li et~al\mbox{.}}{2019a}]%
        {li2019grains}
\bibfield{author}{\bibinfo{person}{Manyi Li}, \bibinfo{person}{Akshay~Gadi
  Patil}, \bibinfo{person}{Kai Xu}, \bibinfo{person}{Siddhartha Chaudhuri},
  \bibinfo{person}{Owais Khan}, \bibinfo{person}{Ariel Shamir},
  \bibinfo{person}{Changhe Tu}, \bibinfo{person}{Baoquan Chen},
  \bibinfo{person}{Daniel Cohen-Or}, {and} \bibinfo{person}{Hao Zhang}.}
  \bibinfo{year}{2019}\natexlab{a}.
\newblock \showarticletitle{Grains: Generative recursive autoencoders for
  indoor scenes}.
\newblock \bibinfo{journal}{\emph{ACM Transactions on Graphics (TOG)}}
  \bibinfo{volume}{38}, \bibinfo{number}{2} (\bibinfo{year}{2019}),
  \bibinfo{pages}{12}.
\newblock


\bibitem[\protect\citeauthoryear{Li, Su, Qi, Fish, Cohen-Or, and Guibas}{Li
  et~al\mbox{.}}{2015}]%
        {Li:2015:JES:2816795.2818071}
\bibfield{author}{\bibinfo{person}{Yangyan Li}, \bibinfo{person}{Hao Su},
  \bibinfo{person}{Charles~Ruizhongtai Qi}, \bibinfo{person}{Noa Fish},
  \bibinfo{person}{Daniel Cohen-Or}, {and} \bibinfo{person}{Leonidas~J.
  Guibas}.} \bibinfo{year}{2015}\natexlab{}.
\newblock \showarticletitle{Joint Embeddings of Shapes and Images via CNN Image
  Purification}.
\newblock \bibinfo{journal}{\emph{ACM Trans. Graph.}} \bibinfo{volume}{34},
  \bibinfo{number}{6}, Article \bibinfo{articleno}{234} (\bibinfo{date}{Oct.}
  \bibinfo{year}{2015}), \bibinfo{numpages}{12}~pages.
\newblock
\showISSN{0730-0301}
\urldef\tempurl%
\url{https://doi.org/10.1145/2816795.2818071}
\showDOI{\tempurl}


\bibitem[\protect\citeauthoryear{Li, Vinyals, Dyer, Pascanu, and Battaglia}{Li
  et~al\mbox{.}}{2018}]%
        {li2018learning}
\bibfield{author}{\bibinfo{person}{Yujia Li}, \bibinfo{person}{Oriol Vinyals},
  \bibinfo{person}{Chris Dyer}, \bibinfo{person}{Razvan Pascanu}, {and}
  \bibinfo{person}{Peter Battaglia}.} \bibinfo{year}{2018}\natexlab{}.
\newblock \showarticletitle{Learning deep generative models of graphs}.
\newblock \bibinfo{journal}{\emph{arXiv preprint arXiv:1803.03324}}
  (\bibinfo{year}{2018}).
\newblock


\bibitem[\protect\citeauthoryear{Liu, Chaudhuri, Kim, Huang, Mitra, and
  Funkhouser}{Liu et~al\mbox{.}}{2014}]%
        {liu2014creating}
\bibfield{author}{\bibinfo{person}{Tianqiang Liu}, \bibinfo{person}{Siddhartha
  Chaudhuri}, \bibinfo{person}{Vladimir~G Kim}, \bibinfo{person}{Qixing Huang},
  \bibinfo{person}{Niloy~J Mitra}, {and} \bibinfo{person}{Thomas Funkhouser}.}
  \bibinfo{year}{2014}\natexlab{}.
\newblock \showarticletitle{Creating consistent scene graphs using a
  probabilistic grammar}.
\newblock \bibinfo{journal}{\emph{ACM Transactions on Graphics (TOG)}}
  \bibinfo{volume}{33}, \bibinfo{number}{6} (\bibinfo{year}{2014}),
  \bibinfo{pages}{211}.
\newblock


\bibitem[\protect\citeauthoryear{Maaten and Hinton}{Maaten and Hinton}{2008}]%
        {maaten2008visualizing}
\bibfield{author}{\bibinfo{person}{Laurens van~der Maaten} {and}
  \bibinfo{person}{Geoffrey Hinton}.} \bibinfo{year}{2008}\natexlab{}.
\newblock \showarticletitle{Visualizing data using t-SNE}.
\newblock \bibinfo{journal}{\emph{Journal of machine learning research}}
  \bibinfo{volume}{9}, \bibinfo{number}{Nov} (\bibinfo{year}{2008}),
  \bibinfo{pages}{2579--2605}.
\newblock


\bibitem[\protect\citeauthoryear{Maddison and Tarlow}{Maddison and
  Tarlow}{2014}]%
        {maddison2014structured}
\bibfield{author}{\bibinfo{person}{Chris Maddison} {and}
  \bibinfo{person}{Daniel Tarlow}.} \bibinfo{year}{2014}\natexlab{}.
\newblock \showarticletitle{Structured generative models of natural source
  code}. In \bibinfo{booktitle}{\emph{International Conference on Machine
  Learning (ICML)}}. \bibinfo{pages}{649--657}.
\newblock


\bibitem[\protect\citeauthoryear{Makadia and Yumer}{Makadia and Yumer}{2014}]%
        {makadia2014learning}
\bibfield{author}{\bibinfo{person}{Ameesh Makadia} {and}
  \bibinfo{person}{Mehmet~Ersin Yumer}.} \bibinfo{year}{2014}\natexlab{}.
\newblock \showarticletitle{Learning 3d part detection from sparsely labeled
  data}. In \bibinfo{booktitle}{\emph{2014 2nd International Conference on 3D
  Vision}}, Vol.~\bibinfo{volume}{1}. IEEE, \bibinfo{pages}{311--318}.
\newblock


\bibitem[\protect\citeauthoryear{Masci, Boscaini, Bronstein, and
  Vandergheynst}{Masci et~al\mbox{.}}{2015}]%
        {masci2015geodesic}
\bibfield{author}{\bibinfo{person}{Jonathan Masci}, \bibinfo{person}{Davide
  Boscaini}, \bibinfo{person}{Michael Bronstein}, {and} \bibinfo{person}{Pierre
  Vandergheynst}.} \bibinfo{year}{2015}\natexlab{}.
\newblock \showarticletitle{Geodesic convolutional neural networks on
  riemannian manifolds}. In \bibinfo{booktitle}{\emph{Proceedings of the IEEE
  international conference on computer vision workshops}}.
  \bibinfo{pages}{37--45}.
\newblock


\bibitem[\protect\citeauthoryear{Mitra, Wand, Zhang, Cohen-Or, Kim, and
  Huang}{Mitra et~al\mbox{.}}{2014}]%
        {Mitra:2014:SSP}
\bibfield{author}{\bibinfo{person}{Niloy~J. Mitra}, \bibinfo{person}{Michael
  Wand}, \bibinfo{person}{Hao Zhang}, \bibinfo{person}{Daniel Cohen-Or},
  \bibinfo{person}{Vladimir Kim}, {and} \bibinfo{person}{Qi-Xing Huang}.}
  \bibinfo{year}{2014}\natexlab{}.
\newblock \showarticletitle{Structure-aware Shape Processing}. In
  \bibinfo{booktitle}{\emph{ACM SIGGRAPH 2014 Courses}}
  \emph{(\bibinfo{series}{SIGGRAPH '14})}. \bibinfo{publisher}{ACM},
  \bibinfo{address}{New York, NY, USA}, Article \bibinfo{articleno}{13},
  \bibinfo{numpages}{21}~pages.
\newblock
\showISBNx{978-1-4503-2962-0}
\urldef\tempurl%
\url{https://doi.org/10.1145/2614028.2615401}
\showDOI{\tempurl}


\bibitem[\protect\citeauthoryear{Mo, Zhu, Chang, Yi, Tripathi, Guibas, and
  Su}{Mo et~al\mbox{.}}{2019}]%
        {mo2018partnet}
\bibfield{author}{\bibinfo{person}{Kaichun Mo}, \bibinfo{person}{Shilin Zhu},
  \bibinfo{person}{Angel Chang}, \bibinfo{person}{Li Yi},
  \bibinfo{person}{Subarna Tripathi}, \bibinfo{person}{Leonidas Guibas}, {and}
  \bibinfo{person}{Hao Su}.} \bibinfo{year}{2019}\natexlab{}.
\newblock \showarticletitle{{PartNet}: A Large-scale Benchmark for Fine-grained
  and Hierarchical Part-level {3D} Object Understanding}. In
  \bibinfo{booktitle}{\emph{Proceedings of the IEEE Conference on Computer
  Vision and Pattern Recognition (CVPR)}}.
\newblock


\bibitem[\protect\citeauthoryear{M{\"u}ller, Wonka, Haegler, Ulmer, and
  Van~Gool}{M{\"u}ller et~al\mbox{.}}{2006}]%
        {muller2006procedural}
\bibfield{author}{\bibinfo{person}{Pascal M{\"u}ller}, \bibinfo{person}{Peter
  Wonka}, \bibinfo{person}{Simon Haegler}, \bibinfo{person}{Andreas Ulmer},
  {and} \bibinfo{person}{Luc Van~Gool}.} \bibinfo{year}{2006}\natexlab{}.
\newblock \showarticletitle{Procedural modeling of buildings}.
\newblock \bibinfo{journal}{\emph{Acm Transactions On Graphics (Tog)}}
  \bibinfo{volume}{25}, \bibinfo{number}{3} (\bibinfo{year}{2006}),
  \bibinfo{pages}{614--623}.
\newblock


\bibitem[\protect\citeauthoryear{Nash and Williams}{Nash and Williams}{2017}]%
        {nash2017shape}
\bibfield{author}{\bibinfo{person}{Charlie Nash} {and}
  \bibinfo{person}{Christopher~KI Williams}.} \bibinfo{year}{2017}\natexlab{}.
\newblock \showarticletitle{The shape variational autoencoder: A deep
  generative model of part-segmented 3D objects}. In
  \bibinfo{booktitle}{\emph{Computer Graphics Forum}},
  Vol.~\bibinfo{volume}{36}. Wiley Online Library, \bibinfo{pages}{1--12}.
\newblock


\bibitem[\protect\citeauthoryear{Paszke, Gross, Chintala, Chanan, Yang, DeVito,
  Lin, Desmaison, Antiga, and Lerer}{Paszke et~al\mbox{.}}{2017}]%
        {paszke2017automatic}
\bibfield{author}{\bibinfo{person}{Adam Paszke}, \bibinfo{person}{Sam Gross},
  \bibinfo{person}{Soumith Chintala}, \bibinfo{person}{Gregory Chanan},
  \bibinfo{person}{Edward Yang}, \bibinfo{person}{Zachary DeVito},
  \bibinfo{person}{Zeming Lin}, \bibinfo{person}{Alban Desmaison},
  \bibinfo{person}{Luca Antiga}, {and} \bibinfo{person}{Adam Lerer}.}
  \bibinfo{year}{2017}\natexlab{}.
\newblock \showarticletitle{Automatic differentiation in PyTorch}.
\newblock  (\bibinfo{year}{2017}).
\newblock


\bibitem[\protect\citeauthoryear{Qi, Su, Mo, and Guibas}{Qi
  et~al\mbox{.}}{2017a}]%
        {qi2017pointnet}
\bibfield{author}{\bibinfo{person}{Charles~R Qi}, \bibinfo{person}{Hao Su},
  \bibinfo{person}{Kaichun Mo}, {and} \bibinfo{person}{Leonidas~J Guibas}.}
  \bibinfo{year}{2017}\natexlab{a}.
\newblock \showarticletitle{Pointnet: Deep learning on point sets for 3d
  classification and segmentation}. In \bibinfo{booktitle}{\emph{Proceedings of
  the IEEE Conference on Computer Vision and Pattern Recognition}}.
  \bibinfo{pages}{652--660}.
\newblock


\bibitem[\protect\citeauthoryear{Qi, Yi, Su, and Guibas}{Qi
  et~al\mbox{.}}{2017b}]%
        {qi2017pointnet++}
\bibfield{author}{\bibinfo{person}{Charles~Ruizhongtai Qi}, \bibinfo{person}{Li
  Yi}, \bibinfo{person}{Hao Su}, {and} \bibinfo{person}{Leonidas~J Guibas}.}
  \bibinfo{year}{2017}\natexlab{b}.
\newblock \showarticletitle{Pointnet++: Deep hierarchical feature learning on
  point sets in a metric space}. In \bibinfo{booktitle}{\emph{Advances in
  Neural Information Processing Systems}}. \bibinfo{pages}{5099--5108}.
\newblock


\bibitem[\protect\citeauthoryear{Sidi, van Kaick, Kleiman, Zhang, and
  Cohen-Or}{Sidi et~al\mbox{.}}{2011}]%
        {sidi2011unsupervised}
\bibfield{author}{\bibinfo{person}{Oana Sidi}, \bibinfo{person}{Oliver van
  Kaick}, \bibinfo{person}{Yanir Kleiman}, \bibinfo{person}{Hao Zhang}, {and}
  \bibinfo{person}{Daniel Cohen-Or}.} \bibinfo{year}{2011}\natexlab{}.
\newblock \bibinfo{booktitle}{\emph{Unsupervised co-segmentation of a set of
  shapes via descriptor-space spectral clustering}}. Vol.~\bibinfo{volume}{30}.
\newblock \bibinfo{publisher}{ACM}.
\newblock


\bibitem[\protect\citeauthoryear{Sinha, Unmesh, Huang, and Ramani}{Sinha
  et~al\mbox{.}}{2017}]%
        {sinha2017surfnet}
\bibfield{author}{\bibinfo{person}{Ayan Sinha}, \bibinfo{person}{Asim Unmesh},
  \bibinfo{person}{Qixing Huang}, {and} \bibinfo{person}{Karthik Ramani}.}
  \bibinfo{year}{2017}\natexlab{}.
\newblock \showarticletitle{Surfnet: Generating 3d shape surfaces using deep
  residual networks}. In \bibinfo{booktitle}{\emph{Proceedings of the IEEE
  conference on computer vision and pattern recognition}}.
  \bibinfo{pages}{6040--6049}.
\newblock


\bibitem[\protect\citeauthoryear{Socher, Huval, Bath, Manning, and Ng}{Socher
  et~al\mbox{.}}{2012}]%
        {socher2012convolutional}
\bibfield{author}{\bibinfo{person}{Richard Socher}, \bibinfo{person}{Brody
  Huval}, \bibinfo{person}{Bharath Bath}, \bibinfo{person}{Christopher~D
  Manning}, {and} \bibinfo{person}{Andrew~Y Ng}.}
  \bibinfo{year}{2012}\natexlab{}.
\newblock \showarticletitle{Convolutional-recursive deep learning for 3d object
  classification}. In \bibinfo{booktitle}{\emph{Advances in neural information
  processing systems}}. \bibinfo{pages}{656--664}.
\newblock


\bibitem[\protect\citeauthoryear{Socher, Lin, Manning, and Ng}{Socher
  et~al\mbox{.}}{2011}]%
        {socher2011parsing}
\bibfield{author}{\bibinfo{person}{Richard Socher}, \bibinfo{person}{Cliff~C
  Lin}, \bibinfo{person}{Chris Manning}, {and} \bibinfo{person}{Andrew~Y Ng}.}
  \bibinfo{year}{2011}\natexlab{}.
\newblock \showarticletitle{Parsing natural scenes and natural language with
  recursive neural networks}. In \bibinfo{booktitle}{\emph{Proceedings of the
  28th international conference on machine learning (ICML-11)}}.
  \bibinfo{pages}{129--136}.
\newblock


\bibitem[\protect\citeauthoryear{Sung, Su, Kim, Chaudhuri, and Guibas}{Sung
  et~al\mbox{.}}{2017}]%
        {sung2017complementme}
\bibfield{author}{\bibinfo{person}{Minhyuk Sung}, \bibinfo{person}{Hao Su},
  \bibinfo{person}{Vladimir~G Kim}, \bibinfo{person}{Siddhartha Chaudhuri},
  {and} \bibinfo{person}{Leonidas Guibas}.} \bibinfo{year}{2017}\natexlab{}.
\newblock \showarticletitle{Complementme: weakly-supervised component
  suggestions for 3D modeling}.
\newblock \bibinfo{journal}{\emph{ACM Transactions on Graphics (TOG)}}
  \bibinfo{volume}{36}, \bibinfo{number}{6} (\bibinfo{year}{2017}),
  \bibinfo{pages}{226}.
\newblock


\bibitem[\protect\citeauthoryear{Tatarchenko, Dosovitskiy, and
  Brox}{Tatarchenko et~al\mbox{.}}{2017}]%
        {tatarchenko2017octree}
\bibfield{author}{\bibinfo{person}{Maxim Tatarchenko}, \bibinfo{person}{Alexey
  Dosovitskiy}, {and} \bibinfo{person}{Thomas Brox}.}
  \bibinfo{year}{2017}\natexlab{}.
\newblock \showarticletitle{Octree generating networks: Efficient convolutional
  architectures for high-resolution 3d outputs}. In
  \bibinfo{booktitle}{\emph{Proceedings of the IEEE International Conference on
  Computer Vision}}. \bibinfo{pages}{2088--2096}.
\newblock


\bibitem[\protect\citeauthoryear{Tian, Luo, Sun, Ellis, Freeman, Tenenbaum, and
  Wu}{Tian et~al\mbox{.}}{2019}]%
        {tian2018learning}
\bibfield{author}{\bibinfo{person}{Yonglong Tian}, \bibinfo{person}{Andrew
  Luo}, \bibinfo{person}{Xingyuan Sun}, \bibinfo{person}{Kevin Ellis},
  \bibinfo{person}{William~T. Freeman}, \bibinfo{person}{Joshua~B. Tenenbaum},
  {and} \bibinfo{person}{Jiajun Wu}.} \bibinfo{year}{2019}\natexlab{}.
\newblock \showarticletitle{Learning to Infer and Execute 3D Shape Programs}.
  In \bibinfo{booktitle}{\emph{International Conference on Learning
  Representations}}.
\newblock


\bibitem[\protect\citeauthoryear{Tulsiani, Su, Guibas, Efros, and
  Malik}{Tulsiani et~al\mbox{.}}{2017}]%
        {tulsiani2017learning}
\bibfield{author}{\bibinfo{person}{Shubham Tulsiani}, \bibinfo{person}{Hao Su},
  \bibinfo{person}{Leonidas~J Guibas}, \bibinfo{person}{Alexei~A Efros}, {and}
  \bibinfo{person}{Jitendra Malik}.} \bibinfo{year}{2017}\natexlab{}.
\newblock \showarticletitle{Learning shape abstractions by assembling
  volumetric primitives}. In \bibinfo{booktitle}{\emph{Proceedings of the IEEE
  Conference on Computer Vision and Pattern Recognition}}.
  \bibinfo{pages}{2635--2643}.
\newblock


\bibitem[\protect\citeauthoryear{Van~Kaick, Xu, Zhang, Wang, Sun, Shamir, and
  Cohen-Or}{Van~Kaick et~al\mbox{.}}{2013}]%
        {van2013co}
\bibfield{author}{\bibinfo{person}{Oliver Van~Kaick}, \bibinfo{person}{Kai Xu},
  \bibinfo{person}{Hao Zhang}, \bibinfo{person}{Yanzhen Wang},
  \bibinfo{person}{Shuyang Sun}, \bibinfo{person}{Ariel Shamir}, {and}
  \bibinfo{person}{Daniel Cohen-Or}.} \bibinfo{year}{2013}\natexlab{}.
\newblock \showarticletitle{Co-hierarchical analysis of shape structures}.
\newblock \bibinfo{journal}{\emph{ACM Transactions on Graphics (TOG)}}
  \bibinfo{volume}{32}, \bibinfo{number}{4} (\bibinfo{year}{2013}),
  \bibinfo{pages}{69}.
\newblock


\bibitem[\protect\citeauthoryear{Veli{\v{c}}kovi{\'c}, Cucurull, Casanova,
  Romero, Lio, and Bengio}{Veli{\v{c}}kovi{\'c} et~al\mbox{.}}{2017}]%
        {velivckovic2017graph}
\bibfield{author}{\bibinfo{person}{Petar Veli{\v{c}}kovi{\'c}},
  \bibinfo{person}{Guillem Cucurull}, \bibinfo{person}{Arantxa Casanova},
  \bibinfo{person}{Adriana Romero}, \bibinfo{person}{Pietro Lio}, {and}
  \bibinfo{person}{Yoshua Bengio}.} \bibinfo{year}{2017}\natexlab{}.
\newblock \showarticletitle{Graph attention networks}.
\newblock \bibinfo{journal}{\emph{arXiv preprint arXiv:1710.10903}}
  (\bibinfo{year}{2017}).
\newblock


\bibitem[\protect\citeauthoryear{Vinyals, Kaiser, Koo, Petrov, Sutskever, and
  Hinton}{Vinyals et~al\mbox{.}}{2015}]%
        {vinyals2015grammar}
\bibfield{author}{\bibinfo{person}{Oriol Vinyals}, \bibinfo{person}{{\L}ukasz
  Kaiser}, \bibinfo{person}{Terry Koo}, \bibinfo{person}{Slav Petrov},
  \bibinfo{person}{Ilya Sutskever}, {and} \bibinfo{person}{Geoffrey Hinton}.}
  \bibinfo{year}{2015}\natexlab{}.
\newblock \showarticletitle{Grammar as a foreign language}. In
  \bibinfo{booktitle}{\emph{Advances in neural information processing
  systems}}. \bibinfo{pages}{2773--2781}.
\newblock


\bibitem[\protect\citeauthoryear{Wang, Schor, Hu, Huang, Cohen-Or, and
  Huang}{Wang et~al\mbox{.}}{2018a}]%
        {wang2018global}
\bibfield{author}{\bibinfo{person}{Hao Wang}, \bibinfo{person}{Nadav Schor},
  \bibinfo{person}{Ruizhen Hu}, \bibinfo{person}{Haibin Huang},
  \bibinfo{person}{Daniel Cohen-Or}, {and} \bibinfo{person}{Hui Huang}.}
  \bibinfo{year}{2018}\natexlab{a}.
\newblock \showarticletitle{Global-to-local generative model for 3d shapes}. In
  \bibinfo{booktitle}{\emph{SIGGRAPH Asia 2018 Technical Papers}}. ACM,
  \bibinfo{pages}{214}.
\newblock


\bibitem[\protect\citeauthoryear{Wang, Sun, Liu, and Tong}{Wang
  et~al\mbox{.}}{2018b}]%
        {wang2018adaptive}
\bibfield{author}{\bibinfo{person}{Peng-Shuai Wang}, \bibinfo{person}{Chun-Yu
  Sun}, \bibinfo{person}{Yang Liu}, {and} \bibinfo{person}{Xin Tong}.}
  \bibinfo{year}{2018}\natexlab{b}.
\newblock \showarticletitle{Adaptive O-CNN: a patch-based deep representation
  of 3D shapes}. In \bibinfo{booktitle}{\emph{SIGGRAPH Asia 2018 Technical
  Papers}}. ACM, \bibinfo{pages}{217}.
\newblock


\bibitem[\protect\citeauthoryear{Wang, Sun, Liu, Sarma, Bronstein, and
  Solomon}{Wang et~al\mbox{.}}{2019}]%
        {wang2018dynamic}
\bibfield{author}{\bibinfo{person}{Yue Wang}, \bibinfo{person}{Yongbin Sun},
  \bibinfo{person}{Ziwei Liu}, \bibinfo{person}{Sanjay~E. Sarma},
  \bibinfo{person}{Michael~M. Bronstein}, {and} \bibinfo{person}{Justin~M.
  Solomon}.} \bibinfo{year}{2019}\natexlab{}.
\newblock \showarticletitle{Dynamic Graph CNN for Learning on Point Clouds}.
\newblock \bibinfo{journal}{\emph{ACM Transactions on Graphics (TOG)}}
  (\bibinfo{year}{2019}).
\newblock


\bibitem[\protect\citeauthoryear{Wang, Xu, Li, Zhang, Shamir, Liu, Cheng, and
  Xiong}{Wang et~al\mbox{.}}{2011a}]%
        {wang2011symmetry}
\bibfield{author}{\bibinfo{person}{Yanzhen Wang}, \bibinfo{person}{Kai Xu},
  \bibinfo{person}{Jun Li}, \bibinfo{person}{Hao Zhang}, \bibinfo{person}{Ariel
  Shamir}, \bibinfo{person}{Ligang Liu}, \bibinfo{person}{Zhiquan Cheng}, {and}
  \bibinfo{person}{Yueshan Xiong}.} \bibinfo{year}{2011}\natexlab{a}.
\newblock \showarticletitle{Symmetry hierarchy of man-made objects}. In
  \bibinfo{booktitle}{\emph{Computer Graphics Forum}},
  Vol.~\bibinfo{volume}{30}. Wiley Online Library, \bibinfo{pages}{287--296}.
\newblock


\bibitem[\protect\citeauthoryear{Wang, Xu, Li, Zhang, Shamir, Liu, Cheng, and
  Xiong}{Wang et~al\mbox{.}}{2011b}]%
        {Wang:2011:Sym}
\bibfield{author}{\bibinfo{person}{Y. Wang}, \bibinfo{person}{K. Xu},
  \bibinfo{person}{J. Li}, \bibinfo{person}{H. Zhang}, \bibinfo{person}{A.
  Shamir}, \bibinfo{person}{L. Liu}, \bibinfo{person}{Z. Cheng}, {and}
  \bibinfo{person}{Y. Xiong}.} \bibinfo{year}{2011}\natexlab{b}.
\newblock \showarticletitle{Symmetry Hierarchy of Man-Made Objects}.
\newblock \bibinfo{journal}{\emph{Computer Graphics Forum}}
  \bibinfo{volume}{30}, \bibinfo{number}{2} (\bibinfo{year}{2011}),
  \bibinfo{pages}{287--296}.
\newblock
\urldef\tempurl%
\url{https://doi.org/10.1111/j.1467-8659.2011.01885.x}
\showDOI{\tempurl}


\bibitem[\protect\citeauthoryear{Wu, Zhang, Xue, Freeman, and Tenenbaum}{Wu
  et~al\mbox{.}}{2016}]%
        {wu2016learning}
\bibfield{author}{\bibinfo{person}{Jiajun Wu}, \bibinfo{person}{Chengkai
  Zhang}, \bibinfo{person}{Tianfan Xue}, \bibinfo{person}{Bill Freeman}, {and}
  \bibinfo{person}{Josh Tenenbaum}.} \bibinfo{year}{2016}\natexlab{}.
\newblock \showarticletitle{Learning a probabilistic latent space of object
  shapes via 3d generative-adversarial modeling}. In
  \bibinfo{booktitle}{\emph{Advances in neural information processing
  systems}}. \bibinfo{pages}{82--90}.
\newblock


\bibitem[\protect\citeauthoryear{Wu, Wang, Lin, Lischinski, Cohen-Or, and
  Huang}{Wu et~al\mbox{.}}{2019}]%
        {wu2018structure}
\bibfield{author}{\bibinfo{person}{Zhijie Wu}, \bibinfo{person}{Xiang Wang},
  \bibinfo{person}{Di Lin}, \bibinfo{person}{Dani Lischinski},
  \bibinfo{person}{Daniel Cohen-Or}, {and} \bibinfo{person}{Hui Huang}.}
  \bibinfo{year}{2019}\natexlab{}.
\newblock \showarticletitle{SAGNet: Structure-aware Generative Network for
  3D-Shape Modeling}.
\newblock \bibinfo{journal}{\emph{ACM Transactions on Graphics (Proceedings of
  SIGGRAPH 2019)}} \bibinfo{volume}{38}, \bibinfo{number}{4}
  (\bibinfo{year}{2019}), \bibinfo{pages}{91:1--91:14}.
\newblock


\bibitem[\protect\citeauthoryear{Xie, Xu, Liu, and Xiong}{Xie
  et~al\mbox{.}}{2014}]%
        {xie20143d}
\bibfield{author}{\bibinfo{person}{Zhige Xie}, \bibinfo{person}{Kai Xu},
  \bibinfo{person}{Ligang Liu}, {and} \bibinfo{person}{Yueshan Xiong}.}
  \bibinfo{year}{2014}\natexlab{}.
\newblock \showarticletitle{3d shape segmentation and labeling via extreme
  learning machine}. In \bibinfo{booktitle}{\emph{Computer graphics forum}},
  Vol.~\bibinfo{volume}{33}. Wiley Online Library, \bibinfo{pages}{85--95}.
\newblock


\bibitem[\protect\citeauthoryear{Xu, Hu, Leskovec, and Jegelka}{Xu
  et~al\mbox{.}}{2019}]%
        {Xu:2019:GIN}
\bibfield{author}{\bibinfo{person}{Keyulu Xu}, \bibinfo{person}{Weihua Hu},
  \bibinfo{person}{Jure Leskovec}, {and} \bibinfo{person}{Stefanie Jegelka}.}
  \bibinfo{year}{2019}\natexlab{}.
\newblock \showarticletitle{How Powerful are Graph Neural Networks?}. In
  \bibinfo{booktitle}{\emph{International Conference on Learning
  Representations (ICLR)}}.
\newblock


\bibitem[\protect\citeauthoryear{Yan, Yang, Yumer, Guo, and Lee}{Yan
  et~al\mbox{.}}{2016}]%
        {yan2016perspective}
\bibfield{author}{\bibinfo{person}{Xinchen Yan}, \bibinfo{person}{Jimei Yang},
  \bibinfo{person}{Ersin Yumer}, \bibinfo{person}{Yijie Guo}, {and}
  \bibinfo{person}{Honglak Lee}.} \bibinfo{year}{2016}\natexlab{}.
\newblock \showarticletitle{Perspective transformer nets: Learning single-view
  3d object reconstruction without 3d supervision}. In
  \bibinfo{booktitle}{\emph{Advances in Neural Information Processing
  Systems}}. \bibinfo{pages}{1696--1704}.
\newblock


\bibitem[\protect\citeauthoryear{Yi, Guibas, Hertzmann, Kim, Su, and Yumer}{Yi
  et~al\mbox{.}}{2017a}]%
        {yi2017learning}
\bibfield{author}{\bibinfo{person}{Li Yi}, \bibinfo{person}{Leonidas Guibas},
  \bibinfo{person}{Aaron Hertzmann}, \bibinfo{person}{Vladimir~G Kim},
  \bibinfo{person}{Hao Su}, {and} \bibinfo{person}{Ersin Yumer}.}
  \bibinfo{year}{2017}\natexlab{a}.
\newblock \showarticletitle{Learning hierarchical shape segmentation and
  labeling from online repositories}.
\newblock \bibinfo{journal}{\emph{arXiv preprint arXiv:1705.01661}}
  (\bibinfo{year}{2017}).
\newblock


\bibitem[\protect\citeauthoryear{Yi, Kim, Ceylan, Shen, Yan, Su, Lu, Huang,
  Sheffer, and Guibas}{Yi et~al\mbox{.}}{2016}]%
        {yi2016scalable}
\bibfield{author}{\bibinfo{person}{Li Yi}, \bibinfo{person}{Vladimir~G Kim},
  \bibinfo{person}{Duygu Ceylan}, \bibinfo{person}{I Shen},
  \bibinfo{person}{Mengyan Yan}, \bibinfo{person}{Hao Su},
  \bibinfo{person}{Cewu Lu}, \bibinfo{person}{Qixing Huang},
  \bibinfo{person}{Alla Sheffer}, {and} \bibinfo{person}{Leonidas~Guibas
  Guibas}.} \bibinfo{year}{2016}\natexlab{}.
\newblock \showarticletitle{A scalable active framework for region annotation
  in 3D shape collections}.
\newblock \bibinfo{journal}{\emph{ACM Transactions on Graphics (TOG)}}
  \bibinfo{volume}{35}, \bibinfo{number}{6} (\bibinfo{year}{2016}),
  \bibinfo{pages}{210}.
\newblock


\bibitem[\protect\citeauthoryear{Yi, Su, Guo, and Guibas}{Yi
  et~al\mbox{.}}{2017b}]%
        {yi2017syncspeccnn}
\bibfield{author}{\bibinfo{person}{Li Yi}, \bibinfo{person}{Hao Su},
  \bibinfo{person}{Xingwen Guo}, {and} \bibinfo{person}{Leonidas~J Guibas}.}
  \bibinfo{year}{2017}\natexlab{b}.
\newblock \showarticletitle{Syncspeccnn: Synchronized spectral cnn for 3d shape
  segmentation}. In \bibinfo{booktitle}{\emph{Proceedings of the IEEE
  Conference on Computer Vision and Pattern Recognition}}.
  \bibinfo{pages}{2282--2290}.
\newblock


\bibitem[\protect\citeauthoryear{You, Ying, Ren, Hamilton, and Leskovec}{You
  et~al\mbox{.}}{2018}]%
        {you2018graphrnn}
\bibfield{author}{\bibinfo{person}{Jiaxuan You}, \bibinfo{person}{Rex Ying},
  \bibinfo{person}{Xiang Ren}, \bibinfo{person}{William~L Hamilton}, {and}
  \bibinfo{person}{Jure Leskovec}.} \bibinfo{year}{2018}\natexlab{}.
\newblock \showarticletitle{Graphrnn: Generating realistic graphs with deep
  auto-regressive models}.
\newblock \bibinfo{journal}{\emph{International Conference on Machine Learning
  (ICML)}} (\bibinfo{year}{2018}).
\newblock


\bibitem[\protect\citeauthoryear{Yu, Liu, Zhang, Zhu, and Xu}{Yu
  et~al\mbox{.}}{2019}]%
        {yu2019partnet}
\bibfield{author}{\bibinfo{person}{Fenggen Yu}, \bibinfo{person}{Kun Liu},
  \bibinfo{person}{Yan Zhang}, \bibinfo{person}{Chenyang Zhu}, {and}
  \bibinfo{person}{Kai Xu}.} \bibinfo{year}{2019}\natexlab{}.
\newblock \showarticletitle{PartNet: A Recursive Part Decomposition Network for
  Fine-grained and Hierarchical Shape Segmentation}. In
  \bibinfo{booktitle}{\emph{Proceedings of the IEEE conference on computer
  vision and pattern recognition (CVPR)}}.
\newblock


\bibitem[\protect\citeauthoryear{Yumer, Chaudhuri, Hodgins, and Kara}{Yumer
  et~al\mbox{.}}{2015}]%
        {yumer2015semantic}
\bibfield{author}{\bibinfo{person}{Mehmet~Ersin Yumer},
  \bibinfo{person}{Siddhartha Chaudhuri}, \bibinfo{person}{Jessica~K Hodgins},
  {and} \bibinfo{person}{Levent~Burak Kara}.} \bibinfo{year}{2015}\natexlab{}.
\newblock \showarticletitle{Semantic shape editing using deformation handles}.
\newblock \bibinfo{journal}{\emph{ACM Transactions on Graphics (TOG)}}
  \bibinfo{volume}{34}, \bibinfo{number}{4} (\bibinfo{year}{2015}),
  \bibinfo{pages}{86}.
\newblock


\bibitem[\protect\citeauthoryear{Zhao, Hu, Guerrero, Mitra, and Komura}{Zhao
  et~al\mbox{.}}{2016}]%
        {zhao2016relationship}
\bibfield{author}{\bibinfo{person}{Xi Zhao}, \bibinfo{person}{Ruizhen Hu},
  \bibinfo{person}{Paul Guerrero}, \bibinfo{person}{Niloy Mitra}, {and}
  \bibinfo{person}{Taku Komura}.} \bibinfo{year}{2016}\natexlab{}.
\newblock \showarticletitle{Relationship templates for creating scene
  variations}.
\newblock \bibinfo{journal}{\emph{ACM Transactions on Graphics (TOG)}}
  \bibinfo{volume}{35}, \bibinfo{number}{6} (\bibinfo{year}{2016}),
  \bibinfo{pages}{207}.
\newblock


\end{thebibliography}

\appendix

\section{Joint Embedding of Shapes and Images}
\begin{figure}[t!]
    \centering
    \includegraphics[width=\linewidth]{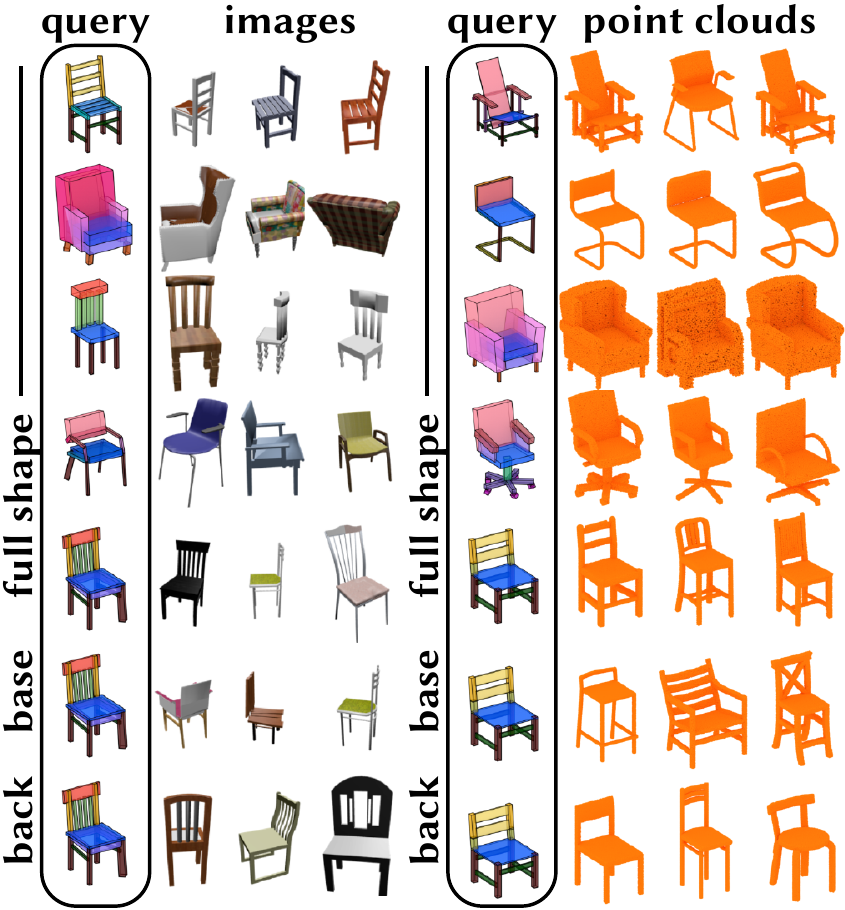}
    \caption{\titlecap{Image and point cloud retrieval.}{Images and point clouds are retrieved using on a query shape based on the distance from the query in the multi-modal latent space. The bottom three rows compare retrieval with the full shape as query to part-based retrieval using the backrest and chair base only. The retrieved shapes are similar to the query, showing that similar shapes have a small distance in latent space, even across modalities.}}
    \label{fig:retrieval}
    \vspace{-3mm}
\end{figure}

The joint embedding of multiple modalities described in Section
6.4 of our paper
%~\ref{sec:shape_abstraction}
can also be used for retrieval. For example, instead of looking up the encoded shape that is closest to an encoded image, we can look up the images that are closest to an encoded shape, and thereby get the top-k image matches for a given shape. In Figure~\ref{fig:retrieval} we show the top-3 images and point clouds for a given query shape. Qualitatively, most of the retrieved results are a good match to the query shape. 

\begin{figure*}[t!]
    \centering
    \includegraphics[width=\linewidth]{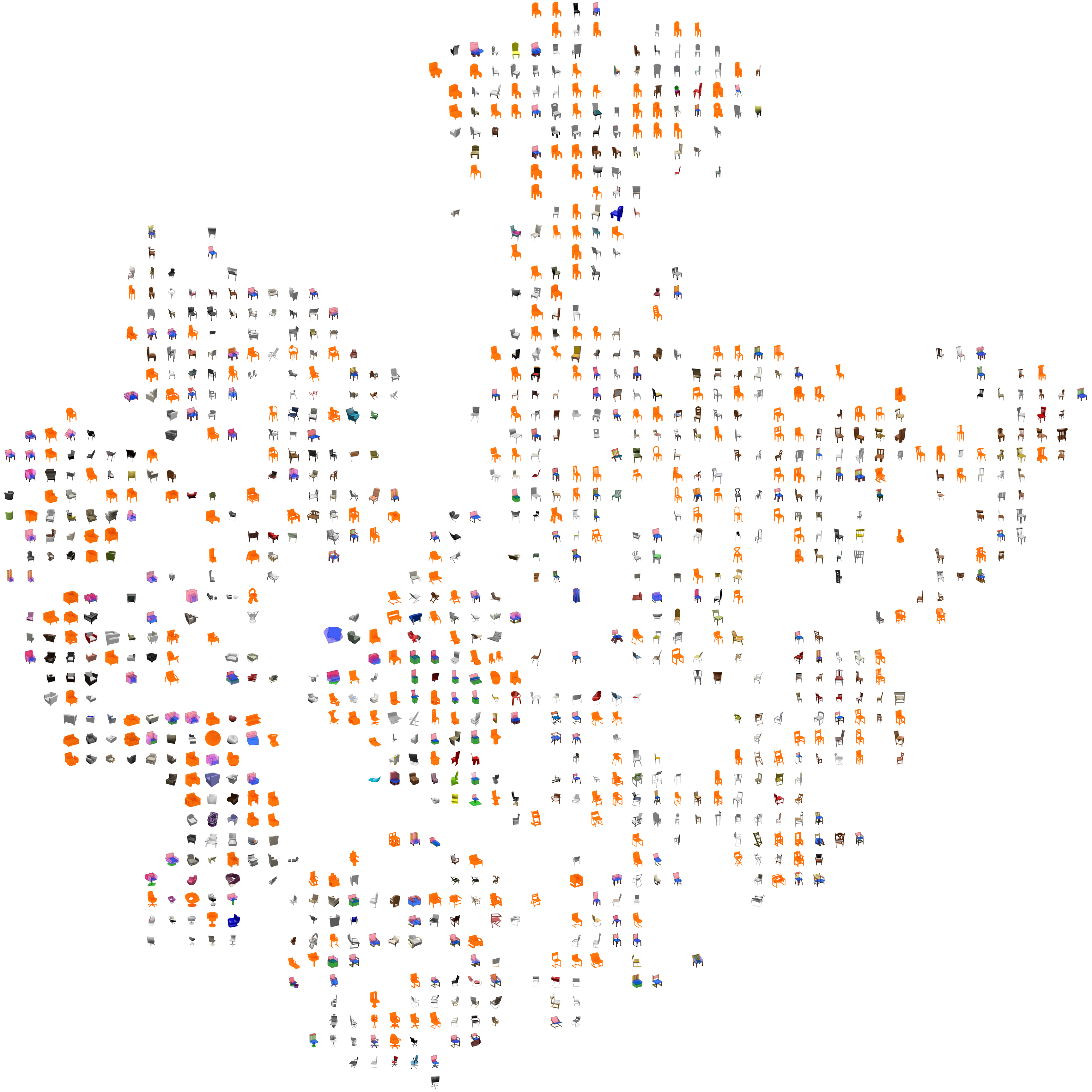}
    \caption{\titlecap{Joint embedding of images, point clouds and shapes.}{We visualize the multi-modal latent space as a two-dimensional embedding. At each grid point, we randomly show one of the modalities.}}
    \label{fig:embedding}
    \vspace{-3mm}
\end{figure*}

\paragraph{Joint embedding}
A two-dimensional t-SNE embedding~\cite{maaten2008visualizing} of the joint multi-modal latent space is shown in Figure~\ref{fig:embedding}. We show representative samples on a grid, choosing at each location randomly one of the modalities: shapes, images or point clouds. We can see that the distributions of the different modalities align well; nearby samples tend to represent similar shapes. Sofa chairs, for example, are clustered on the left side of the diagram for all modalities, and on the right side, we find chairs with backrests that have multiple vertical bars. Furthermore, 
%we can see that
%, even with the non-variational autoencoder used here \kaichun{We use VAE here! I train a VAE for retrieval and joint embedding. Train AEs for abstraction application.}, 
the learned latent space is `structurally smooth' that nearby regions tend to be connected by natural transitions between the structures of the chairs, which is also confirmed by the interpolation experiments in Section
6.3. of the paper.
%~\ref{sec:experiments_interp}.

\paragraph{Part-based retrieval}
Retrieval based on individual parts, for example, retrieving chairs with backrests similar to a query shape, can be done by training a separate encoder for each part type that we want to retrieve. The bottom three rows of Figure~\ref{fig:retrieval} show part-based retrieval results for the base and backrest of chairs, compared to performing a retrieval based on the full shape. To retrieve images with similar bases, for example, we train an encoder similar to Section
6.4 of our paper,
%~\ref{sec:shape_abstraction},
but trained to using the latent space of chair bases only instead of full chairs. Unlike the latent space of the full shape, the latent space of parts is not specifically regularized to be smooth. Still, we can see from the successful retrieval results, that the latent space of individual parts tends to be meaningful, where feature vectors that have a small distance in latent space correspond to similar parts.

\section{Ablation study}
We performed an ablation study to evaluate the contribution of individual components to our method for the shape reconstruction experiments. Results are shown in Table~\ref{tab:ablation}. Specifically, we trained $5$ variations of our method, removing a combination of components in each. Components we examined are the \emph{message passing} performed in the decoder, where, different from the encoder, it is not strictly necessary to handle relationship edges, the \emph{normal reconstruction loss} $\mathcal{L}_{\text{normal}}$, and the \emph{structure consistency loss} $\mathcal{L}_{\text{sc}}$. The normal reconstruction loss noticeably decreases the geometry reconstruction error $E_{\mathbf{P}}$ and together with the structure consistency loss $\mathcal{L}_{\text{sc}}$, lowers the consistency errors. The normal and structure consistency losses come at the cost of a slightly increased hierarchy error $E_{\mathbf{H}}$, presumably since these losses encourage the network to focus more resources on the part geometry, as opposed to the hierarchy. This cost is reduced by message passing, which significantly lowers the hierarchy error.
Finally, we also compare to removing edges all-together, which results in a significant increase in the geometry reconstruction error.

\begin{table}[h!]
\caption{\titlecap{Ablation study.}{We compare our full method (bottom row) to a version without combinations of message passing (mp), the normal loss $\mathcal{L}_{\text{normal}}$ (nl), and the structure consistency loss $\mathcal{L}_{\text{sc}}$ (scl). In the top row we show a version that does not use relationship edges. The normal and edge loss both increase consistency significantly, at a small cost in the hierarchy reconstruction. Message passing improves coordination between parts, reducing this cost.}}
\label{tab:ablation}
\begin{tabular}{@{}lrrrrrr@{}}
\toprule
\multirow{2}{*}{}              & \multicolumn{3}{c}{reconstruction err.} & \phantom{abc} & \multicolumn{2}{c}{consistency err.} \\
\cmidrule{2-4} \cmidrule{6-7}
                               & \multicolumn{1}{c}{$E_{\mathbf{P}}$}  & \multicolumn{1}{c}{$E_{\mathbf{H}}$} & \multicolumn{1}{c}{$E_{\mathbf{R}}$} && \multicolumn{1}{c}{$E_{\text{rc}}$}          & \multicolumn{1}{c}{$E_{\text{gc}}$}           \\
\midrule
no edges             & 0.0662   & 0.194    &          &&              & 0.0288      \\
- (mp, scl, nl) & 0.0649   & 0.192    & 0.240    && 0.0323       & 0.0365      \\
- (mp, scl)        & 0.0616   & 0.198    & 0.243    && 0.0216       & 0.0259      \\
- (nl)             & 0.0631   & 0.201    & 0.254    && 0.0323       & 0.0380      \\
- (scl)             & 0.0649   & 0.201    & 0.249    && 0.0194       & 0.0242      \\
- (mp)         & 0.0621   & 0.212    & 0.250    && 0.0186       & 0.0223      \\
\name                & 0.0620   & 0.200    & 0.246    && 0.0183       & 0.0226     \\
\bottomrule
\end{tabular}
\vspace{-3mm}
\end{table}

\section{Implementation}

We implement \name in PyTorch~\cite{paszke2017automatic}. All sub-networks of our hierarchical graph networks are implemented as simple Multilayer Perceptrons (MLPs) with ReLU non-linearities, and without batch normalization~\cite{ioffe2015batch}, except for the specialized encoders for images and unannotated point clouds, and the pre-trained point cloud autoencoder for the part geometry. We use a batch size of $32$ shapes. Due to the difficulty of batched training with recursive networks, we compute the loss for each shape separately before summing the per-shape losses up to obtain the loss for the batch. Back-propagation is performed on the batch loss.
%We use a batch size of one, due to the difficulty of batched training with recursive networks.
Typically, our networks for bounding box geometry converge in $1-2$ days, whereas the networks for point cloud geometry require $2-4$ days to train on a single GeForce RTX 2080 Ti and an Intel i9-7940X CPU. Memory consumption is at approximately $1-2$ GB.
\section{Semantic Hierarchies}

\begin{figure*}[t!]
    \centering
    \includegraphics[width=\linewidth]{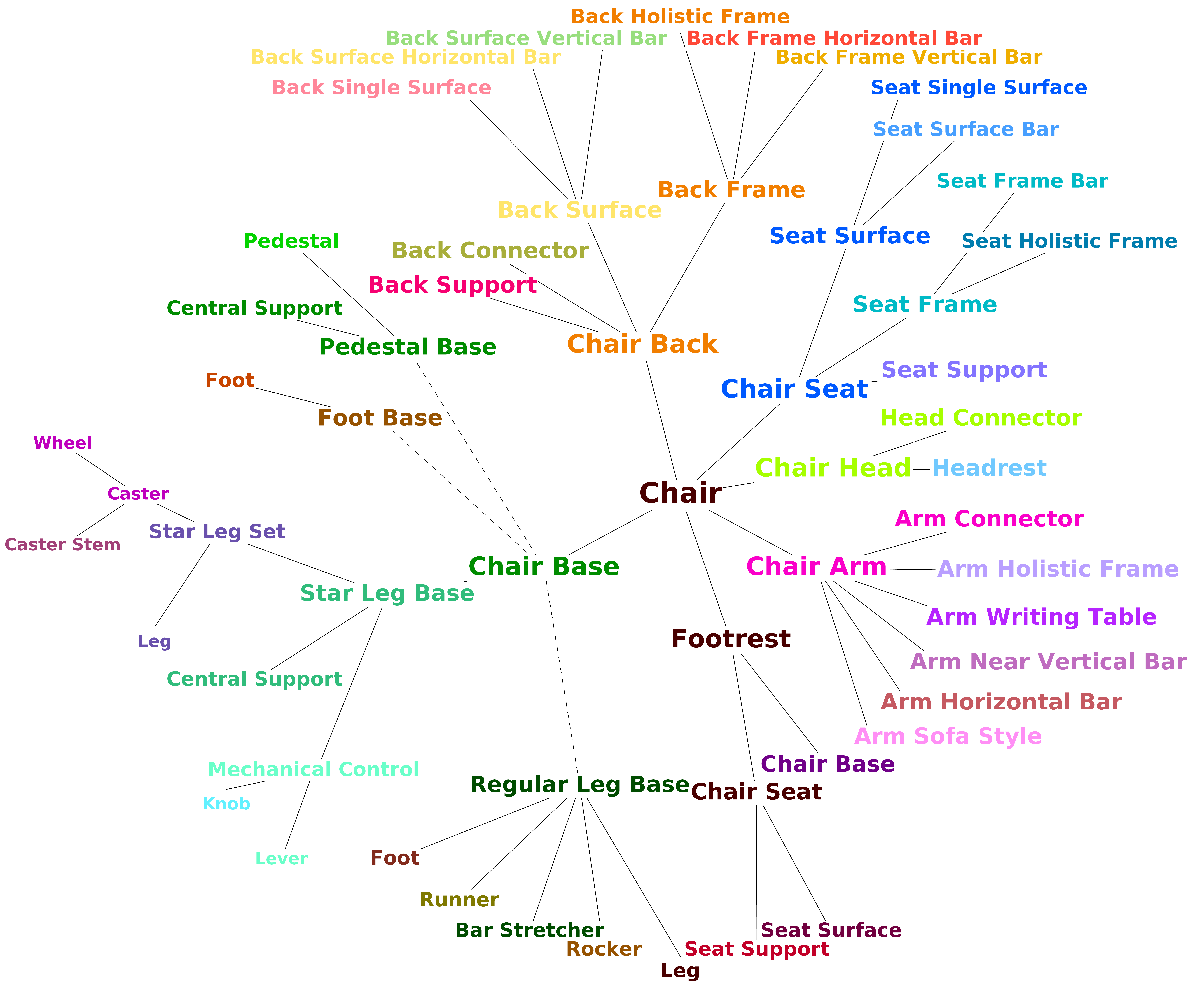}
    \caption{\titlecap{PartNet semantic hierarchy for chairs.}{Dash lines show the OR-nodes and solid lines show the AND-node in PartNet. We assign the semantic labels in the figures with the colors that we use for box-shape and point cloud visualization in the main paper.}}
    \label{fig:chair_semhier}
    \vspace{-3mm}
\end{figure*}

\begin{figure*}[t!]
    \centering
    \includegraphics[width=\linewidth]{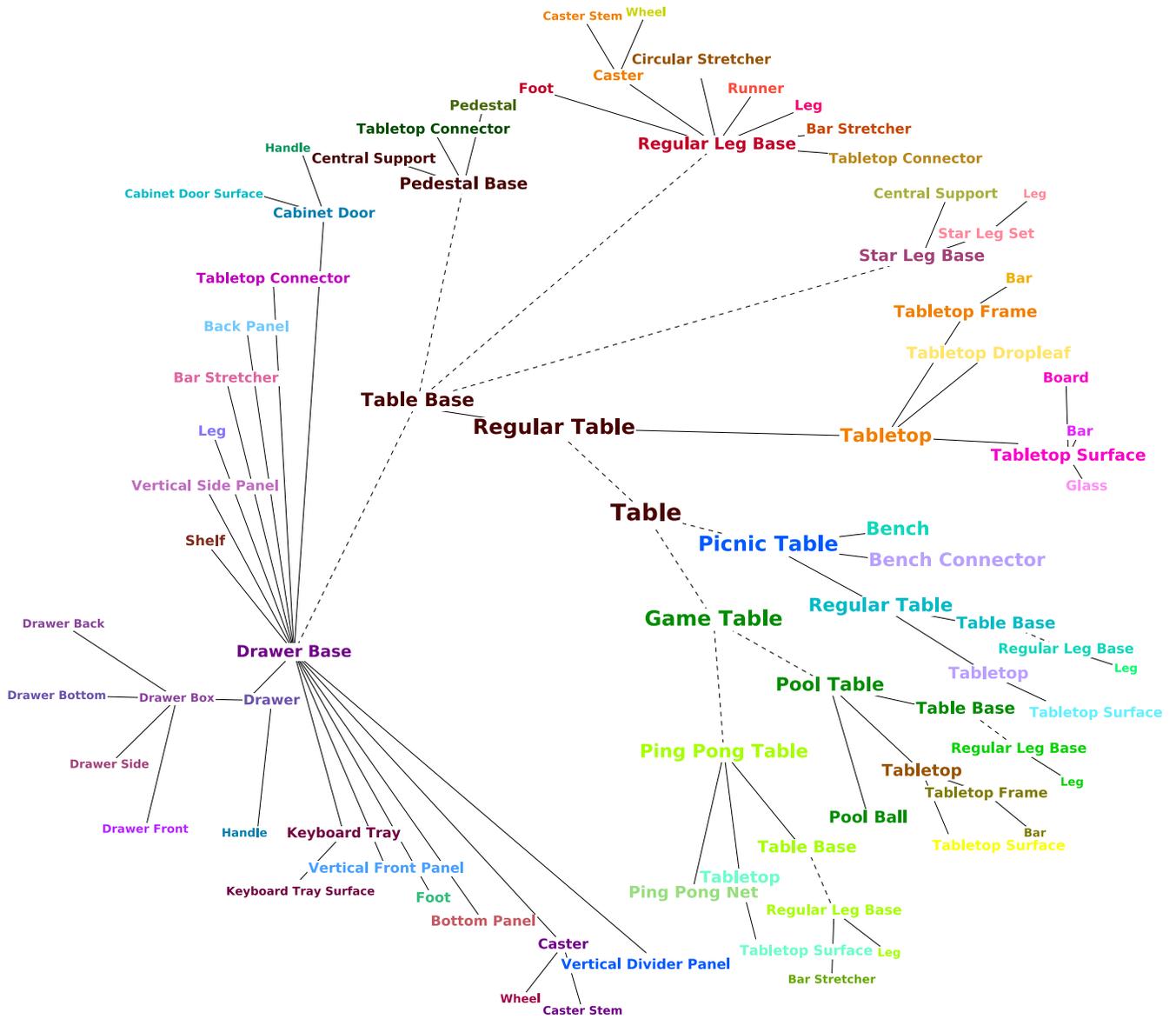}
    \caption{\titlecap{PartNet semantic hierarchy for tables.}{Dash lines show the OR-nodes and solid lines show the AND-node in PartNet. We assign the semantic labels in the figures with the colors that we use for box-shape and point cloud visualization in the main paper.}}
    \label{fig:table_semhier}
    \vspace{-3mm}
\end{figure*}

\begin{figure*}[t!]
    \centering
    \includegraphics[width=\linewidth]{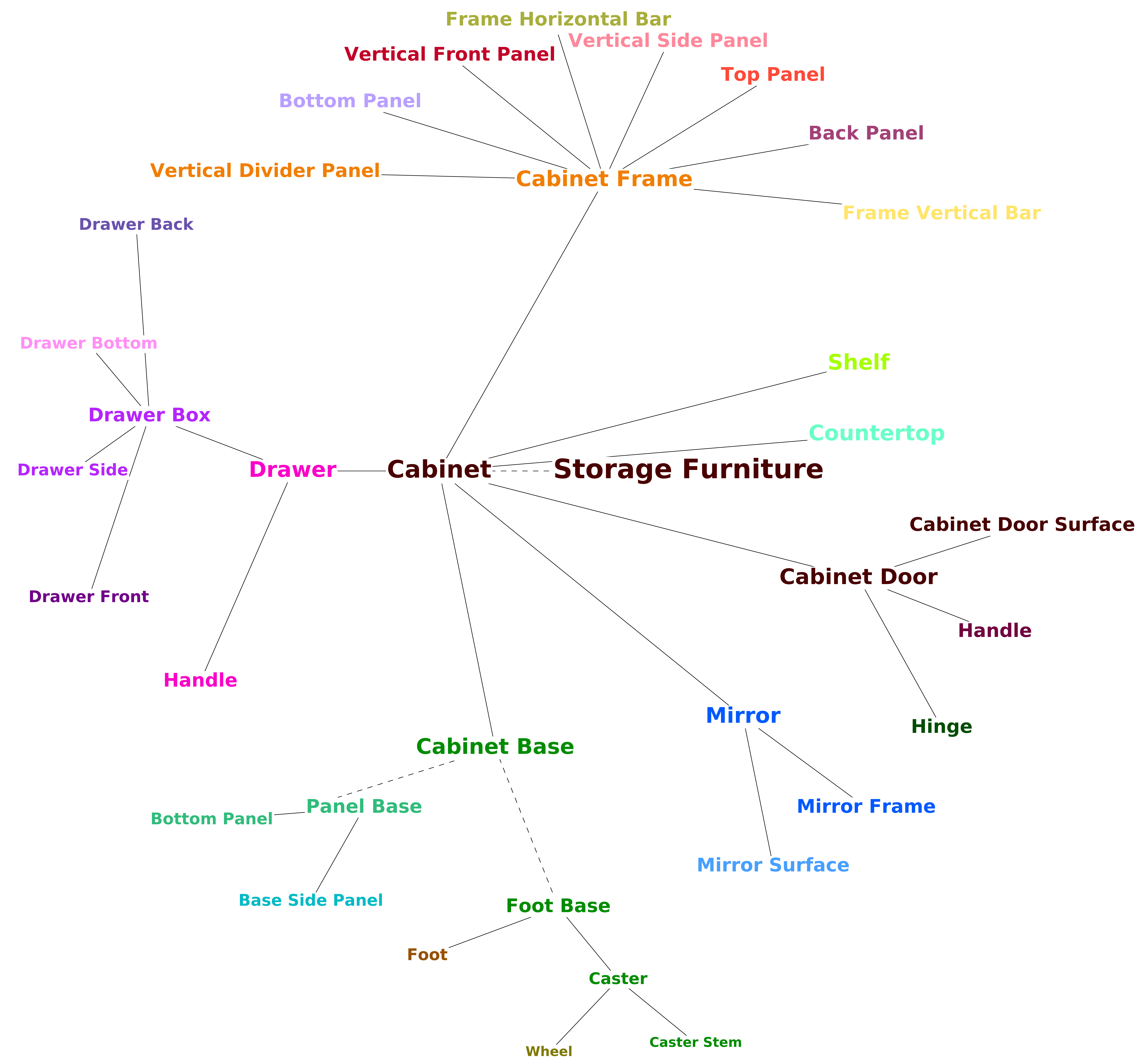}
    \caption{\titlecap{PartNet semantic hierarchy for storage furnitures.}{Dash lines show the OR-nodes and solid lines show the AND-node in PartNet. We assign the semantic labels in the figures with the colors that we use for box-shape and point cloud visualization in the main paper.}}
    \label{fig:store_semhier}
    \vspace{-3mm}
\end{figure*}

We present the PartNet~\cite{mo2018partnet} semantics hierarchies for chairs (Figure~\ref{fig:chair_semhier}), tables (Figure~\ref{fig:table_semhier}) and storage furnitures (Figure~\ref{fig:store_semhier}) that we use in this paper. We assign the semantic labels in the figures with the colors that we use for box-shape and point cloud visualization in the main paper.

\section{More Object Categories}

\begin{figure*}[t!]
    \centering
    \includegraphics[width=\linewidth]{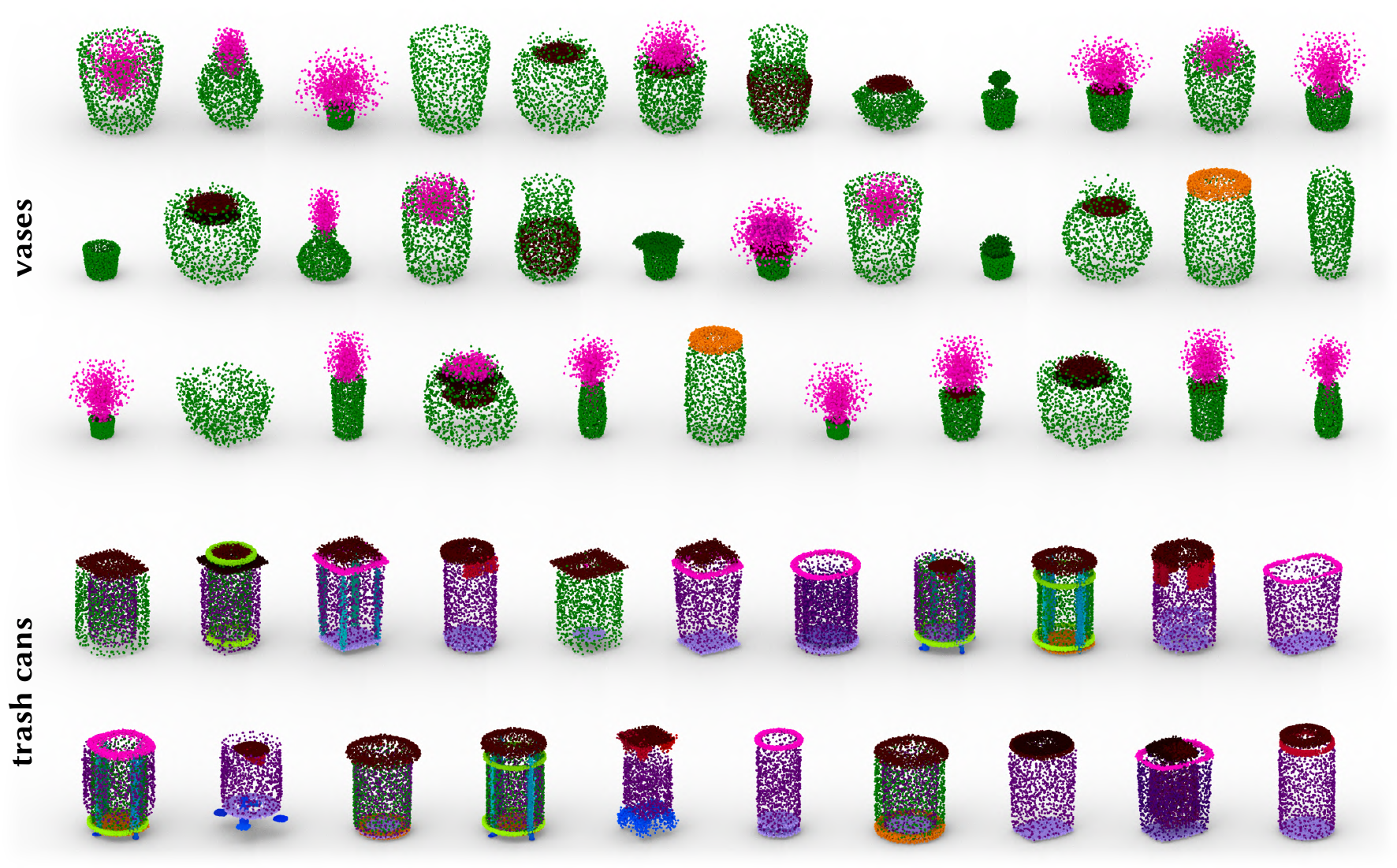}
    \caption{\titlecap{Shape generation results for vases and trash cans.}{The datasets for these categories are smaller than for our main categories: $505$ samples for vases and $83$ for trashcans. Vases have a less complex structure compared to the other categories, making the quality of the generated geometry more important, while trashcans have a wider range of structures.}}
    \label{fig:more_cats}
    \vspace{-3mm}
\end{figure*}

\begin{figure*}[t!]
    \centering
    \includegraphics[width=\linewidth]{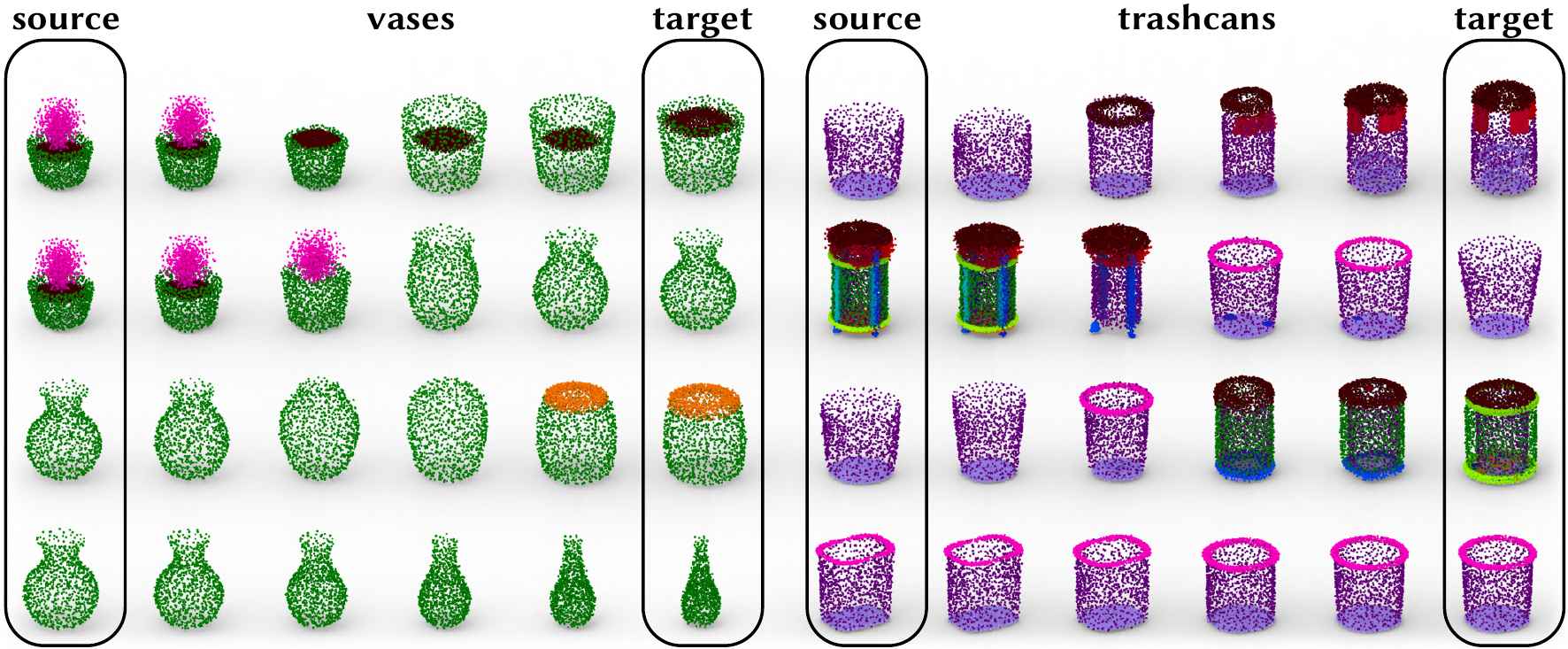}
    \caption{\titlecap{Shape interpolation results for vases and trash cans.}{Similar to our main categories, structure is interpolated smoothly. The last rows for vases and trash cans show that the part geometry is interpolated smoothly as well.}}
    \label{fig:interp_more_cats}
    \vspace{-3mm}
\end{figure*}

\begin{figure*}[t!]
    \centering
    \includegraphics[width=\linewidth]{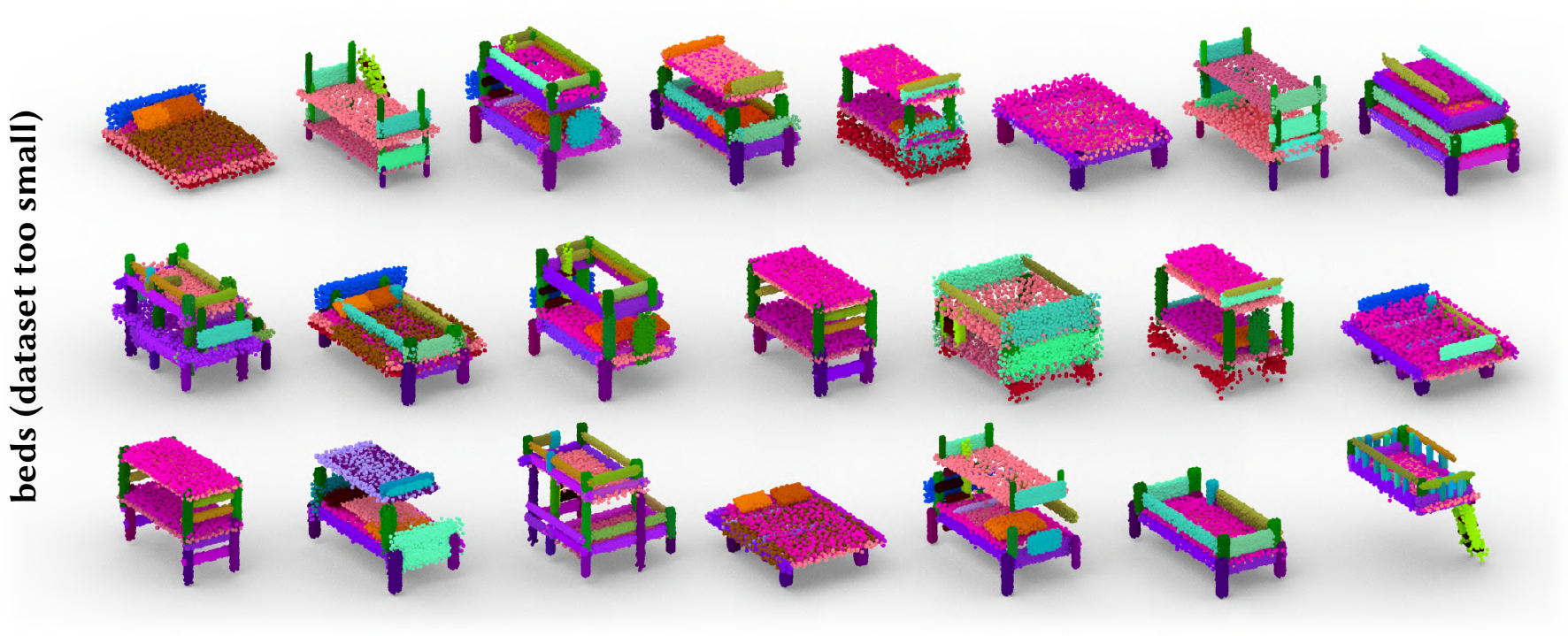}
    \caption{\titlecap{Shape generation results for beds.}{We also test on this third, severely under-sampled category, with a training set size of $54$. As we can see in Figure~\ref{fig:beds}, the network is experimenting with different structures, but the size of our dataset is not large enough for the network to reliably distinguish between realistic and unrealistic beds.}}
    \label{fig:beds}
    \vspace{-3mm}
\end{figure*}

Figures~\ref{fig:more_cats} and \ref{fig:interp_more_cats} show shape generation and interpolation results for two additional object categories in PartNet: vases and trash cans. Additionally, we show a training attempt on a severely under-sampled dataset in Figure~\ref{fig:beds}. See the captions for more detailed descriptions.

\section{Additional Generated Shapes}

\begin{figure*}[t!]
    \centering
    \includegraphics[width=\linewidth]{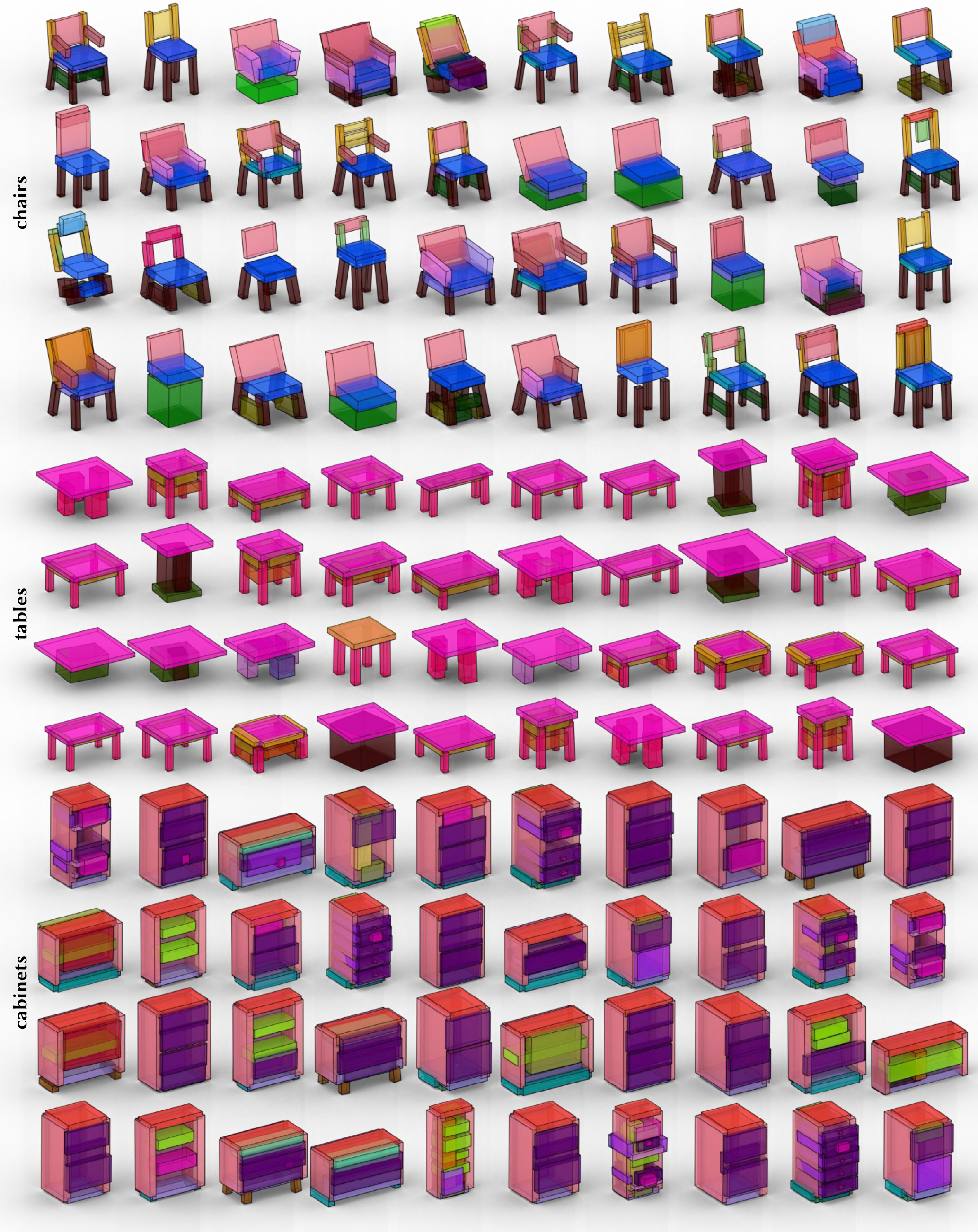}
    \caption{More Box-shape Generation Results.}
    \label{fig:box_gen}
    \vspace{-3mm}
\end{figure*}

\begin{figure*}[t!]
    \centering
    \includegraphics[width=\linewidth]{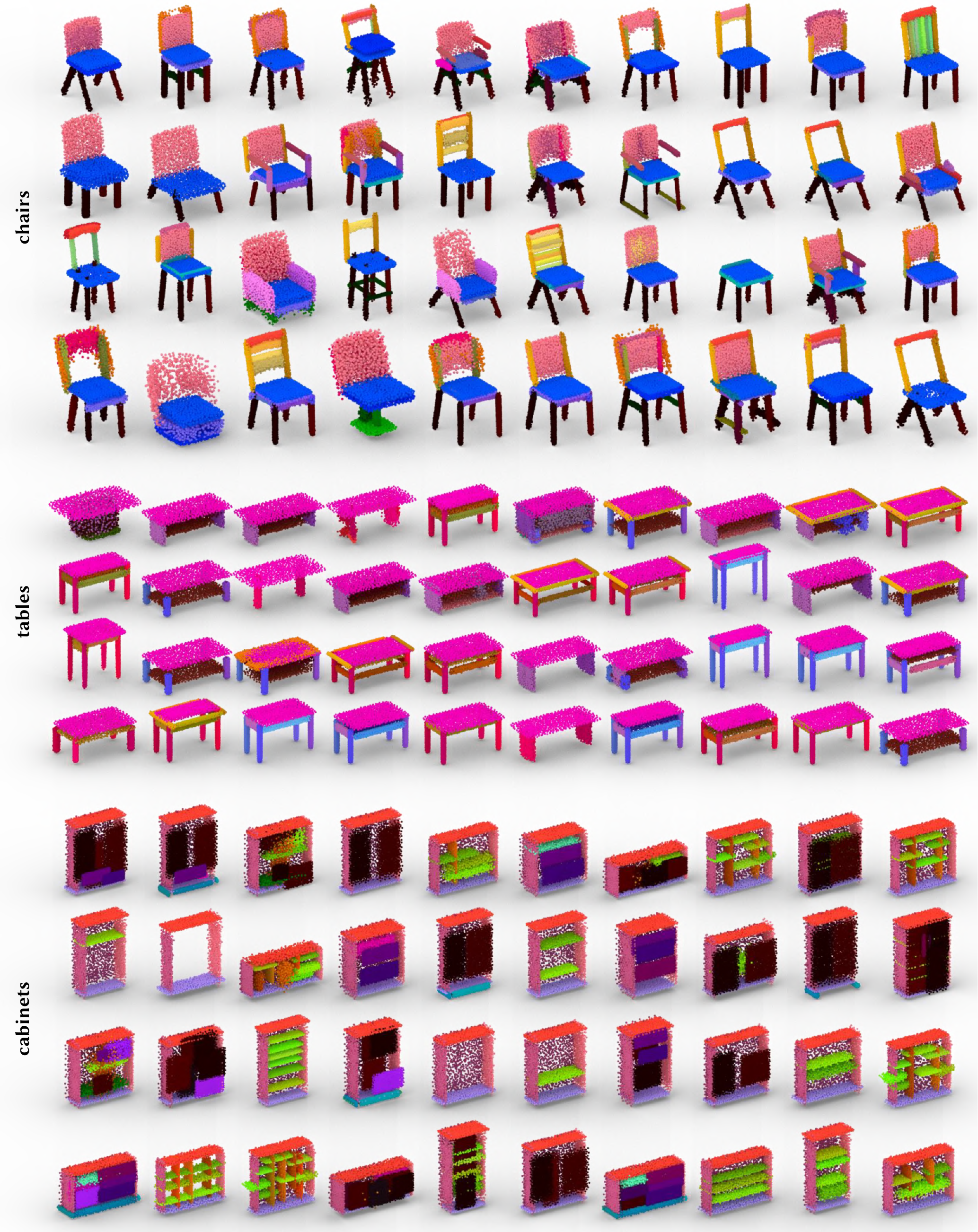}
    \caption{More Point Cloud Generation Results.}
    \label{fig:pc_gen}
    \vspace{-3mm}
\end{figure*}

We show more \name VAE generation results for box-shapes in Figure~\ref{fig:box_gen} and for point cloud shapes in Figure~\ref{fig:pc_gen}.

\section{Additional Shape Interpolations}

\begin{figure*}[t!]
    \centering
    \includegraphics[width=\linewidth]{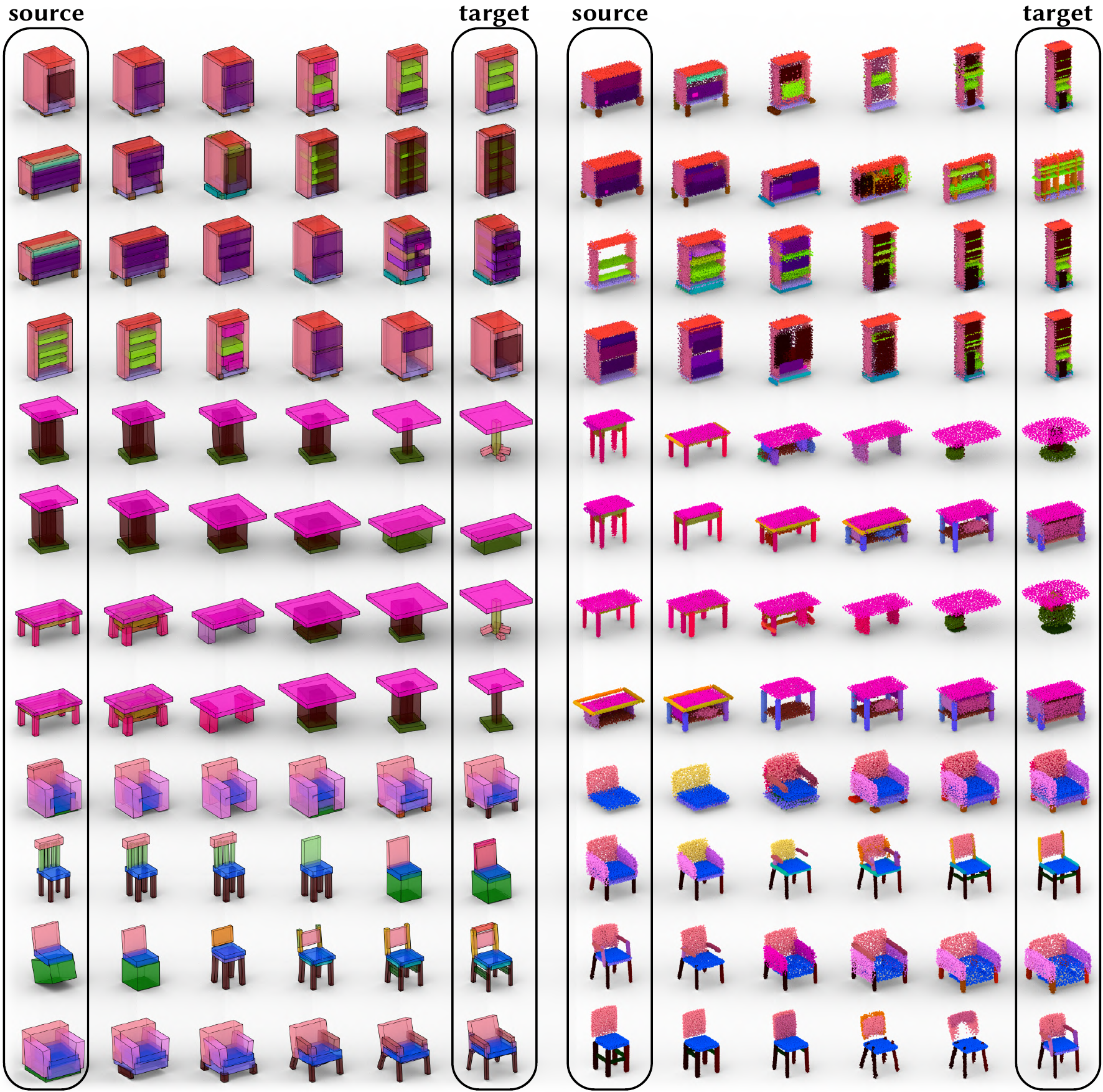}
    \caption{More Shape Interpolation Results.}
    \label{fig:supp_interp}
    \vspace{-3mm}
\end{figure*}

We show more \name VAE interpolation results for box-shapes and point cloud shapes in Figure~\ref{fig:supp_interp}.

\section{Additional Shape Abstractions}

\begin{figure*}[t!]
    \centering
    \includegraphics[width=\linewidth]{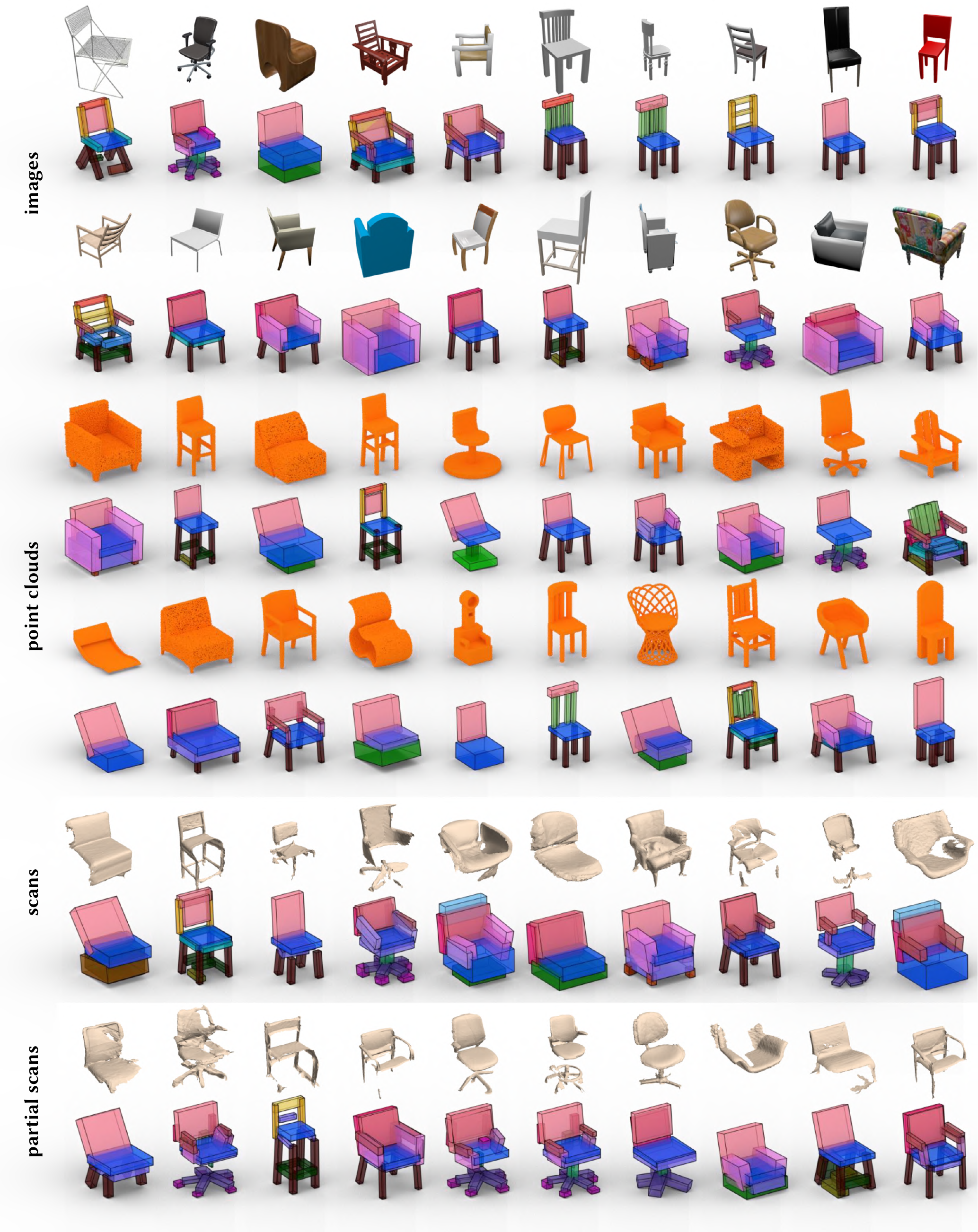}
    \caption{More Shape Abstraction Results.}
    \label{fig:supp_abstraction}
    \vspace{-3mm}
\end{figure*}

We show more \name shape abstraction results from 2D images, 3D point clouds or partial scans in Figure~\ref{fig:supp_abstraction}.

\end{document}